\documentclass[american,aps,pra,reprint,superscriptaddress, longbibliography,floatfix]{revtex4-1}
\usepackage[T1]{fontenc}
\usepackage{inputenc}
\usepackage{color}
\usepackage{babel}
\usepackage{physics}
\usepackage{mathtools}
\usepackage{bbding}
\usepackage{pifont}
\usepackage{textcomp}
\usepackage{wasysym}
\usepackage{amsthm}
\usepackage[normalem]{ulem}
\usepackage{nicefrac}
\usepackage{wasysym}
\usepackage{dsfont}
\usepackage{amsmath}
\usepackage{amssymb}
\usepackage{graphicx}
\usepackage {hyperref}
\usepackage{ulem} 
\usepackage{soul}
\setstcolor{red} 

\hypersetup{
     colorlinks   = true,
     linkcolor    = blue,
     citecolor    = blue,
     urlcolor     = blue,
     }

\usepackage{appendix}
\renewcommand{\selectlanguage}[1]{}
\makeatletter
\@ifundefined{textcolor}{}
{%
	\definecolor{BLACK}{gray}{0}
	\definecolor{WHITE}{gray}{1}
	\definecolor{RED}{rgb}{1,0,0}
	\definecolor{GREEN}{rgb}{0,1,0}
	\definecolor{BLUE}{rgb}{0,0,1}
	\definecolor{CYAN}{cmyk}{1,0,0,0}
	\definecolor{MAGENTA}{cmyk}{0,1,0,0}
	\definecolor{YELLOW}{cmyk}{0,0,1,0}
}
\theoremstyle{plain}

\theoremstyle{plain}

\ifx\proof\undefined
\newenvironment{proof}[1][\protect\proofname]{\par
	\normalfont\topsep6\p@\@plus6\p@\relax
	\trivlist
	\itemindent\parindent
	\item[\hskip\labelsep
	\scshape
	#1]\ignorespaces
}{%
	\endtrivlist\@endpefalse
}
\providecommand{\proofname}{Proof}
\fi
\theoremstyle{plain}

\providecommand{\lemmaname}{Lemma}
\providecommand{\definitionname}{Definition}
\providecommand{\propositionname}{Proposition}

\usepackage{babel}
\usepackage{txfonts}
\usepackage{colortbl}\definecolor{myurlcolor}{rgb}{0,0,0.7}

\def\ket#1{| #1 \rangle}
\def\bra#1{\langle  #1 |}

\newcommand{\haH}

\definecolor{bka}{rgb}{1,0,0.5647}

\definecolor{orange}{RGB}{255,127,0}

\begin{document}

\title{Enhancing the Performances of Autonomous Quantum Refrigerators via Two-Photon Transitions}

\author{Brij Mohan}
\affiliation{Department of Physical Sciences, Indian Institute of Science Education and Research (IISER), Mohali, Punjab 140306, India}
\affiliation{Nano and Molecular Systems Research Unit, University of Oulu, Oulu FI-90014, Finland}

\author{Bijay Kumar Agarwalla}
\affiliation{Department of Physics, Indian Institute of Science Education and Research, Pune 411008, India}

\author{Manabendra Nath Bera}
\email{mnbera@gmail.com}
\affiliation{Department of Physical Sciences, Indian Institute of Science Education and Research (IISER), Mohali, Punjab 140306, India}

\begin{abstract}
Conventional autonomous quantum refrigerators rely on uncorrelated heat exchange between the working system and baths via two-body interactions enabled by single-photon transitions and positive-temperature work baths, inherently limiting their cooling performance. Here, we introduce distinct qutrit refrigerators that exploit correlated heat transfer via two-photon transitions with the hot and cold baths, yielding a genuine enhancement in performance over conventional qutrit refrigerators that employ uncorrelated heat transfer. These refrigerators achieve at least a twofold enhancement in cooling power and reliability compared to conventional counterparts. Moreover, we show that cooling power and reliability can be further enhanced simultaneously by several folds, even surpassing existing cooling limits, by utilizing a synthetic negative-temperature work bath. Such refrigerators can be realized by combining correlated heat transfer and synthetic work baths, which consist of a four-level system coupled to hot and cold baths and two conventional work baths via two independent two-photon transitions. Here, the composition of two work baths effectively creates a synthetic negative-temperature work bath under suitable parameter choices. Additionally, our autonomous refrigerators with negative temperature baths significantly outperform previously studied autonomous and non-autonomous refrigerators in terms of cooling ability without requiring any additional energy resources, as they cool the cold bath to much lower temperature, which is forbidden for others refrigerators. Our results demonstrate that correlated heat transfers and baths with negative temperatures can yield thermodynamic advantages in quantum devices. Finally, we discuss the experimental feasibility of the proposed refrigerators across various existing platforms.

\end{abstract}

\maketitle

In recent years, quantum thermal devices have become a focal point of research as they are playing a crucial role in advancing our understanding of thermodynamics at the quantum scale and in the development of emerging quantum technologies~\cite{Binder2018, DeffnerBook2019, Alexia2022, Myers2022, Mukherjee2024, Vinjanampathy01102016, Alicki2018, Chattopadhyay2025}. These devices offer valuable insights into managing and controlling energy flows utilizing the second law of thermodynamics at the microscopic level, which is essential for effectively implementing practical quantum technologies. Quantum thermal devices can be categorized based on their functionality, including heat engines~\cite{Scovil1959, Kosloff2014, Mohan2024}, refrigerators~\cite{Palao2001, Linden2010, Levy2012, Correa2014, Kosloff2014, Singh2020T, Ghoshal2021, Mohanta2022}, diode~\cite{Yang2023}, transistors~\cite{Joulain2016, Ghosh2021, Gupt2022}, clocks~\cite{Erker2017}, rectifies~\cite{Santiago-García2025}, batteries~\cite{Sergi2020, Mohan2021, Mohan2022} etc, each serving distinct roles in energy conversion and regulation. By leveraging quantum properties, these devices become pivotal for improving the efficiency, power, and reliability of quantum thermal devices, making them indispensable for the progress of quantum technologies. For a detailed review on various thermal devices, see Refs.~\cite{Kosloff2014, Mitchison2019, Myers2022, Cangemi2024}.

Recent studies in quantum thermodynamics focused on autonomous thermal devices, as controlling quantum systems can often become challenging in various experimental platforms \cite{Guzmán2024}. Autonomous thermal devices operate in a steady-state regime while continuously interacting with the thermal baths, and such devices do not require any external control, energy, or any additional resources. One such thermal device is an autonomous quantum absorption refrigerator where quantum systems achieve cooling by harnessing natural thermal gradients without requiring any external control or work input. These nanoscale thermal devices can establish a steady-state energy transfer from a cold bath ($c$) to a hot bath ($h$), assisted by residual heat from an additional work bath ($w$). These quantum refrigerators can be employed to reset qubits, cool quantum processors, and efficiently manage the heat generated during computational tasks. In previous studies, several models of autonomous refrigerators have been widely studied, which differ by their working medium~\cite{AR_PRL_2012, Palao2001, Linden2010, Correa2014, Mitchison2019, Mohanta2022, Cangemi2024, Mu_2017, ISS_PRE_2022, KS_PRE_2018,Gonzalez_2017,Nimmrichter2017quantumclassical,SEJ_PRB_2020,Brask2015,Correa2013}. However, all these models are operationally equivalent. Many recent experimental works have successfully realized autonomous refrigerators on experimental platforms, such as trapped ions~\cite{Maslennikov2019} and superconducting circuits~\cite{Aamir2025}.

In this work, we consider the simplest autonomous refrigerator model, consisting of a qutrit working system (medium) interacting with three distinct heat baths—hot, cold, and work at unequal temperatures~\cite{Correa2014, Mohanta2022}. In this conventional qutrit model, all the baths interact with qutrit independently via two-body interactions due to the transitions induced in the qutrit system being independent; thus, heat transfer between the qutrit system and each bath is uncorrelated. Due to such uncorrelated heat transfer, these refrigerators lead to low cooling power and significantly high fluctuation in cooling power output. Additionally, in such a refrigerator model, due to the absence of unitary driving and self-interaction within the working system, one cannot exploit the quantum resources such as energetic coherence or entanglement for the refrigeration process. Moreover, one cannot exploit causal or temporal correlations to enhance the cooling performance, as used in Ref.~\cite{Felce2020}, since doing so would render the refrigerator non-autonomous due to the requirement of measurements and outcome selection, which incur a thermodynamic work cost. Apart from this, existing autonomous refrigerators have narrow cooling windows due to strict cooling limits set by hot and work bath temperatures and their frequency modes. As shown in a recent experiment, the existing autonomous quantum refrigerator can cool the target system down to 22 mK~\cite{Aamir2025}. To this end, the fundamental question arises: How can the performance as well as cooling limits of autonomous refrigerators be enhanced?

In this article, we affirmatively address the above question by utilizing the concept of a correlated heat transfer mechanism (proposed in Ref.~\cite{Mohan2024}) in autonomous refrigerators, where the baths induce correlated (or mutually dependent) transitions in the working system, thereby reducing the stochastic nature of transitions. Continuous autonomous refrigerators operating with this mechanism can be termed quantum refrigerators with correlated heat transfer (QRCs). These refrigerators can be physically realized by considering a qutrit coherently interacting with hot and cold baths through two-photon transitions (Raman interactions~\cite{Gerry1990, Gerry1992, Wu1996}, i.e., three-body interactions between the system and baths) in the presence of an additional work bath attached via two-body interaction (one-photon transition). In contrast, analogous refrigerators with uncorrelated heat transfer (QRIs) correspond to standard autonomous quantum absorption refrigerators~\cite{Correa2014, Mohanta2022}, where a qutrit interacts incoherently (independently, through one-photon transitions) with the hot, cold, and work baths. For the same set of qutrit and bath parameters, QRCs deliver significantly higher cooling power and reliability (i.e., much lower relative fluctuations) in power compared to QRIs. In fact, the performance of QRCs can be enhanced by a minimum of two folds compared to QRIs. This enhancement is directly attributed to the presence of much higher photon flux in QRCs, which is a consequence of correlated heat transfer. Moreover, we further enhance the cooling power while minimizing its relative fluctuation (noise-to-signal ratio) for QRCs by exploiting a synthetic negative temperature work bath. Interestingly, such negative temperature baths can widen the cooling window, thus enabling the cooling of the target system (cold bath) beyond the limits imposed by positive temperature work bath. We show that such refrigerators can be realized when "two work baths" and "hot and cold baths" are attached to four-level systems via two independent three-body interactions (i.e., employing two independent two-photon transitions). The reliability of a refrigerator is often gets constrained by the inverse of entropy production, and this bound is now termed as the Thermodynamic Uncertainty Relation (TUR)~\cite{Barato2015,Gingrich2016,Seifert2019,AS_2018,Kalaee2021,Guarnieri2019,Falasco2020,Hasegawa2021,Sushant_TUR_thermal,expt_TUR}. Therefore, we also discuss the implication of TUR in the context of refrigerator models \cite{LS_PRE_2021} considered in this work. It is worthwhile to mention that by reversing the direction of heat flow, the continuous heat engine utilizing correlated heat transfer, as proposed in Ref.~\cite{Mohan2024}, can potentially be turned into a refrigerator that outperforms its counterparts due to its high coherence-generating capability. However, such refrigerators would be non-autonomous, as they rely on periodic driving to reverse the direction of heat flow. In contrast, QRCs utilize a work bath to reverse the direction of heat flow, making them autonomous and fundamentally different from the former.

The rest of the article is organized as follows. In section~\ref{model}, we introduce the generic models of qutrit autonomous quantum refrigerators with correlated and uncorrelated heat transfers, respectively. In section~\ref{sec:Advs}, we demonstrate the genuine enhancements in performances by refrigerators with correlated heat transfer than the refrigerators with uncorrelated heat transfer. In section~\ref{NQRC}, we further discuss the enhancement in performance of autonomous quantum refrigerators with correlated heat transfer in the presence of synthetic negative temperature work bath.  Finally, we discuss the implications of TUR for refrigerator models considered in this work and summarize our results in section~\ref{RTUR}and section~\ref{sec:Summary}, respectively. 

\begin{figure}
\centering    

\includegraphics[width=8cm,height=4.9cm]{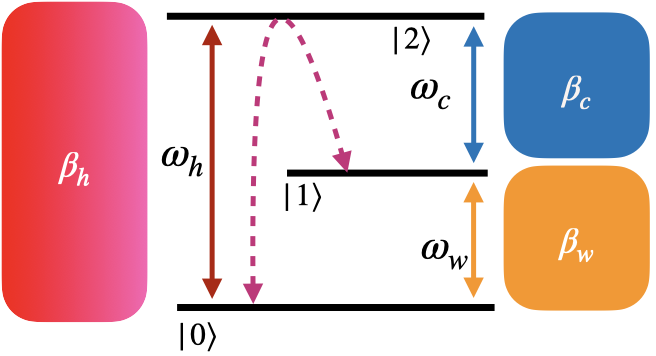}
\caption{Schematic of an autonomous quantum refrigerator with correlated and uncorrelated heat transfer. The refrigerator is constituted by a three-level quantum system (qutrit), which weakly interacts with hot, cold, and work baths with the inverse temperatures $\beta_h$, $\beta_c$, and $\beta_w$, respectively. In refrigerator with uncorrelated heat transfer (QRIs), the energy transfer takes place via (independent) single photon transitions, i.e., energy levels $\ket{0}$ and $\ket{2}$ interact with the hot bath, levels $\ket{1}$ and $\ket{2}$ interact with the cold bath and levels $\ket{1}$ and $\ket{0}$ interact with the work bath, governed by the interaction Hamiltonian, given in Eq.~\eqref{QRIsI}. Solid (red, blue, and yellow) arrows indicate these independent or uncorrelated energy transfers. In refrigerator with correlated heat transfer (QRCs), the energy transfer takes place between qutrit-hot bath-cold bath via two-photon transitions, where effectively energy levels $\ket{0}$ and $\ket{1}$ participate in the process, and absorption of a photon from the hot bath is associated with the release of a photon to the cold bath and vice versa. This correlated heat transfer is governed by the interaction Hamiltonian, given in Eq.~\eqref{RTT} and indicated here by the dashed pink arrow.}
\label{fig:QRCs}
\end{figure}

\section{Models of qutrit autonomous refrigerator with correlated and uncorrelated heat transfer}\label{model}
A continuous autonomous refrigerator is a quantum thermal device consisting of a working system that weakly interacts with three heat baths at different temperatures. One of the simplest and widely studied models of a refrigerator utilizes a qutrit system, characterized by the Hamiltonian \(H_{S} = \omega_{h} \ketbra{2} + (\omega_{h} - \omega_{c}) \ketbra{1}\),  interacting with three thermal baths, each at inverse temperatures \(\beta_c\), \(\beta_h\), and \(\beta_w\) (see Fig.~\ref{fig:QRCs}). In this configuration, the hot bath (with inverse temperature $\beta_h$) is coupled to levels $|0\rangle$ and $|2\rangle$ with energy spacing \(\omega_h\), while the cold bath (with inverse temperature $\beta_c$) is connected to levels $|1\rangle$ and $|2\rangle$ with spacing \(\omega_c\). The work bath (with inverse temperature $\beta_w$) is coupled to levels $|0\rangle$ and $|1\rangle$, with energy spacing defined by \(\omega_w = \omega_h - \omega_c\). To operate as a refrigerator, the system must transfer heat from the cold bath to the hot bath, assisted by a work bath, effectively cooling the cold bath further. For this refrigeration process to occur, the following conditions must be satisfied \cite{FS_PRE_2019, S_PRE_2018}
\begin{equation}
 \beta_w < \beta_h < \beta_c < \beta_s = \frac{\beta_h \omega_h - \beta_w \omega_w}{\omega_h - \omega_w}.
 \label{ref-cond-conve}
\end{equation}
When this condition is fulfilled, the qutrit enables energy exchanges between the baths in a manner that effectively extracts heat from the cold bath, thus lowering its temperature. This process occurs autonomously, driven purely by thermal gradients and the internal structure of the qutrit, without any external work input or control. The total Hamiltonian of the qutrit-baths composite is
\begin{align}
\label{H:QRCNs}
H=H_{S} + H_{B_h} + H_{B_c} + H_{B_w} + H_{SB_hB_cB_w}^X,  
\end{align}
where the hot, cold, and work baths are photonic (bosonic) thermal baths that are modeled as a collection of infinite dimensional systems and can be described by the Hamiltonians  
$ H_{B_h} = \sum_k \Omega_{k,h} \ a^\dagger_{k,h} a_{k,h} $,  $ H_{B_c} = \sum_{k'} \Omega_{k',c} a^\dagger_{k',c} a_{k',c} $, and $ H_{B_w} = \sum_{k''} \Omega_{k'',w} a^\dagger_{k'',w} a_{k'',w} $, respectively, where \( \Omega_{k,h} \), \( \Omega_{k',c} \), and \( \Omega_{k'',w} \) are the mode frequencies of the respective baths. The interaction between the qutrit and the baths is represented by \( H_{SB_hB_cB_w}^X \), which can be modeled in two different ways ($X=\{C, I\}$, see below). We assume $\hbar = k_B = 1$ throughout this work.

{\it Autonomous Quantum Refrigerators with correlated heat transfer (QRCs)}---
In this model, hot and cold baths interact with the qutrit (working system) via three-body interaction (e.g., Raman interaction enabled via two-photon transition~\cite{Gerry1990, Gerry1992, Wu1996}), and the work bath interacts with the qutrit independently via two-body interaction (one-photon transition), i.e., 
\begin{equation}\label{RTT}
H_{SB_hB_cB_w}^C= H_{SB_hB_c}+H_{SB_w}, 
\end{equation}
where $H_{SB_hB_c}= g_{hc} \sum_{k,k'} (a_{k,h} a_{k',c}^\dag b_{hc}^\dag + a_{k,h}^\dag a_{k',c} b_{hc})$ and $H_{SB_w}= g_w \sum_k (a_{k'',w} b^\dag_w + a_{k'',w}^\dag b_{w})$ with $b_{hc}=\ketbra{0}{1}$ and $b_{w} = \ketbra{0}{1}$ are the ladder operator acting on the qutrit space. Here, $g_{x}$'s are system-bath coupling strengths. It is important to note that when the working system (qutrit) interacts with hot and cold baths via Raman interaction, the heat transfers (energy exchanges) between "working system and hot bath" and "working system and cold bath" become dependent, i.e., correlated. Important to note that the term "correlated" stands for a simultaneous (or associated) process of photon absorption from a hot bath and photon emission to a cold bath or vice versa.
Thus, we term this simultaneous heat exchange process as correlated heat exchange between the system, the hot bath, and the cold bath. It should be noted that the work bath does not participate in this correlated heat transfer between the system and the hot and cold baths.
For very weak system-baths coupling ($g_{hc}$), the local dynamics of the qutrit can be described via a Lindblad quantum master equation given as 
\begin{align}\label{Lqrc}
    \dot{\rho} = i \ [\rho, \ H_{S}] + \mathcal{D}_{hc} (\rho)+ \mathcal{D}_{w} (\rho),
\end{align}
for a qutrit state $\rho$, where the dissipators are given by,
\begin{align}
    \mathcal{D}_{hc}(\rho) &=\! \gamma_{1} (b_{hc} \rho b_{hc}^{\dag} \! -\!\frac{1}{2}\{b_{hc}^{\dag}b_{hc},\rho\}) \!+\! \gamma_{2} (b_{hc}^{\dag} \rho b_{hc} \!-\! \frac{1}{2}\{b_{hc} b_{hc}^{\dag},\rho\}),\nonumber \\
    \ \mbox{and} \  \nonumber\\
      \mathcal{D}_{w}(\rho) &=\! \gamma_{3} (b_{w} \rho b_{w}^{\dag}  \!-\!\frac{1}{2}\{b_{w}^{\dag}b_{w},\rho\}) \!+\! \gamma_{4} (b_{w}^{\dag} \rho b_{hc} - \frac{1}{2}\{b_{w} b_{w}^{\dag},\rho\}),\nonumber
\end{align}
with decay rates $\gamma_1=\gamma_{hc} n_c(n_h+1)$, $\gamma_2=\gamma_{hc}n_h(n_c+1)$, $\gamma_3=\gamma_{w} (n_w+1)$ and $\gamma_4=\gamma_{w}n_w$ (see Appendix~\ref{appsec:QRCsteaystate} for the details). Here, $\{Y,Z\} = YZ + ZY$ is the anti-commutator, $n_{x} = 1/(e^{\beta_{x}\omega_x}-1)$ is the average number of photons in the bath with frequency $\omega_x$, and $\gamma_{hc}$ and $\gamma_{w}$ are the Weiskopf-Wigner decay constants. The dissipator $\mathcal{D}_{hc}$ involves the parameters of both hot and cold baths and induces dissipation utilizing the levels $\ket{0}$ and $\ket{1}$. Effectively, the level $\ket{2}$ is never ``engaged'' in the process. Due to the nature of the interaction between qutrit and hot and cold baths, the heat exchange with baths (hot and cold) is correlated. Hence, the energy (heat) transfer between the baths and the qutrit is less random (i.e., involves less stochastic transitions).

{\it Autonomous Quantum Refrigerators with uncorrelated heat transfer (QRIs)}---The conventional qutrit quantum heat refrigerators can be regarded as the counterparts of QRCs because they utilize uncorrelated energy transfers between working systems and baths, i.e., heat exchanges between qutrit with each bath are independent (uncorrelated), for example, see~\cite{Correa2014, Kosloff2014, Mitchison2019, S_PRE_2018, FS_PRE_2019, Mohanta2022}. In this model, all the baths interact with the working system independently via two-body interaction (one-photon transition), i.e.,
\begin{equation}\label{QRIsI}
H_{SB_hB_hB_w}^I= H_{SB_h}+H_{SB_c}+H_{SB_w},
\end{equation}
where $H_{SB_x}= g_x \sum_k (a_{k,h} b^\dag_x + a_{k,h}^\dag b_{x})$ (with  $x$ = \{$h$, $c$, $w$\}), $b_{h} = \ketbra{0}{2}$, $b_{c} =\ketbra{1}{2}$ and $b_{w} = \ketbra{0}{1}$ are the ladder operator acting on the qutrit space. The coefficients $g_h$, $g_c$, and $g_w$ are the interaction strengths with the hot, cold, and work baths, respectively. The interaction drives uncorrelated energy (heat) transfer in the sense that the energy exchange between the "working system and hot bath" is independent of the energy exchange between the "working system and cold bath", unlike QRCs. In the limit of weak system-baths couplings ($g_h$, $g_c$, and $g_w$), the local dynamics of the qutrit is expressed by the Lindblad master equation with three independent dissipators corresponding to hot, cold, and work baths. The appearance of three dissipators in the Lindblad master equation reflects that the heat exchange between qutrit and one bath is independent (or uncorrelated) of the heat exchange between qutrit and other baths. These refrigerators have previously been studied extensively in the literature. In Appendix~\ref{appsec:QRIsteaystate}, we provide certain details for the completeness.

The average currents corresponding to each bath can be calculated using  \(
\langle {J}_{x}^X \rangle = \Tr(\mathcal{D}_{x}(\sigma_X)H_{S})  
\), if the corresponding dissipators are uncorrelated, where $\sigma_X$ is the steady state density matrix. However, for the correlated case, the average heat currents \(\langle {J}_x^X \rangle\) cannot be directly quantified because there are no independent dissipators associated with each bath. To address this, we employ the full counting statistics of the steady-state dynamics (see Appendix~\ref{appsec:FCS} for the details) \cite{Friedman_2018}. This approach also allows us to compute the fluctuations in currents (\(\Delta J_x^X\)) corresponding to each bath. As per our convention, the average currents entering the working system are positive, \(\langle {J}_x^X \rangle > 0\), while those leaving the system are negative, \(\langle {J}_x^X \rangle < 0\). Moreover, in the considered models of refrigerators, the average and variance of currents are proportional to the average photon flux and its variance, respectively.

\section{Enhancements in Refrigeration via correlated heat transfer \label{sec:Advs}}
In this section, we will consider these two fundamentally different models, QRCs and QRIs, which utilize distinct heat transfer mechanisms, and compare their figures of merits in the steady state regime.

The evaluation of the performance of autonomous quantum refrigerators requires a comprehensive analysis of three metrics:  (i) cooling power (current corresponding to cold bath), which is the rate of cooling; (ii) reliability of a refrigerator, measured by noise-to-signal ratio (NSR) in cooling power and therefore signifies the relative fluctuation or inverse of precision in the cooling power, and (iii) coefficient of performance, which signifies how efficiently heat is being withdrawn from the cold bath. We compare these metrics for QRCs and QRIs and demonstrate that the former have substantial enhancements in performance over the latter. In our analysis, we set $\gamma_x =\gamma_0$ (where $x\in\{hc,h,c,w\}$) for fair comparison between QRCs and QRIs.

{\it Average Cooling Power} -- The cooling power (current) delivered by a steady-state qutrit autonomous refrigerators with correlated and uncorrelated heat transfer are directly proportional to the photon flux exchange with the cold bath (however, photon flux exchange with each bath is the same), and it is given by (for $X=I, C$)
\begin{equation}
\langle J_{c}^{X} \rangle= \langle \dot{N}^{X} \rangle~ \omega_c,
\end{equation}
where $\omega_c$ is the energy spacing beween the levels $|1\rangle$ and $|2\rangle$. The average photon fluxes $\langle \dot{N}^{X} \rangle$ for  QRCs and QRIs, respectively, are given as (see Appendix~\ref{appsec:FCS} for the details) 
\begin{align}
& \langle\dot{N}^{C} \rangle=\frac{\gamma_0   \Big(n_c n_w-n_h (n_c+n_w+1)\Big)}{2 n_c n_h+n_c+n_h+2 n_w+1},\nonumber\,\, \ \mbox{and}\\ 
&\langle\dot{N}^{I} \rangle=\frac{ \gamma_0 \Big(n_c n_w-n_h (n_c+n_w+1)\Big)}{  n_c \Big(2+3 (n_h+n_w)\Big)+3 n_h (1+  n_w)+2 (2 n_w+1) }.
\end{align}
Note that the currents corresponding to the cold bath and the work bath have a positive sign, while the current corresponding to the hot bath has a negative sign in the refrigeration regime.
Interestingly, using the detailed balance and refrigeration conditions, we find the following lower bound on the ratio of cooling power of QRCs and QRIs (see Appendix~\ref{comQRIvsQRC} for the details)
\begin{equation}\label{PFR}
\frac{\langle {J}_{c}^{C} \rangle}{\langle {J}_{c}^{I} \rangle }\equiv\frac{\langle \dot{N}^{C} \rangle}{\langle \dot{N}^{I} \rangle } >2.
\end{equation}
The above condition suggests that the cooling power (current) of QRCs is always greater than twice the cooling power of QRIs; this advantage is attributed to correlated heat transfer that increases the average photon flux.

{\it Noise-to-signal ratio (NSR) of cooling power} -- 
Ideally, a good and reliable refrigerator is expected to deliver high cooling power output and low cooling power output fluctuations. This quality is characterized by the NSR (relative or scaled fluctuations) in cooling power, i.e., the ratio between the fluctuation in power $\Delta J^{X}_c$, and the square of the average cooling power output $\langle J^{X}_c \rangle^{2}$, given by (for $X=I, C$)
\begin{equation}
  \mathcal{N}^{X}_c := \frac{\Delta J^{X}_{c}}{\langle J^{X}_{c} \rangle^2} \equiv \frac{\Delta\dot{N}^{X}}{\langle\dot{N}^{X}\rangle^2},
\end{equation}
where $\Delta J^{X}_{c}=\Delta\dot{N}^{X} \omega_c^2$. 

It is important to note that highly reliable refrigerators have a low noise-to-signal ratio in cooling power output. The expression for the noise-to-signal ratio for QRCs and QRIs can be written as (see Appendix~\ref{appsec:FCS} for the details)
\begin{equation}
    \mathcal{N}^{C}_c = \frac{\alpha}{\langle\dot{N}^{C}\rangle} \Big(1- \frac{2}{p}\langle\dot{N}^{C}\rangle^2\Big),\ \mbox{and} \ \mathcal{N}^{I}_{c} = \frac{\alpha}{\langle\dot{N}^{I}\rangle} \Big(1- \frac{2 k}{p}\langle\dot{N}^{I}\rangle^2\Big),
\end{equation}
respectively, where $\alpha=p/m$, ${m}={n_h (n_c+n_w+1)-n_c n_w}$, $p=n_h (2 n_c n_w+n_c+n_w+1)+n_c n_w$, and $k=2 (n_c+ n_h+ n_w)+3$.

\begin{figure}
\centering    
\includegraphics[width=7cm,height=6.5cm]{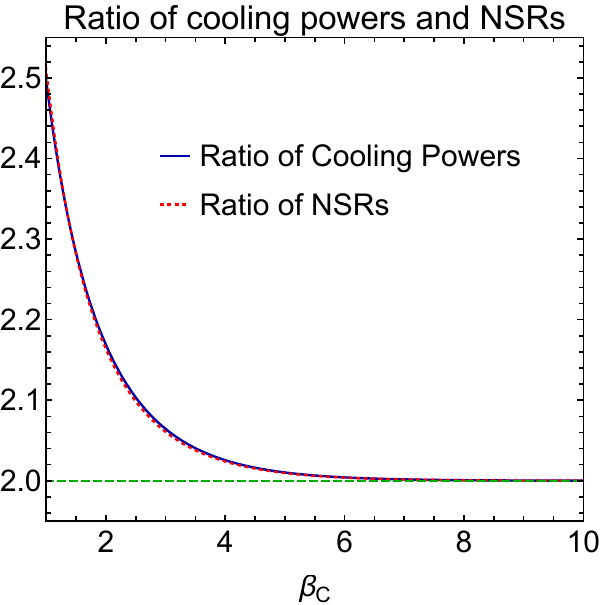}
\caption{Plot for comparison of cooling power and noise-to-signal ratio for QRIs and QRCs.
The calculations use parameters: $\omega_h = 10$, $\omega_c = 0.90$, $\omega_w = 9.10$, $\gamma_0 = 0.01$, $\beta_h = 1.00$, $\beta_w = 0.09$, and $\beta_s = 10.20$. The plot shows the ratio of cooling power (${\langle {J}^{C}_c \rangle}/{\langle {J}^{I}_c \rangle }$) (solid blue) and the ratio of noise-to-signal ratios (NSRs) in cooling power (${\mathcal{N}^{I}_c}/{\mathcal{N}^{C}_c}$) (dotted red) for QRCs and QRIs. The dashed green line marks the lower bounds for the cooling power and NSR ratios, as given by Eqs.~\eqref{PFR} and \eqref{R:NSR}. The plot confirms that these bounds are respected. See the main text for more details.}
\label{fig:Bounds}
\end{figure}

Interestingly, using again the detailed balance, refrigeration conditions, and Eq.~\eqref{PFR}, we find the lower bounds on the ratio of NSRs of cooling power for QRCs and QRIs as
(see Appendix~\ref{comQRIvsQRC}):
\begin{equation}\label{R:NSR}
\frac{\mathcal{N}^{I}_c}{ \mathcal{N}^{C}_c}> \frac{\langle \dot{N}^{C} \rangle}{\langle \dot{N}^{I}\rangle} >2.
\end{equation}
The advantage in the precision of cooling power in QRCs arises yet again due to correlated heat transfers, which not only increase the average photon flux but also reduce photon flux fluctuations by minimizing the stochasticity of transitions in the working system induced by hot and cold baths. As a result, QRCs are {\it at least} twofold more reliable than QRIs.
In Fig.~\ref{fig:Bounds}, we show that the bounds obtained in Eq.~\eqref{PFR} and Eq.~\eqref{R:NSR} are respected for the appropriate choice of system and bath parameters.

\begin{figure*}
\centering    
\includegraphics[width=2\columnwidth]{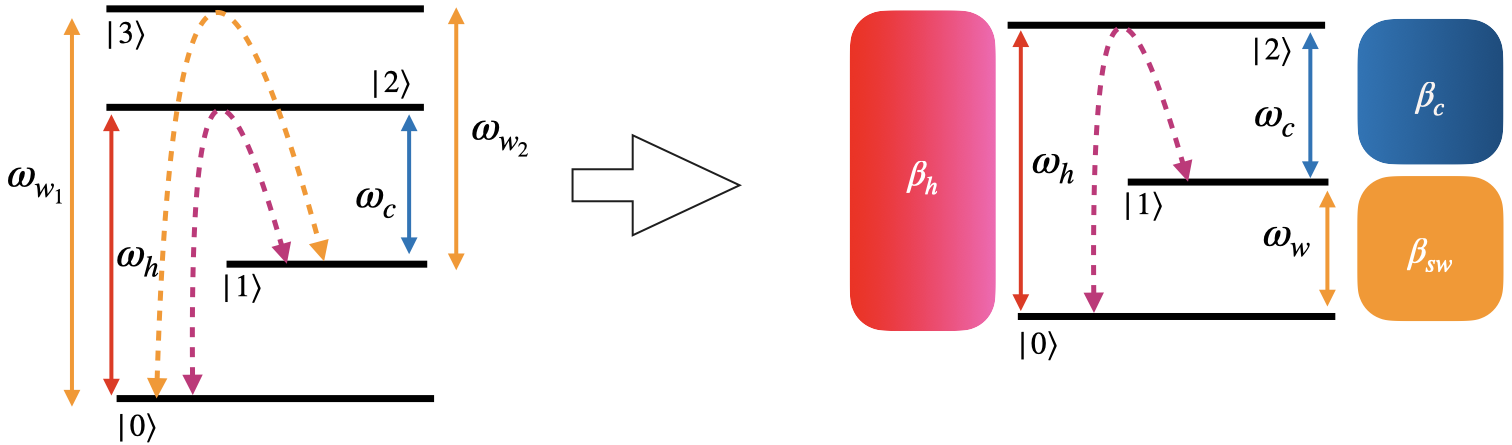}
\caption{Schematic of an autonomous quantum refrigerator with correlated heat transfer and synthetic negative temperature work bath. The figure on the left side displays the refrigerator consists of a four-level quantum system weakly interacting with hot, cold, and two work baths at inverse temperatures $\beta_h$, $\beta_c$, $\beta_{w_1}$, and $\beta_{w_2}$, respectively (see main text). In these quantum refrigerators with correlated heat transfer (QRCs), energy exchange occurs via two-photon transitions involving levels $\ket{0}$ and $\ket{1}$ (indicated by the dashed pink arrow in the left figure). A photon absorbed from the hot bath is released into the cold bath, and vice versa. Simultaneously, energy transfer also takes place between the system and the two work baths through two-photon transitions, where absorption from one work bath corresponds to emission into the other (indicated by the dashed yellow arrow in the left figure). The combined effect of the two work baths can be modeled as a synthetic work bath with inverse temperature $ \beta_{sw} = ({\beta_{w_1} \omega_{w_1} - \beta_{w_2} \omega_{w_2}})/({\omega_{w_1} - \omega_{w_2}})$, effectively coupled to levels $\ket{0}$ and $\ket{1}$. This refrigerator can thus be regarded as a QRCs with synthetic work baths see figure on the right side. Important to note that the temperature of synthetic work bath $ \beta_{sw} = ({\beta_{w_1} \omega_{w_1} - \beta_{w_2} \omega_{w_2}})/({\omega_{w_1} - \omega_{w_2}})<0$ can be negative for the appropriate choice of system-bath parameters, if ${\beta_{w_1} \omega_{w_1} - \beta_{w_2} \omega_{w_2}}<0$. The QRCs with synthetic work bath (right side) are similar to QRCs displayed in Fig.~\ref{fig:QRCs} except the formar utilizes the synthetic work bath, which can be negative. See main text for details.}
\label{fig:QRCNs}
\end{figure*}

{\it Coefficient of performance} --
As we know, due to photon flux, the heat currents in QRCs are higher than in QRIs. In other words, the QRCs have a higher capacity to draw heat from the cold bath than the QRIs. However, the former also draws more heat from the work bath than the latter to cool the cold bath. Consequently, the coefficient of performance $\eta_X=\langle J_{c}^X \rangle/\langle {J}^X_w \rangle$ remains same for both the refrigerators, i.e.,
\begin{align}
\eta_I=\eta_C=\frac{\omega_c}{\omega_w}.    
\end{align}
 Thus, as far as cooling efficiency is concerned, both QRCs and QRIs perform the same. It is worthwhile to mention that for both types of refrigerators, if one interchanges the work bath for a cold bath, all the figures of metrics remain the same.

`\section{Autonomous refrigeration with synthetic negative temperature work bath}\label{NQRC}

In the previous section, we discussed how three-level refrigerators with correlated heat transfer (QRCs) outperform refrigerators with uncorrelated heat transfer (QRIs). In this section, we demonstrate that the performance of QRCs can be further enhanced by replacing the conventional work bath with a synthetic temperature work bath. The synthetic temperature work bath is composed of two equilibrium baths. Interestingly, this composition of two baths can effectively act as a single work bath with synthetic temperature and, for appropriate system-baths parameters, it can serve as a negative temperature work bath (see Appendix~\ref{NT} for the details). Moreover, such a negative temperature bath does not require any external resources.

\begin{figure*}
\centering    
\includegraphics[width=4.7cm,height=4.65cm]{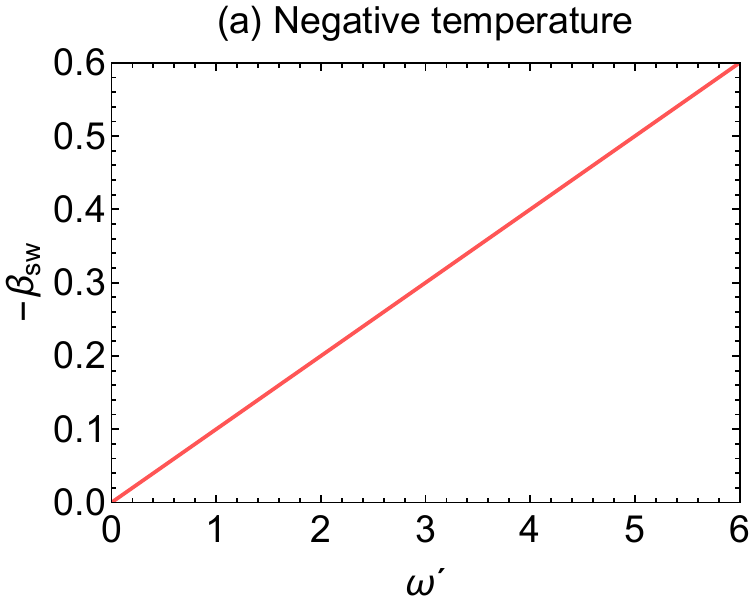}
\includegraphics[width=6.45cm,height=4.7cm]{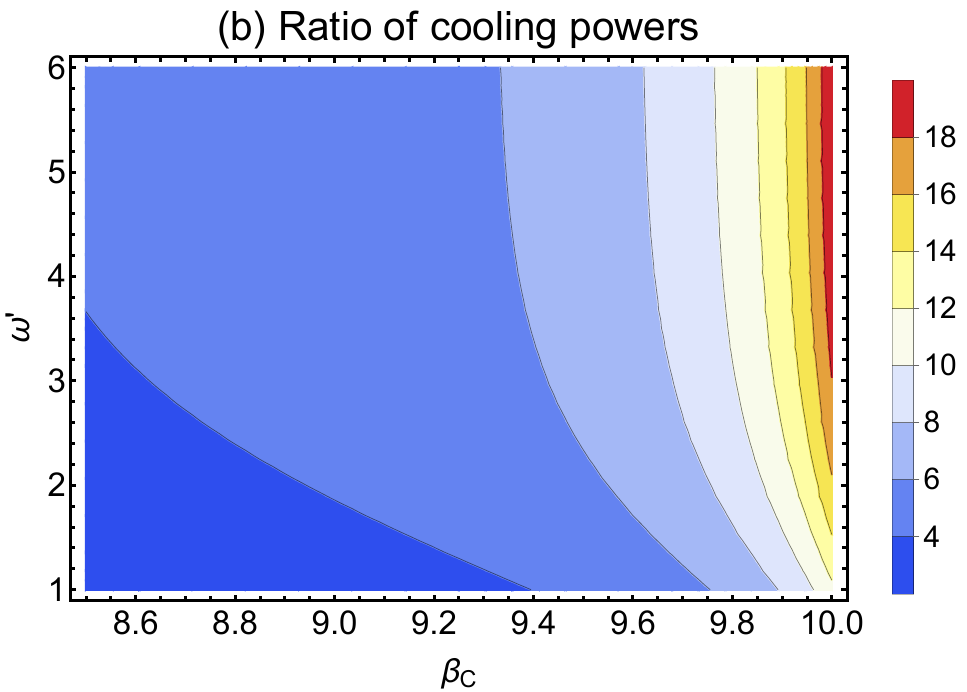}
\includegraphics[width=6.45cm,height=4.7cm]{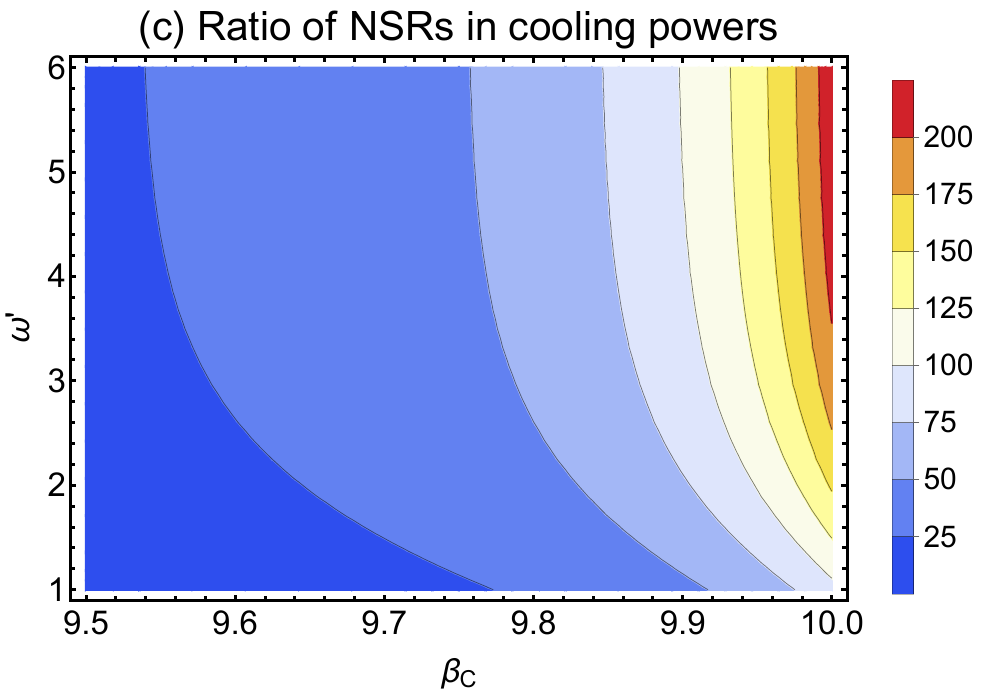}
\caption{Plot for the comparison of cooling power and noise-to-signal ratio in QRCs and QRCNs.
The calculations use parameters: $\omega_h= 10$, $\omega_c=0.90$, $\omega_w=9.10$, $\gamma_0 =0.01$, $\beta_h=\beta_{w_2}=1.00$, $\beta_w=\beta_{w_1} =0.09$, and $\beta_s=10.20$. (a) The figure displays $-\beta_{sw}$ against $\omega'$, to identify the region, where $\beta_{sw}$ is negative.  (b) The figure in the middle shows the ratio of cooling power (${\langle {J}^{SC}_c \rangle}/{\langle {J}^{C}_c \rangle }$) corresponding to QRCs and QRNs against $\beta_c$ and $\omega'$. Note, ${\langle {J}^{SC}_c \rangle} \geq {\langle {J}^{C}_c \rangle }$ signifies that the QRCNs have more cooling power than the QRCNs, and the ratio can reach up to ${\langle {J}^{C}_c \rangle}/{\langle {J}^{I}_c \rangle } \geq 20$ for the considered scenario. (c) The figure on the right displays the ratio $\mathcal{N}^C_{c}/\mathcal{N}^{SC}_{c}$ of NSRs in cooling power corresponding to QRCs and QRNs against $\beta_c$ and $\omega'$. Note, $\mathcal{N}^{C}_{c}>\mathcal{N}_{c}^{SC} $ signifies that the QRCNs produce less NSR in power than the QRCs and the ratio can reach up to $\mathcal{N}^C_{c}/\mathcal{N}^{SC}_{c} \geq 200$ for the considered scenario. See the main text for more details.}
\label{fig:CAIQHE}
\end{figure*}

As discussed in section~\ref{model}, the working principle of the refrigerator relies on reversing the temperature gradient between the hot and cold baths with the assistance of a work bath that is usually hotter than the hot bath. However, the temperature gradient (between the cold bath and
the composition of the hot and work baths) achievable with positive temperature baths is inherently limited. As we know, baths with negative temperatures are `hotter' than conventional baths, including baths with positive infinite temperatures, and exhibit exotic thermodynamic properties (for details, see Refs.~\cite{Bera2024}). Therefore, we incorporate (synthetic) negative temperature work baths into the model of QRCs. Utilizing a synthetic temperature work bath alongside correlated heat transfer with the hot and cold baths requires a four-level working system rather than a three-level system (see Fig.~\ref{fig:QRCNs}). In this model of the refrigerator, a four-level system characterized by the Hamiltonian \(H_{S} =\omega_{w_1} \ketbra{3} +\omega_{h} \ketbra{2} +  (\omega_{h} -\omega_{c})\ketbra{1}\) interacting with four thermal baths, each at inverse temperatures \(\beta_c\), \(\beta_h\), \(\beta_{w_{1}}\) and \(\beta_{w_{2}}\). Here, the hot bath (with inverse temperature $\beta_h$) is coupled to levels $|0\rangle$ and $|2\rangle$ with energy spacing \(\omega_h\), while the cold bath (with inverse temperature $\beta_c$) is connected to levels $|1\rangle$ and $|2\rangle$ with spacing \(\omega_c\). The work baths (with inverse temperatures $\beta_{w_1}$ and $\beta_{w_2}$) are coupled to levels $|0\rangle$ and $|3\rangle$, and $|1\rangle$ and $|3\rangle$, respectively, with energy spacings \(\omega_{w_{1}}\) and \(\omega_{w_{2}} = \omega_{w_{1}}-(\omega_h - \omega_c)\), respectively. Moreover, the four-level system interacts with two work baths through a three-body Raman interaction and independently with the hot and cold baths via another three-body interaction. As a result, the system undergoes two independent two-photon transitions, each facilitated by a three-body interaction (see Appendix~\ref{QRCsNT} for the details). The reduced dynamics of the four-level working system under weak system-baths couplings is given as 
\begin{align}
    \dot{\rho} = i \ [\rho, \ H_{S}] + \mathcal{D}_{hc} (\rho)+ \mathcal{D}_{w_1w_2} (\rho),
\end{align}
where the dissipater $\mathcal{D}_{hc}(\rho)$ is defined below Eq.~\eqref{Lqrc} and the work dissipator $\mathcal{D}_{w_1w_2} (\rho)$ is given as 
\begin{align*}
    \mathcal{D}_{w_1w_2}(\rho) &=  \gamma_{3}' (b_{w_1w_2} \rho b_{w_1w_2}^{\dag}  -\{b_{w_1w_2}^{\dag}b_{w_1w_2},\rho\}/2)  \nonumber \\
    &+ \gamma_{4}' (b_{w_1w_2}^{\dag} \rho b_{w_1w_2} - \{b_{w_1w_2} b_{w_1w_2}^{\dag},\rho\}/2),
\end{align*}
with $b_{w_1w_2}=\ketbra{0}{1}$, decay rates $\gamma_3'=\gamma_0 n_{w_2}(n_{w_1}+1)$, $\gamma_4'=\gamma_0n_{w_1}(n_{w_2}+1)$, and $\gamma_0$ is Weiskopf-Wigner decay constant. Here, the dissipator $\mathcal{D}_{hc}$ involves the parameters of both the hot and cold baths and induces dissipation utilizing the levels $\ket{0}$ and $\ket{1}$. Similarly, the dissipator $\mathcal{D}_{w_1w_2}$ involves the parameters of both work baths and induces dissipation utilizing the same levels, $\ket{0}$ and $\ket{1}$. Effectively, only the levels $\ket{0}$ and $\ket{1}$ are actively involved in the process, while the levels $\ket{2}$ and $\ket{3}$ are never ``engaged.''

The composition of two work baths can be thought of as effective single work bath attached to energy levels $\ket{0}$ and $\ket{1}$. Thus, this model of the refrigerator can be  regarded as QRCs with synthetic temperature work bath defined as 
\begin{equation}
    \beta_{sw} = \frac{\beta_{w_1} \omega_{w_1} - \beta_{w_2} \omega_{w_2}}{\omega_{w_1} - \omega_{w_2}}.
    \label{beta-synthetic}
\end{equation}
It is important to note that $\beta_{sw}$ can be negative if $\beta_{w_1} \omega_{w_1} - \beta_{w_2} \omega_{w_2}<0$. For a detailed description of this model of refrigerator, see Appendix~\ref{QRCsNT}. In the rest of the text, we refer to this model as QRCNs (QRCs with synthetic negative temperature work baths) and compare its performance with QRCs with positive work baths. Note that the QRCs with synthetic positive work baths are equivalent to QRCs studied in the previous section.
The average cooling power and noise-to-signal ratio of QRCNs are given by
\begin{align}
 \langle J^{SC}_c \rangle &= \frac{  \gamma_1 \gamma'_4-\gamma_2 \gamma'_3}{\gamma_1 + \gamma_2 + \gamma'_3 + \gamma'_4} \omega_c,  \nonumber
   \quad \mbox{and} \\
    \mathcal{N}^{SC}_c &= \frac{\alpha'}{\langle\dot{N}^{SC}\rangle} \left(1- \frac{2}{p'}\langle\dot{N}^{SC}\rangle^2 \right).
\end{align}
where \( \alpha' = p'/m' \), \( {m}' = \gamma_1 \gamma_4'-\gamma_2 \gamma_3' \), and \( p' = \gamma_1 \gamma_4'+\gamma_2 \gamma_3' \). The superscript "$SC$" in $\langle J^{SC}_c \rangle$ and $\mathcal{N}^{SC}_c$ refers to refrigerators with synthetic work bath.

If the decay rates of QRCs and QRCNs satisfy the conditions \(\gamma'_4\geq\gamma_4\) and \(\gamma'_3\leq\gamma_3\), we obtain the following inequality between the respective cooling powers (see Appendix~\ref{compQRCsVsQRCsNT} for the details)
\begin{equation}
\langle J_{c}^{SC} \rangle \geq \langle J_{c}^{C} \rangle.
\end{equation}  

Moreover, if the decay rates also satisfy an additional condition \(  \frac{\gamma_2}{\gamma_1}\leq\frac{\gamma'_4-\gamma_4}{\gamma_3-\gamma'_3}\) along with the previous conditions, then the noise-to-signal ratios of QRCs and QRCNs obey the following inequality (see Appendix~\ref{compQRCsVsQRCsNT} for the details) 
\begin{equation}
 {\mathcal{N}^{SC}_c}\leq{ \mathcal{N}^{C}_c}.
\end{equation}  

The inequalities \(\gamma'_4\geq\gamma_4\) and \(\gamma'_3\leq\gamma_3\) lead to the condition \({\gamma'_4}/{\gamma'_3}\geq{\gamma_4}/{\gamma_3}\), which can be equivalently rewritten as  
\begin{equation}
e^{\beta_{sw}\omega_{w}} \leq e^{\beta_{w}\omega_{w}} \implies \beta_{sw} \leq \beta_{w},
\end{equation}  
where $\beta_{sw}$ is the inverse temperature of the `synthetic' work bath in QRCNs, and $\beta_{w}$ is the inverse temperature conventional work bath in QRCs. The above conditions suggest that utilization of a negative temperature work bath increases the cooling power as well as suppresses its relative fluctuation further. With the assistance of a negative temperature work bath, the synthetic (virtual) temperature of the composition of the hot and synthetic work baths can be significantly smaller than that of a positive temperature work bath. As a result, this creates a higher temperature gradient between the cold bath and the composition of the hot and synthetic work baths. Note, this enhancement is enrooted in the fact that baths with negative temperatures are hotter than baths with positive temperatures~\cite{Ramsey1956, Bera2024}.

Therefore, the conditions for QRCNs in the refrigeration process can be modified as 

\begin{equation}\label{SWL}
     \beta_{sw}<\beta_h <\beta_c <\beta_s<\beta_s' = \frac{\beta_h \omega_h - \beta_{sw} \omega_w}{\omega_h - \omega_w},
\end{equation}
where $\beta_s$ (defined in Eq.~\eqref{ref-cond-conve}) and $\beta_s'$ are the cooling limits as they provide the minimum possible temperature a cold bath can attain in a refrigeration process in corresponding refrigerator models. For negative temperature synthetic work bath, i.e., $\beta_{sw}=-|\beta_{sw}|$ ($\beta_{sw}<0$), the upper bound on cooling limits for QRCNs (given in Eq.~\eqref{SWL}) becomes 
\begin{equation}\label{SNT}
\beta_s' = \frac{\beta_h \omega_h + |\beta_{sw}| \omega_w}{\omega_h - \omega_w},
\end{equation}
The above modification in cooling limits allows us to cool the cooler baths even further (because $\beta_s <\beta_s'$), which is not possible by utilizing positive temperature work baths. Moreover, it widens up the cooling window of the refrigeration process.

Now, we compare the performances of QRCs and QRCNs for the appropriate choice of parameters. First, let us consider that the inverse temperatures of hot and cold baths, i.e., $\beta_h$ and $\beta_c$, and the energy spacing with which these baths are attached are the same. Since the work bath in QRCNs is composed of two work baths. We assume one of the work baths has the same temperature as a hot bath, i.e., $\beta_{w_2}=\beta_{h}$, and the other work bath has the same as the work bath of QRCs $\beta_{w_1}=\beta_{w}$. Thus the only extra parameter QRCNs has $\omega'$ (spacing between energy levels $|3\rangle$ and $|4\rangle$). We consider the value of $\omega'$ such that $\beta_{sw}$ is negative for considered parameters.  In Fig.~\ref{fig:CAIQHE}(a), we plot the $\beta_{sw}$ vs. $\omega'$ to identify that region where the synthetic temperature of the work bath is negative in QRCNs model. The Figs.~\ref{fig:CAIQHE}(b) and ~\ref{fig:CAIQHE}(c) demonstrate that, for the considered scenario and appropriate choice of parameter, the cooling power and reliability of QRCNs can be enhanced over QRCs by many folds. It is important to note that we compare the QRCNs and QRCs in common refrigeration regions, while the refrigeration window of QRCNs is wider than QRCs. This means that for some ranges of parameters, while the QRCNs can operate as refrigerators, the QRCs cannot. The observed advantage is attributed to the negative temperature of the (synthetic) work bath.

\section{Discussion on Fundamental Limits given on precision Cooling Power given by TUR}\label{RTUR}
Classical steady-state thermal processes always exhibit a trade-off relationship between relative fluctuation in current and the thermodynamic cost (quantified by the rate of entropy production $\dot{S}$), which is known as the thermodynamic uncertainty relation (TUR). In the context of the refrigeration process, the TUR can be utilized as the lower bound on the noise-to-signal ratio of cooling power. The TUR provides fundamental limits on the precision of cooling power, given by~\cite{Barato2015}
\begin{equation}
     \frac{\Delta J_{c}^{C}}{\langle J_{c}^{C} \rangle^2} \geq \frac{2}{\dot{S}}.
\end{equation}
Note that this classical thermodynamic uncertainty relation (cTUR) is derived for classical Markovian stochastic dynamics. However, in the considered refrigeration process, although it utilizes quantum Markovian stochastic dynamics, it still holds because, in the steady state, the density matrix carries no off-diagonal entries when expressed in energy bases. Therefore, classical TUR can be applied to autonomous refrigerator models discussed in this work. The above bound can be written in terms of average photon flux and bath temperatures (for $X=I, C, SC$)
\begin{equation}\label{cTUR}
 \mathcal{Q}_{X} =  f \frac{\Delta \dot{N}^{X}}{\langle \dot{N}^{X} \rangle} \geq 2,
\end{equation}
where $f=(\beta^{k}_s -\beta_c) \omega_c$ and $\beta^{k}_s\in\{\beta_{s},\beta'_{s}\}$ (see Appendix~\ref{CTUR} for the details).
\begin{figure}
\centering    
\includegraphics[width=7cm,height=5.5cm]{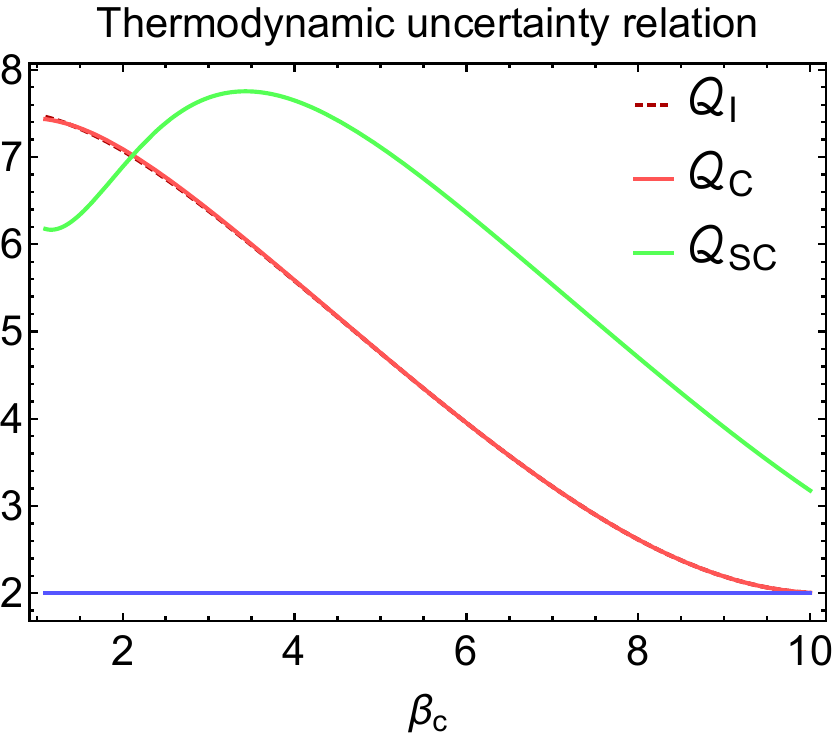}
\caption{Plot for the attainability of TUR for QRIs, QRCs, and QRCNs. The calculations use parameters: $\omega_h= 10$, $\omega_c=.90$, $\omega_w=9.1$, $\gamma_0 =.01$, $\beta_h=\beta_{w_2}=1.00$, $\beta_w=\beta_{w_1} =0.09$, $\beta_s=10.20$ and $\omega'=2$. In the above figure, we plotted the $Q_{X}$ (given in Eq.~\eqref{cTUR}) against $\beta_c$, with $X\in \{I, C,SC\}$. We observed that all the refrigerator models respect the TUR (given in Eq.~\eqref{cTUR}). See the main text for more details.}
\label{fig:TUR}
\end{figure}

In Fig.~\ref{fig:TUR}, we found that the TUR for QRCs and QRIs is almost equally tight and saturates when $\beta_{c}$ approaches $\beta_{s}$. However, it is very loose for QRCNs. The reason for the looseness of TUR for QRCNs is the negative temperature of the work bath. Since the negative temperature work bath can suppress the relative fluctuations in cooling power, simultaneously, it enhances the entropic cost. Thus, all refrigerator models respect the TUR, which states that enhancing the current precision requires an entropic cost. It is important to note that the classical TUR is derived under the assumption of classical stochastic dynamics with the detailed balance condition. All the refrigerators considered here respect these conditions, since in the steady state regime, the dynamics is equivalent to classical stochastic dynamics that obeys detailed balance, and the steady-state density matrix has only diagonal elements. Therefore, the TUR is respected, and no violation occurs here.

{\it \bf Comparison of the cooling ability of QRCNs with previous autonomous and non-autonomous refrigerators: } 
The SSD engine model proposed in Ref.~\cite{Mohan2024} can function as a refrigerator by reversing the direction of heat flow through external periodic driving. Importantly, this refrigerator may be regarded as the non-autonomous counterpart of the refrigerators due to the presence of external driving.  The cooling ability of such a non-autonomous refrigerator, as well as other previously studied various non-autonomous refrigerators, including SSD type three-level refrigerators (e.g., Ref.~\cite{Kosloff2014}), is limited by the condition:
\begin{equation}
    \beta_{c} \leq \frac{\omega_{h}}{\omega_{c}} \beta_{h}.
\end{equation}
The above bound is obtained using the refrigeration condition of non-autonomous refrigerators,i.e. $\beta_{c}\omega_{c}\leq \beta_{h}\omega_{h}$ or using the fact that the coefficient of performance (COP) of non-autonomous refrigerators is bounded by the Carnot COP~\cite{Kosloff2014}, i.e.,
\(
\eta = \frac{\omega_{c}}{\omega_{h} - \omega_{c}} \leq \eta^{C} = \frac{T_{c}}{T_{h} - T_{c}}.
\)

In contrast, the cooling ability of the QRCNs is limited by following condition,
\begin{equation}\label{BM}
    \beta_{c} \leq \frac{\omega_{h}}{\omega_{c}} \beta_{h} - \frac{\omega_{w}}{\omega_{c}} \beta_{sw} =\frac{\omega_{h}}{\omega_{c}} \beta_{h} + \frac{\omega_{w}}{\omega_{c}} |\beta_{sw}|,
\end{equation}
where we have utilized Eqs.~\eqref{SWL} and \eqref{SNT} and $\omega_{c}=\omega_{h}-\omega_{w}$ to obtain above inequality. Note, here we consider $\beta_{sw} < 0$.

From the inequality above, it is evident that QRCNs can cool the cold bath to significantly lower temperatures than those achievable with previously studied non-autonomous refrigerators. Moreover, it is worth while to mention that negative synthetic temperature of work bath in QRCNs ie., $-\beta_{sw}$ can be made sufficiently large for suitable choice of system and work bath parameters (for example, see Fig.~\ref{fig:CAIQHE}(a)), thus it can cool the cold bath to much lower temperature. Therefore, QRCNs significantly outperform all previously studied autonomous (with positive temperature work baths) and non-autonomous refrigerators in terms of cooling capability. It is important to note, however, that QRCs, QRIs, and other models of autonomous refrigerators cannot surpass the cooling ability of general non-autonomous refrigerators as the positive sign between two terms in the left-hand side of Eq.~\eqref{BM} becomes negative.

\section{Conclusions \label{sec:Summary}}
 In this work, we first introduce distinct three-level autonomous refrigerators that utilize correlated heat transfer with the hot and cold baths via a two-photon transition to cool the cold bath with the assistance of a work bath. The correlated heat transfer between the working system and the hot and cold baths is facilitated via two-photon transitions, which is a consequence of three-body interactions between the working system and hot and cold baths. We demonstrate that these refrigerators deliver significantly higher cooling power with much greater reliability, i.e., a lower signal-to-noise ratio in power, by a factor of two compared to their conventional three (qutrit) refrigerators, which interacts with hot, cold, and work bath via two body interaction enabled via one photon transition. To further significantly enhance the performance of autonomous refrigerators, we replace the conventional work bath with a (synthetic) negative temperature work bath, where the latter is created utilizing two conventional equilibrium work baths. We show that refrigerators with synthetic work baths can be realized using a four-level quantum system coupled to two work baths via three-body interactions. Similarly, hot and cold baths are also coupled via another three-body interaction. As a result, it utilizes two two-photon transitions. Furthermore, we found that synthetic work baths enhance cooling power and improve cooling precision (reliability) by many folds. Notably, our refrigerators with negative temperature work baths significantly outperform previously studied autonomous and non-autonomous refrigerators in cooling ability without requiring additional energy resources or control, as they can autonomously cool the cold bath to significantly lower temperatures, which is forbidden for other refrigerators to attain.

 If we utilize synthetic work bath with negative temperature, we may expect enhancement in refrigeration performances of the conventional qutrit refrigerator (QRIs). Moreover, cooling limits also can be modified similarly, as we discussed previously. Additionally, this could also be true for the three-qubit refrigerators theoretically studied in Ref.~\cite{Mitchison2019}; if we attach a qubit with negative temperature work bath (qutrit attached with two work baths via two-photon transition is equivalent to a qubit attached to synthetic work bath, see Appendix~\ref{NT}), we expected to get a similar enhancement in the performance and modification in cooling limits. Moreover, a three-qubit refrigerator with a negative work bath can cool the target qubit to a further lower temperature than what is attained in the recent experimental work~\cite{Aamir2022}. Our proposed models are experimentally feasible as two-photon transitions have already been realized on various platforms (see Appendix~\ref{exp} for the details). The methodology presented in this work can be utilized to enhance the performance of other autonomous and non-autonomous quantum thermal devices.

\section*{Acknowledgments}
B.M. thanks Tanmoy Pandit for the fruitful discussion. B.K.A. acknowledges CRG Grant No.~CRG/2023/003377 from the Science and Engineering Research Board (SERB), Government of India. B. K. A. would also like to acknowledge funding from the National Mission on Interdisciplinary  Cyber-Physical  Systems (NM-ICPS)  of the Department of Science and Technology,  Govt.~of  India through the I-HUB  Quantum  Technology  Foundation, Pune, India. B. M. acknowledges funding by the Research Council of Finland by Grant No.~355824.

\onecolumngrid
\section*{Appendix}
Here, we include the derivations and analytical calculations to supplement the results presented in the main text.

\appendix

\section{Description of three-level autonomous quantum refrigerators with uncorrelated heat transfer (QRIs)\label{appsec:QRIsteaystate}}
For QRIs, the total Hamiltonian of the qutrit system and three photonic (bosonic) thermal baths can be written as
\begin{equation}
    H = H_{S} + H_{B_{h}} + H_{B_{c}} + H_{B_{w}}+ H_{SB_hB_cB_w}^I,
\end{equation}
where the Hamiltonians and of the qutrit system and baths are given by 
\begin{equation}
    H_S = \omega_h \ketbra{2} + (\omega_h - \omega_c) \ketbra{1}, \quad 
    H_{B_h} = \sum_k \Omega_{k,h} \ a^\dagger_{k,h} a_{k,h}, \quad 
    H_{B_c} = \sum_{k'} \Omega_{k',c} a^\dagger_{k',c} a_{k',c}, \quad 
    H_{B_w} = \sum_{k''} \Omega_{k'',w} a^\dagger_{k'',w} a_{k'',w},
\end{equation}
with $\omega_{h}$ and  $\omega_{h}-\omega_{c}$ being the frequencies corresponding to the energy gaps. The interaction Hamiltonian($H_{SB_hB_cB_w}^I$ ) between qutrit with the photonic thermal baths given by

\begin{align}\label{TQC}
H_{SB_hB_cB_w}^I = g_h \sum_k (a_{k,h} b^\dag_h + a_{k,h}^\dag b_{h}) 
+ g_c \sum_{k'} (a_{k',c} b_{c}^\dag + a_{k',c}^\dag b_{c}) 
+ g_w \sum_{k''} (a_{k'',w} b_{w}^\dag + a_{k'',w}^\dag b_{w}). 
\end{align}
For very weak system-baths couplings ($g_h$, $g_c$, and $g_w$), the local dynamics of the qutrit (described by the state $\rho$) is expressed by the Lindblad master equation ($\hbar=1$)
\begin{align}\label{IQHER}
    \dot{\rho} =& -i[H_{S},\rho] + \mathcal{D}_{h}(\rho) + \mathcal{D}_{c}(\rho)+\mathcal{D}_{w}(\rho).
\end{align}

where $\rho$ is the density matrix representing the state of the qutrit. The dissipators $\mathcal{D}_{h}(\rho)$, $\mathcal{D}_{c}(\rho)$ and $\mathcal{D}_{w}(\rho)$ represent dissipative dynamics due to the interactions with the hot, cold and work baths and are given by (for $x$ = $h$, $c$, $w$)
\begin{align}\label{eq:IMasEq}
\mathcal{D}_{x}(\rho)=\gamma_{x}&(n_{x}+1)(b_{x} \rho b^{\dag}_{x} - \{b^{\dag}_{x}b_{x},\rho\}/2)
+ \gamma_{x}n_{x} (b^{\dag}_{x} \rho b_{x} - \{b_{x}b^{\dag}_{x},\rho\}/2),
\end{align}
where the anti-commutator $\{Y,Z\} = YZ + ZY$, the coefficient $\gamma^x$ is the Weiskopf-Wigner decay constant, and $n_{x} = 1/(e^{\beta_{x}\omega_x}-1)$ is the average number of photons in the bath with frequency $\omega_x$. The appearance of three dissipators, $\mathcal{D}_{h}(\rho)$, $\mathcal{D}_{h}(\rho)$ and $\mathcal{D}_{w}(\rho)$, in the master equation \eqref{eq:IMasEq} reflects that the heat energy exchange between working with each bath are independent (or uncorrelated). The steady-state solution of the above master equation can be obtained by solving $\dot{\rho}=0$ (we denote the steady state by $\sigma_{I}$), and it is 
\begin{align}
    \sigma_{I} &=  \frac{\gamma_c \gamma_h n_c (n_{h}+1)+\gamma_w (n_{w}+1) (\gamma_c+\gamma_h+\gamma_c n_c+\gamma_h n_{h})}{\gamma_c n_c (\gamma_h+\gamma_w+3 \gamma_h n_{h}+3 \gamma_w n_{w})+\gamma_h n_{h} (\gamma_c+2 \gamma_w+3 \gamma_w n_{w})+\gamma_w (2 n_{w}+1) (\gamma_c+\gamma_h)}\op{0} \nonumber\\
&+  \frac{\gamma_c \gamma_h (n_c+1) n_{h}+\gamma_w n_{w} (\gamma_c+\gamma_h+\gamma_c n_c+\gamma_h n_{h})}{\gamma_c n_c (\gamma_h+\gamma_w+3 \gamma_h n_{h}+3 \gamma_w n_{w})+\gamma_h n_{h} (\gamma_c+2 \gamma_w+3 \gamma_w n_{w})+\gamma_w (2 n_{w}+1) (\gamma_c+\gamma_h)}\op{1} \nonumber  \\
&+ \frac{\gamma_c n_c (\gamma_h n_{h}+\gamma_w n_{w})+\gamma_h \gamma_w n_{h} (n_{w}+1)}{\gamma_c n_c (\gamma_h+\gamma_w+3 \gamma_h n_{h}+3 \gamma_w n_{w})+\gamma_h n_{h} (\gamma_c+2 \gamma_w+3 \gamma_w n_{w})+\gamma_w (2 n_{w}+1) (\gamma_c+\gamma_h)}\op{2} .
\end{align}

Now the average heat currents $\langle \dot{J}^{x}_I \rangle$ corresponding to the either baths are given by 
\begin{align}
\langle \dot{J}^{x}_I \rangle = \Tr(\mathcal{D}_{x}(\sigma_I)H_{S}).
\end{align}
The average heat currents in QRIs corresponding to the hot, cold and work baths are given by
\begin{align}
&\langle J^{I}_{w} \rangle = \Tr[\mathcal{D}_w(\sigma_I) H_S] =\frac{\gamma_0 \left(n_c n_w-n_h (n_c+n_w+1)\right)}
{ n_c \left(2+ {3} (n_h+ n_w)\right) + 3n_h (1+ n_w) + 2(2 n_w+1) }\omega_w, \\
&\langle \dot{J}_{h}^{I} \rangle= \Tr[\mathcal{D}_h(\sigma_I) H_S]= -\frac{\gamma_0 \left(n_c n_w-n_h (n_c+n_w+1)\right)}
{ n_c \left(2+ {3} (n_h+ n_w)\right) + 3n_h (1+ n_w) + 2(2 n_w+1) }\omega_h, \\
 \mbox{and} \ \ & \langle \dot{J}_{c}^{I} \rangle= \Tr[\mathcal{D}_c(\sigma_I) H_S] = \frac{\gamma_0 \left(n_c n_w-n_h (n_c+n_w+1)\right)}
{ n_c \left(2+ {3} (n_h+ n_w)\right) + 3n_h (1+ n_w) + 2(2 n_w+1) }\omega_c,
\end{align}
respectively. Note that the currents corresponding to the cold bath and the work bath are positive as they are coming out of the bath, while the current corresponding to the hot bath has a negative as flowing into the bath in the refrigeration regime. We have considered $\gamma_w=\gamma_h=\gamma_c=\gamma_0$ in this work. To determine the fluctuations in currents corresponding to each bath (\(\Delta J_x^I\)), we employ the full counting statistics of the steady-state dynamics (see Appendix~\ref{appsec:FCS} for the details).

\section{Steady state solution of autonomous quantum refrigerators with correlated heat transfer (QRCs) \label{appsec:QRCsteaystate}}
For QRCs, the total Hamiltonian of the qutrit system and two photonic (bosonic) thermal baths can be written as
\begin{equation}
       H = H_{S} + H_{B_{h}} + H_{B_{c}} + H_{B_{w}}+ H_{SB_hB_cB_w}^C,
\end{equation}
where the Hamiltonian and of the qutrit system is given by 
\begin{equation}\label{RI}
    H_{S}= \omega_{h} \ketbra{2} + (\omega_{h}-\omega_{c})\ketbra{1},
\end{equation}
with $\omega_{h}$ and  $\omega_{h}-\omega_{c}$ being the frequencies corresponding to the energy gaps. The Hamiltonians of the photonic thermal baths $H_{B_{h}}$ and $H_{B_{c}}$ and the interaction $H_{SB_hB_c}$ given in the main text. The corresponding Lindblad master equation describing the local dynamics of the qutrit (described by state $\rho$) is given by (see Eq.~\eqref{Lqrc} of the main text)
\begin{align}\label{QHER}
    \dot{\rho} =& -i[H_{S},\rho] + \mathcal{D}_{hc}(\rho) + \mathcal{D}_{w}(\rho).
\end{align}
where 
\begin{align}
    \mathcal{D}_{hc}(\rho) &=  \gamma_{1} (b_{hc} \rho b_{hc}^{\dag}  -\frac{1}{2}\{b_{hc}^{\dag}b_{hc},\rho\}) + \gamma_{2} (b_{hc}^{\dag} \rho b_{hc} - \frac{1}{2}\{b_{hc} b_{hc}^{\dag},\rho\}),\nonumber \\
    \ \mbox{and} \  \nonumber\\
      \mathcal{D}_{w}(\rho) &=  \gamma_{3} (b_{w} \rho b_{w}^{\dag}  -\frac{1}{2}\{b_{w}^{\dag}b_{w},\rho\}) + \gamma_{4} (b_{w}^{\dag} \rho b_{hc} - \frac{1}{2}\{b_{w} b_{w}^{\dag},\rho\}),\nonumber
\end{align}
with $\gamma_1=\gamma_{hc}  n_c (n_h+1)$, $\gamma_2=\gamma_{hc}  (n_c+1)n_{h}$, $\gamma_3=\gamma_w  (n_w+1)$, $\gamma_4=\gamma_w n_{w}$. 
It is important to note that, in the above, the Lindblad master equation has a single dissipator corresponding to hot and cold baths. This is due to the three-body interaction between system + hot bath + cold bath, for details see Ref.~\cite{Mohan2024}. The steady-state solution of the above master equation can be obtained by solving $\dot{\rho}=0$ (we denote the steady state by $\sigma_{C}$), and it is 
\begin{equation}
  \sigma_{C}= 
\frac{\gamma_2 + \gamma_4}{\gamma_1 + \gamma_2 + \gamma_3 + \gamma_4} \op{1} + \frac{\gamma_1 + \gamma_3}{\gamma_1 + \gamma_2 + \gamma_3 + \gamma_4} \op{0}.
\end{equation}

Now, the average heat current in QRCs corresponding to the work baths is given by
\begin{align}
&\langle J^{C}_{w} \rangle = \frac{\gamma_{0}  \omega_w (n_c n_w-n_h (n_c+n_w+1))}{2 n_c n_h+n_c+n_h+2 n_w+1}\omega_w.
\end{align}
Note that we have considered $\gamma_{w}=\gamma_{hc}=\gamma_0$ in this work. The average heat currents corresponding to hot and cold baths cannot be directly obtained using \(
\langle {J}_{x}^C \rangle = \Tr(\mathcal{D}_{x}(\sigma_C)H_{S})  
\) because there are no independent dissipators associated with each bath. To determine them, we employ the full counting statistics of the steady-state dynamics (see Appendix~\ref{appsec:FCS} for the details). This approach also allows us to compute the fluctuations in currents (\(\Delta J_x^C\)) corresponding to each bath.

\section{Full Counting Statistics for quantum absorption refrigerator for QRIs and QRCs \label{appsec:FCS}}
Full Counting Statistics (FCS) provides an analytical approach to determine the statistics of the quantity of interests $M$, i.e., currents corresponding to distinct  baths, and their fluctuations in an open quantum system dynamics~\cite{Esposito2009}. This approach incorporates counting fields into the Linblad master equation. Suppose $\rho(\chi,t)$ represents the solution of the dressed Lindblad master equation. In that case, we define the moment-generating function $\mathcal{M}(\chi,t)$ and the cumulant-generating function $\mathcal{F}(\chi,t)$ as follows:
\begin{align}
\mathcal{M}(\chi,t)=\mathrm{Tr}{\{\rho(\chi,t)\}}, \ \mbox{and} \ \mathcal{F}(\chi,t)=\ln \mathcal{M}(\chi,t).
\end{align}
It is important to note that, often, the description in terms of cumulants is more convenient and transparent. The advantage lies in the fact that the dominant eigenvalue of the Liouvillian usually determines the long-time evolution of the cumulant-generating function:
\begin{equation}
 \mathcal{C}(\chi,t) \approx \lambda(\chi)t,
\end{equation}
where $\lambda(\chi)$is the eigenvalue of $\mathcal{L}(\chi)=\mathcal{L}(\chi,0)$ with the largest real part (uniqueness assumed) and it vanishes when $\chi = 0$.

In the long-time limit, the cumulants of the quantity of interest $M$ in the steady state can be obtained using the following formula:
\begin{equation}
\langle \langle M^{k} \rangle \rangle = \bigg(\frac{d}{d(i\chi)}\bigg)^{k} \lambda(\chi)\bigg|_{\chi=0}.
\end{equation}
The first and second cumulants correspond to the mean and variance of the quantity of interest $M$, respectively:
\begin{align}
\langle M \rangle = \bigg(\frac{d}{d(i\chi)}\bigg) \lambda(\chi)\bigg|_{\chi=0}, \ \ \mbox{and} \ \ \Delta M = \langle \langle M^{2} \rangle \rangle = \bigg(\frac{d}{d(i\chi)}\bigg)^{2} \lambda(\chi)\bigg|_{\chi=0}.
\end{align}

A direct computation of $\lambda(\chi)$ is not straightforward. To analytically determine the mean and variance from the derivatives, we follow the method outlined in Refs.~\cite{Bruderer2014, Kalaee2021, Prech2023, Landi2024}. Consider the characteristic polynomial of $\cal{L}(\chi)$
\begin{equation}
    \sum_{n}a_{n} \lambda(\chi)^{n} =0,
\end{equation}
where the terms $a_n$ are functions of $\chi$. Derivatives of $a_n$ are defined as 
\begin{align}
    a'_{n} = i\frac{d}{d\chi} a_{n}|_{\chi=0},  \ \mbox{and} \  a''_{n} = \big(i\frac{d}{d\chi}\big)^{n} a_{n}|_{\chi=0}.
\end{align}
With a little analysis, we can express mean and variance as (for more details, see appendices of Refs.~\cite{Kalaee2021, Prech2023, Landi2024}):
\begin{align}
\langle M \rangle =-\frac{a'_{0}}{a_{1}},  \ \mbox{and} \ \Delta M = \bigg(\frac{a''_0}{a'_0}-\frac{2a'_1}{a_1}\bigg)\langle M \rangle -\frac{2a_2}{a_1} \langle M \rangle^2.
\end{align}
Note that the above formalism hold for all systems with Lindblad dynamics with a unique steady state.

To determine the current statistics in QRIs, we are using the Full Counting Statistics (FCS) technique, which includes counting fields in the master equation. Let $\chi_{h}$, $\chi_{c}$, and $\chi_{w}$ be counting fields for the hot, cold, and work baths, respectively. The dressed Lindblad master equation~\eqref{IQHER} of QRIs can be written as \cite{Kalaee2021,Agarwalla2015}

\begin{align}
    \frac{d{\rho}(\chi,t)}{dt} =& -i[H_{S},\rho]  +  \gamma_{h}(n_{h}+1)(e^{-i\omega_h\chi_h}b_{h}{\rho}_R b^{\dagger}_{h} - \frac{1}{2}\{b^{\dagger}_{h}     b_{h},{\rho}_R\})
    +  \gamma_{h}n_{h}( e^{i\omega_h\chi_h}b^{\dagger}_{h}{\rho}_R b_{h} 
    - \frac{1}{2}\{b_{h}  b^{\dagger}_{h},{\rho}\})\\
    &+\gamma_{c}(n_{c}+1)(e^{-i\omega_c\chi_c}b_{c}{\rho} b^{\dagger}_{c} - \frac{1}{2}\{b^{\dagger}_{c}  b_{c},{\rho}_R\})
    +  \gamma_{c}n_{c}( e^{i\omega_c\chi_c}b^{\dagger}_{c}{\rho}_R b_{c} - \frac{1}{2}\{b_{c}b^{\dagger}_{c},{\rho}\})\\
    &+\gamma_{w}(n_{w}+1)(e^{-i\omega_w\chi_w}b_{w}{\rho} b^{\dagger}_{w} - \frac{1}{2}\{b^{\dagger}_{w}  b_{w},{\rho}_R\}) +\gamma_{w}n_{w}( e^{i\omega_w\chi_w}b^{\dagger}_{w}{\rho}_R b_{w} - \frac{1}{2}\{b_{w}b^{\dagger}_{w},{\rho}\}). \nonumber
\end{align}
It is important to note that we can neglect the commutator term as do not affect the population dynamics. The full Liouvillian super-operator $\mathcal{L}(\chi_{h},\chi_{c},\chi_{w})$ with counting fields can be easily constructed. As we are interested in the cold bath current statistics, we write the Liouvillian $\mathcal{L}(\chi_{c})$ by setting $\chi_{h}=\chi_{w}=0$. We write,
\[
\mathcal{L}(\chi_{h},\chi_{c},\chi_{c})=
\begin{bmatrix}
    k_1 & 0 & 0 & 0 & e^{i \chi_c \omega_c} g_4 & 0 & 0 & 0 & g_2 \\
    0 & k_2 & 0 & 0 & 0 & 0 & 0 & 0 & 0 \\
    0 & 0 & k_3 & 0 & 0 & 0 & 0 & 0 & 0 \\
    0 & 0 & 0 & k_4 & 0 & 0 & 0 & 0 & 0 \\
    e^{-i \chi_c \omega_c} g_3 & 0 & 0 & 0 & k_5 & 0 & 0 & 0 & g_6 \\
    0 & 0 & 0 & 0 & 0 & k_6 & 0 & 0 & 0 \\
    0 & 0 & 0 & 0 & 0 & 0 & k_7 & 0 & 0 \\
    0 & 0 & 0 & 0 & 0 & 0 & 0 & k_8 & 0 \\
    g_1 & 0 & 0 & 0 & g_5 & 0 & 0 & 0 & k_9
\end{bmatrix}
\]
with
\[
\begin{aligned}
    k_1 &= -g_1 - g_3, \quad
    k_2 = -\left(\frac{g_1}{2} + \frac{g_3}{2} + \frac{g_4}{2} + \frac{g_5}{2} \right), 
    k_3 = -\left(\frac{g_1}{2} + \frac{g_2}{2} + \frac{g_3}{2} + \frac{g_6}{2} \right), \quad
    k_4 = -\left(\frac{g_1}{2} + \frac{g_3}{2} + \frac{g_4}{2} + \frac{g_5}{2} \right), \\
    k_5 &= -g_4 - g_5, \quad
    k_6 = -\left(\frac{g_2}{2} + \frac{g_4}{2} + \frac{g_5}{2} + \frac{g_6}{2} \right), 
    k_7 = -\left(\frac{g_1}{2} + \frac{g_2}{2} + \frac{g_3}{2} + \frac{g_6}{2} \right), \quad
    k_8 = -\left(\frac{g_2}{2} + \frac{g_4}{2} + \frac{g_5}{2} + \frac{g_6}{2} \right), \\
    k_9 &= -g_2 - g_6.
\end{aligned}
\]
where $g_{1}=\gamma_{h}(n_{h}+1)$, $g_{2}=\gamma_{h}n_{h}$, $g_{3}=\gamma_{c}(n_{c}+1)$, $g_{4}=\gamma_{c}n_{c}$, $g_{5}=\gamma_{w}(n_{w}+1)$, $g_{6}=\gamma_{w}n_{w}$. 
Following the previous discussion in this section, we find the polynomial factors with respective derivatives as
\begin{align*}
    a_1 &= -\frac{1}{64} (g_1 + g_3 + g_4 + g_5)^2 (g_1 + g_2 + g_3 + g_6)^2 (g_2 + g_4 + g_5 + g_6) \notag \\
    &\quad \times (-4 g_2 g_3 g_5 + g_2^2 (g_3 + g_4 + g_5) - 4 g_1 g_4 g_6 + (g_4 + g_5 + g_6) (g_4 g_6 + g_3 (g_5 + g_6)) \notag \\
    &\quad + g_2 (g_4^2 + 2 g_4 g_5 + g_5^2 + 2 g_4 g_6 + g_5 g_6 + g_1 (g_4 + g_5 + g_6) + g_3 (g_4 + 6 g_5 + 2 g_6)) \notag \\
    &\quad + g_1 (g_4^2 + (g_5 + g_6)^2 + 2 g_4 (g_5 + 3 g_6))) \notag 
\end{align*}
\begin{align*}    
    a_2 &= \left( \left( -\frac{g_1}{2} - \frac{g_3}{2} - \frac{g_4}{2} - \frac{g_5}{2} \right)^2 \right. + 4 \left( -\frac{g_1}{2} - \frac{g_3}{2} - \frac{g_4}{2} - \frac{g_5}{2} \right) 
    \left( -\frac{g_1}{2} - \frac{g_2}{2} - \frac{g_3}{2} - \frac{g_6}{2} \right) \notag \\
    &\quad \left. + \left( -\frac{g_1}{2} - \frac{g_2}{2} - \frac{g_3}{2} - \frac{g_6}{2} \right)^2 \right) \notag 
    \quad \times \left( -\frac{g_2}{2} - \frac{g_4}{2} - \frac{g_5}{2} - g_6 \right)^2 \notag \\
    &\quad \times \left((-g_3 g_4 - (-g_1 - g_3) (g_4 + g_5)) (-g_2 - g_6) \right. - g_5 (-g_2 g_3 + (-g_1 - g_3) g_6) + g_1 (g_2 (g_4 + g_5) + g_4 g_6)) \notag \\
    a_0' &= -\frac{1}{64} (g_1 + g_3 + g_4 + g_5)^2 (g_1 + g_2 + g_3 + g_6)^2 (g_2 + g_4 + g_5 + g_6)^2 \times (-g_2 g_3 g_5 + g_1 g_4 g_6) \omega_c \notag \\
    a_0'' &= \frac{1}{32} g_1 g_4 (g_1 + g_3 + g_4 + g_5)^2 g_6 (g_1 + g_2 + g_3 + g_6)^2 (g_2 + g_4 + g_5 + g_6)^2 \omega_c^2 \notag \\
    &\quad - \frac{1}{64} (g_1 + g_3 + g_4 + g_5)^2 (g_1 + g_2 + g_3 + g_6)^2 (g_2 + g_4 + g_5 + g_6)^2  \times (-g_2 g_3 g_5 + g_1 g_4 g_6) \omega_c^2 \notag 
    \end{align*}

\begin{align}
  a_1'   & =-\frac{1}{16} (g_1 + g_3 + g_4 + g_5) (g_1 + g_2 + g_3 + g_6) (g_2 + g_4 + g_5 + g_6) (-g_2 g_3 g_5 + g_1 g_4 g_6) \notag \\
    & \quad \times \Big( g_1^2 + g_2^2 + g_3^2 + 3 g_3 g_4 + g_4^2 + 3 g_3 g_5 + 2 g_4 g_5 + g_5^2 + 3 g_2 (g_3 + g_4 + g_5) + 2 g_2 g_6\notag \\
    & \quad + 3 (g_3 + g_4 + g_5) g_6 + g_6^2 + g_1 (3 g_2 + 2 g_3 + 3 (g_4 + g_5 + g_6)) \Big) \omega_c\notag
\end{align}

Utilizing these expressions, the average and the variance of cold bath current QRIs become
\begin{align}
\langle J^{I}_{c} \rangle &=
\langle \dot{N}^{I} \rangle \omega_c
, \,\,\,\mbox{and}  \,\,\,   \Delta J^{I}_{c}=\Delta \dot{N}^{I} \omega_c^2,
\end{align}

where the average and variance of photon flux are given by 
\begin{align}
\langle N^{I} \rangle &=
\frac{(n_c n_w  - n_h (1 + n_c + n_w)) \gamma_0}
{(1 + n_c) (2 + 3 n_h) + (4 + 3 n_c + 3 n_h) n_w},\,\,\, \mbox{and}\,\,\,
\  \Delta \dot{N}^{I} = {\alpha}{\langle\dot{N}^{I}\rangle} (1- \frac{2 k}{p}\langle\dot{N}^{I}\rangle^2),
\end{align}
 where $\alpha=p/m$, ${m}={n_h (n_c+n_w+1)-n_c n_w}$, $p=n_h (2 n_c n_w+n_c+n_w+1)+n_c n_w$ and $k=2 (n_c+ n_h+ n_w)+3$.
 It is important to note that the currents and variance corresponding to work and hot baths for QRIs can be obtained similarly, given as
\begin{align}
\langle J^{I}_{h} \rangle &=
-\langle \dot{N}^{I} \rangle \omega_h
, \,\,\, \mbox{and}  \,\,\,  \Delta J^{I}_{h}=\Delta \dot{N}^{I} \omega_h^2, \nonumber \\
\langle J^{I}_{w} \rangle &=
\langle \dot{N}^{I} \rangle \omega_w
, \,\,\, \mbox{and}  \,\,\,    \Delta J^{I}_{w}=\Delta \dot{N}^{I} \omega_w^2.
\end{align}

To determine the current statistics in QRCs, we again use the Full Counting Statistics (FCS) technique, which includes counting fields in the master equation. Let $\chi_{h}$, $\chi_{c}$, and $\chi_{w}$ be counting fields for the hot, cold, and work baths, respectively. The dressed Lindblad master equation~\eqref{QHER} of QRCs can be written as 
\begin{align}
    & \frac{d{\rho}(\chi,t)}{dt} = -i[H_{S},\rho] +  \gamma_{1}(e^{i(\omega_h\chi_h-\omega_c\chi_c)}b_{hc}{\rho}(t) b_{hc}^{\dagger} 
    - \frac{1}{2}\{b_{hc}^{\dagger}    b_{hc} ,{\rho}(t)\}) +  \gamma_{2}(e^{-i(\omega_h\chi_h-\omega_c\chi_c)}b_{hc}^{\dagger}{\rho}(t) b_{hc} - \frac{1}{2}\{b_{hc}b_{hc}^{\dagger},{\rho}(t)\}) \\
     &+\gamma_{w}(n_{w}+1)(e^{-i\omega_w\chi_w}b_{w}{\rho} b^{\dagger}_{w} - \frac{1}{2}\{b^{\dagger}_{w}  b_{w},{\rho}_R\}) +\gamma_{w}n_{w}( e^{i\omega_w\chi_w}b^{\dagger}_{w}{\rho}_R b_{w} - \frac{1}{2}\{b_{w}b^{\dagger}_{w},{\rho}\}).
\end{align}

Utilizing FCS like QRIs, the average and variance of the  baths (cold, hot and work) currents in QRCs can be obtained and are given as
\begin{align}
\langle J^{C}_{c} \rangle &=
\langle \dot{N}^{C} \rangle \omega_c
, \mbox{and}  \   \Delta J^{C}_{c}=\Delta \dot{N}^{C} \omega_c^2, \nonumber \\
\langle J^{C}_{h} \rangle &=
-\langle \dot{N}^{C} \rangle \omega_h
, \,\,\, \mbox{and}  \,\,\,    \Delta J^{C}_{h}=\Delta \dot{N}^{C} \omega_h^2, \nonumber \\
\langle J^{C}_{w} \rangle &=
\langle \dot{N}^{C} \rangle \omega_w
, \,\,\, \mbox{and}  \,\,\,    \Delta J^{C}_{w}=\Delta \dot{N}^{C} \omega_w^2.
\end{align}

where the average and variance of photon flux are given by 
\begin{align}
\langle \dot{N}^{C} \rangle &=
\frac{ \gamma_1 \gamma_4-\gamma_2 \gamma_3}{\gamma_1 + \gamma_2 + \gamma_3 + \gamma_4},\ \mbox{and}
\  \Delta \dot{N}^{C} = {\alpha}{\langle\dot{N}^{C}\rangle} (1- \frac{2 }{p}\langle\dot{N}^{C}\rangle^2),
\end{align}
 where $\alpha=p/m$, ${m}=\gamma_1 \gamma_4-\gamma_2 \gamma_3={n_h (n_c+n_w+1)-n_c n_w}$, and $p=\gamma_1 \gamma_4+\gamma_2 \gamma_3=n_h (2 n_c n_w+n_c+n_w+1)+n_c n_w$.
 
 Note that the currents corresponding to the cold bath and the work bath are positive, as they represent energy flowing out of the baths, while the current corresponding to the hot bath is negative, as it represents energy flowing into the bath in the refrigeration regime. Moreover, this fact yields the refrigeration condition given in Eq.~\eqref{ref-cond-conve} of the main text.

\section{Comparison of cooling powers and noise-to-signal ratios for QRCs and QRIs}\label{comQRIvsQRC}
The cooling power (cold bath current) and its variance are proportional to the average photon flux and its fluctuation, respectively. Therefore, we are required to compare the performance of QRCs and QRIs, we first need to compare the average photon flux and its fluctuation. Let us write the photon flux of QRCs and QRIs in terms of the steady state density matrix elements as 
\begin{equation}\label{photonN}
   \langle\dot{N}^{C} \rangle= \sigma_{C}^{00} \Big(n_{w}-(n_{w}+1)\frac{\sigma_{C}^{11}}{\sigma_{C}^{00}}\Big), \,\,\, \mbox{and} \,\,\,  \langle\dot{N}^{I} \rangle = \sigma_{I}^{00} \Big(n_{w}-(n_{w}+1)\frac{\sigma_{I}^{11}}{\sigma_{I}^{00}}\Big),
\end{equation}
where the population ratio of the first exited state and grounds state can be written as
\begin{equation}\label{DME}
 \frac{\sigma^{11}_{C}}{\sigma_{C}^{00}}= \frac{\gamma_{hc}  (n_{c}+1) n_{h}+\gamma_{w} n_{w}}{\gamma_{hc}  n_{c} (n_{h}+1)+\gamma_{w} (n_{w}+1)} , \,\,\, \mbox{and} \,\,\,   \frac{\sigma^{11}_{I}}{\sigma^{00}_{I}}= \frac{\gamma_c \gamma_h (n_c+1) n_{h}+\gamma_w n_{w} (\gamma_c+\gamma_h+\gamma_c n_c+\gamma_h n_{h})}{\gamma_c \gamma_h n_c (n_{h}+1)+\gamma_w (n_{w}+1) (\gamma_c+\gamma_h+\gamma_c n_c+\gamma_h n_{h})}.
\end{equation}
For fair comparison, we consider $\gamma_h=\gamma_c=\gamma_{hc}=\gamma_w=\gamma_0$ for further calculations in this section as well as in this work. Moreover, to compare $ \langle\dot{N}_{C} \rangle$ and $ \langle\dot{N}_{I} \rangle$, we first need to compare the ground state population and its ratio with the exited state population of QRCs and QRIs. To begin with, let us recall the condition of refrigeration, given as
\begin{equation}
    \beta_s = \frac{ \beta_h \omega_h- \beta_w \omega_w}{\omega_h-\omega_w}> \beta_c.
\end{equation}
Using the above condition and using the fact that $\omega_c=\omega_h-\omega_w$, we obtain
\begin{equation}
    \beta_h \omega_h- \beta_w \omega_w > \beta_c \omega_c.
\end{equation}
The above condition can be equivalently written as
\begin{equation}
    e^{-\beta_h \omega_h+ \beta_c \omega_c}< e^{-\beta_w \omega_w},\,\,\, \mbox{or}  \,\,\, \frac{n_h(n_c+1)}{n_c(n_h+1)} < \frac{n_w}{n_w+1}.
\end{equation}
Now, using the fact that if $a/b<c/d$ then we can write $a/b<(a+c)/(b+d)<(a+mc)/(b+md)<c/d$ with $m>1$, we obtain following condition
\begin{equation}
    \frac{n_h(n_c+1)+n_w}{n_c(n_h+1)+n_{w}+1} <  \frac{n_h(n_c+1)+xn_w}{n_c(n_h+1)+x(n_{w}+1)},
\end{equation}
where, we assume $x = (n_h + n_c +2)>2$.
Using the above condition and Eq.~\eqref{DME}, we obtain the following conditions on the density matrix elements
\begin{equation}\label{DD}
    \frac{\sigma_C^{11}}{\sigma_C^{00}} < \frac{\sigma_{I}^{11}}{\sigma_{I}^{00}} \Rightarrow  {\sigma^{00}_I}< \frac{\sigma^{00}_{I}}{1-\sigma^{22}_{I}} < {\sigma_C^{00}}.
\end{equation}
where the last condition is obtained by adding both sides of the first condition and using the fact that $\sigma_{00} + \sigma_{11}=1-\sigma_{22}$. Now, using the Eq.~\eqref{photonN}, the ratio of photon flux can be written as 
\begin{equation}
\frac{\langle\dot{N}^{C} \rangle}{\langle\dot{N}^{I} \rangle} =\frac{ \sigma_{C}^{00} (n_{w}-(n_{w}+1)\frac{\sigma_{C}^{11}}{\sigma_{C}^{00}})}{\sigma_{I}^{00} (n_{w}-(n_{w}+1)\frac{\sigma_{I}^{11}}{\sigma_{I}^{00}})} =\frac{\sigma_{C}^{00}}{\sigma_{I}^{00}}\frac{  (1- \frac{\frac{n_h(n_c+1)}{n_w}+1}{\frac{n_c(n_h+1)}{n_{w}+1}+1})}{ (1- \frac{\frac{n_h(n_c+1)}{n_w}+x}{\frac{n_c(n_h+1)}{n_{w}+1}+x}))},
\end{equation}
where last equality obtained by taking common the term $\frac{n_{w}}{n_{w}+1}$ from numerator and denominator and canceling it. For further simplification, let us assume $a=\frac{n_h(n_c+1)}{n_w}$ and $b=\frac{n_c(n_h+1)}{n_{w}+1}$, then we can rewrite above expression as
\begin{equation}
\frac{\langle\dot{N}^{C} \rangle}{\langle\dot{N}^{I} \rangle} =\frac{\sigma_{C}^{00}}{\sigma_{I}^{00}}\frac{  (1- \frac{a+1}{b+1})}{ (1- \frac{a+x}{b+x}) }.
\end{equation}

Let us add and subtract the term $\frac{a+x}{b+x}$ in the numerator on the right-hand side of the above equation, and then we obtain
\begin{align}
\frac{\langle\dot{N}^{C} \rangle}{\langle\dot{N}^{I} \rangle} &=\frac{\sigma_{C}^{00}}{\sigma_{I}^{00}}\frac{  (1- \frac{a+1}{b+1}+\frac{a+x}{b+x}-\frac{a+x}{b+x})}{ (1- \frac{a+x}{b+x}) } =\frac{\sigma_{C}^{00}}{\sigma_{I}^{00}}(1+\frac{  (\frac{a+x}{b+x}- \frac{a+1}{b+1})}{ (1- \frac{a+x}{b+x}) } )\nonumber\\
&=\frac{\sigma_{C}^{00}}{\sigma_{I}^{00}}(1+\frac{1}{b+1}\frac{  ((a+x)(b+1)- (a+1)(b+x))}{ ({b}- {a}) } )\nonumber\\
 &=\frac{\sigma_{C}^{00}}{\sigma_{I}^{00}}(1+\frac{1}{b+1}\frac{  ((b-a)(x-1))}{ ({b}- {a}) } )\nonumber\\
&=\frac{\sigma_{C}^{00}}{\sigma_{I}^{00}}(1+\frac{x-1}{b+1})\nonumber\\
&=\frac{\sigma_{C}^{00}}{\sigma_{I}^{00}}(1+\frac{1+n_h+n_c}{1+\frac{n_c(n_h+1)}{n_{w}+1}})
\end{align}

Since we observe that ${n_h+n_c}>\frac{n_c(n_h+1)}{n_{w}+1}$ which implies $(n_{w}+1)n_h+n_{w}n_c>{n_cn_h)}$ and this is trivially satisfying because $n_w>n_h$, i.e., $\beta_w \omega_w < \beta_h \omega_h$), then we can obtain a following lower bound given as
\begin{equation}\label{RTP}
\frac{\langle\dot{N}^{C} \rangle}{\langle\dot{N}^{I} \rangle} >2\frac{\sigma_{C}^{00}}{\sigma_{I}^{00}}>2,
\end{equation}
where the last inequality is obtained using
$\frac{ \sigma^{C}_{00} }{\sigma^{I}_{00}} >1$, which is follows from Eq.~\ref{DD}. As mentioned in the  main text, this condition suggests that the cooling power (current) of QRCs is always greater than twice the cooling power of QRIs.

The noise-to-signal ratio in cooling powers of QRCs and QRIs given as
\begin{equation}
    \mathcal{N}^{C}_c = \frac{\alpha}{\langle\dot{N}^{C}\rangle} (1- \frac{2}{p}\langle\dot{N}^{C}\rangle^2),\ \mbox{and} \ \mathcal{N}^{I} = \frac{\alpha}{\langle\dot{N}^{I}\rangle} (1- \frac{2 k}{p}\langle\dot{N}^{I}\rangle^2),
\end{equation}
where $\alpha=p/m$, ${m}= \gamma_1 \gamma_4-\gamma_2 \gamma_3={n_h (n_c+n_w+1)-n_c n_w}$, $p=\gamma_1 \gamma_4+\gamma_2 \gamma_3=n_h (2 n_c n_w+n_c+n_w+1)+n_c n_w$ and $k=2 (n_c+ n_h+ n_w)+3$. 
The ratio of noise-to-signal ratios can be written as 
\begin{equation}
\frac{ \mathcal{N}_c^{I}}{  \mathcal{N}_c^{C}}= \frac{\langle\dot{N}^{C} \rangle}{\langle\dot{N}^{I} \rangle}\frac{1- \frac{2 k}{p}(\langle\dot{N}_{I} \rangle)^2}{1- \frac{2}{p}( \langle\dot{N}_{C} \rangle)^2}.
\end{equation}
Let us assume the $c=\frac{1- \frac{2 k}{p}(\langle\dot{N}_{I} \rangle)^2}{1- \frac{2}{p}( \langle\dot{N}_{C} \rangle)^2}>1$, which implies $\frac{\langle\dot{N}_{C} \rangle}{\langle\dot{N}_{I} \rangle}\geq \sqrt{k}$. Note that $k$ must follow the condition $k>4$. Otherwise, it leads to a violation of Eq.~\eqref{RTP}. Hence, $c>1$ is true. Thus, we obtain the following lower bound the ratio of noise-to-signal ratio of QRCs and QRIs given as
\begin{equation}
\frac{ \mathcal{N}_c^{I}}{  \mathcal{N}_c^{C}}> \frac{\langle\dot{N}^{C} \rangle}{\langle\dot{N}^{I} \rangle}  >2 \implies \frac{ \mathcal{N}_c^{I}}{ 2} >\mathcal{N}_c^{C}.
\end{equation}
The proof therefore implies that QRCs are {\it at least} twofold more reliable than QRIs.

\section{Engineering a synthetic negative temperature thermal bath}
\label{NT}
Let us consider a three-level quantum system coupled with the two bosonic (photonic) thermal baths having unequal temperatures, namely $R$ and $L$, where the system and baths interact via two-photon transitions (Raman interactions, i.e., three-body interactions). The total Hamiltonian of the qutrit system and two thermal baths can be written as
\begin{equation}\label{totalEq}
    H = H_{S} + H_{B_{L}} + H_{B_{R}} + H_{SB_LB_R}.
\end{equation}
where suffixes $L$ and $R$ correspond to hot and cold baths, respectively. We assume $\hbar = k_{B} = 1$ throughout this work. In Eq.~\eqref{totalEq}, the system Hamiltonian $H_S$ describes a three-level system (qutrit), given by
\begin{equation}
    H_{S}= \omega_{L} \ketbra{2} + (\omega_{L}-\omega_{R})b_{LR}^{\dagger}b_{LR} = \omega_{L} \ketbra{2} + (\omega_{L}-\omega_{R})\ketbra{1}.
\end{equation}
where $\omega_{L}$ and  $\omega_{L}-\omega_{R}$ refers to the frequencies corresponding to the energy gaps, and $b_{LR}^{\dagger}=\ket{1}\bra{0}$ and $b_{LR}=\ket{0}\bra{1}$. In Eq.~\eqref{totalEq}, the photonic baths are a collection of infinite dimensional systems whose total Hamiltonian is given as
\begin{equation}
 H_{B_{L}} +   H_{B_{R}} = \sum_{k} \Omega_{k,L} a^{\dagger}_{k,L}a_{k,L} + \sum_{k'} \Omega_{k',R} a^{\dagger}_{k',R}a_{k',R}.
\end{equation}
Furthermore, in Eq.~\eqref{totalEq}, the interaction Hamiltonian between the system and the baths has the following form~\cite{Gerry1990, Gerry1992, Wu1996}
\begin{equation}
    H_{SB_LB_R}= g_0 \sum_{kk'}  (a_{k,L}a^{\dagger}_{k',R}b_{LR}^{\dagger}+a^{\dagger}_{k,L}a_{k',R}b_{LR}).
\end{equation}
Note that the qutrit system interacts with two thermal baths, each at inverse temperatures \(\beta_L\) and \(\beta_R\). In this configuration, the bath $L$ (with inverse temperature $\beta_L$) is coupled to levels $|1\rangle$ and $|2\rangle$ with energy spacing \(\omega_L\), while the bath R (with inverse temperature $\beta_R$) is connected to levels $|1\rangle$ and $|2\rangle$ with spacing \(\omega_R\).
Here, we consider system-baths coupling ($g_0$) to be weak. The Lindblad master equation can be derived~\cite{Mohan2024}, which is given by
\begin{align}
     \dot{\rho}&= -i [H_{S},\rho] +  \gamma_{1}^{LR}\left(b_{LR}{\rho}(t) b_{LR}^{\dagger} 
    - \frac{1}{2}\{b_{LR}^{\dagger}    b_{LR} ,{\rho}\}\right) +  \gamma_{2}^{LR}\left(b_{LR}^{\dagger}{\rho} b_{LR} - \frac{1}{2}\{b_{LR}b_{LR}^{\dagger},{\rho}\}\right)
\end{align}
with $\gamma_1^{LR}=\gamma_0  n_R (n_L+1)$ and $\gamma_2^{LR}=\gamma_0  (n_R+1)n_{L}$. 
This dynamics leads to a steady state $\rho_{ss}$, i.e., $\dot{\rho}_{ss}=0$, given by 
\begin{equation}
    \rho_{ss} = \frac{\text{$\gamma $}_1^{LR} }{(\text{$\gamma $}_1^{LR}+\text{$\gamma $}_2^{LR})}\op{0} + \frac{\text{$\gamma $}_2^{LR} }{(\text{$\gamma $}_1^{LR}+\text{$\gamma $}_2^{LR})}\op{1}.
\end{equation}
For the steady state, the ratio of populations of exited state $\ket{1}$ and ground state $\ket{0}$ is given as 
\begin{equation}\label{inv}
    \frac{\rho^{(11)}_{ss}}{\rho^{(00)}_{ss}} = \frac{\gamma_2^{LR}}{\gamma_1^{LR}}= e^{{-(\beta_{L}\omega_{L}-\beta_{R}\omega_{R})}}=  e^{-\frac{(\beta_{L}\omega_{L}-\beta_{R}\omega_{R})}{(\omega_{L}-\omega_{R})}(\omega_{L}-\omega_{R})},
\end{equation}
where $\rho^{(ij)}_{ss}=\bra{i} \rho_{ss} \ket{j}$.

To have a population inversion by utilizing two-photon transitions, we need the condition $\frac{\rho^{(11)}_{ss}}{\rho^{(00)}_{ss}} >1$. For this, the required condition is $\beta_{L}\omega_{L}-\beta_{R}\omega_{R} <0$. This also implies $n_L>n_R$. Since the third energy level $|2\rangle$ got adiabatic eliminated, therefore by observing Eq.~\eqref{inv}, we can think of the total system as an effective two-level system attached with effective thermal bath at inverse temperature $\beta_{\rm eff}=\frac{(\beta_{L}\omega_{L}-\beta_{R}\omega_{R})}{(\omega_{L}-\omega_{R})}$. Since we can have a situation $\beta_{L}\omega_{L}-\beta_{R}\omega_{R} <0$ which correspond to population inversion, in such situation $\beta_{\rm eff}$ becomes negative, which is not possible with conventional thermal bath. The steady-state thermodynamics of quantum systems attached to synthetic temperature baths is studied in detail in Ref.~\cite{Bera2024}.

\section{Autonomous refrigeration with synthetic temperature work bath}\label{QRCsNT}
In this section, we discuss the refrigerator model that utilizes the synthetic temperature work bath instead of the conventional work bath. The synthetic work bath consists of two conventional work baths, which are attached to two distinct energy spacings, which we describe explicitly described in the section. Let us consider models of a refrigerator utilizing a four-level system (characterized by the Hamiltonian \(H_{S} =\omega_{w_1} \ketbra{3} +\omega_{h} \ketbra{2} +  (\omega_{h} -\omega_{c})\ketbra{1}\) ) interacting with four thermal baths, each at inverse temperatures \(\beta_c\), \(\beta_h\), \(\beta_{w_{1}}\) and \(\beta_{w_{2}}\). In this configuration, the hot bath (with inverse temperature $\beta_h$) is coupled to levels $|0\rangle$ and $|2\rangle$ with energy spacing \(\omega_h\), while the cold bath (with inverse temperature $\beta_c$) is connected to levels $|1\rangle$ and $|2\rangle$ with spacing \(\omega_c\). The work baths (with inverse temperatures $\beta_{w_1}$ and $\beta_{w_2}$) is coupled to levels $|0\rangle$ and $|3\rangle$, and $|1\rangle$ and $|3\rangle$ respectively, with energy spacings \(\omega_{w_{1}}\) and \(\omega_{w_{2}} = \omega_{w_{1}}-(\omega_h - \omega_c)\), respectively.
The total Hamiltonian of the four-level system + baths is given as

\begin{align}
H=H_{S} + H_{B_h} + H_{B_c} + H_{B_{w_1}} + H_{B_{w_2}}+ H_{SB_hB_cB_{w_1}B_{w_2}},    
\end{align}

where $H_{S}$ is the Hamiltonian of the four-level system, $H_{B_h}=\sum_k \Omega_{k,h} \ a^\dag_{k,h} a_{k,h}$, $H_{B_c}= \sum_{k'} \Omega_{k',c} a^\dag_{k',c} a_{k',c}$, $H_{B_{w_{1}}}= \sum_{k''} \Omega_{k'',w_{1}} a^\dag_{k'',w_{1}} a_{k'',w_{1}}$ and $H_{B_{w_{2}}}= \sum_{k'''} \Omega_{k''',w_{2}} a^\dag_{k''',w_{2}} a_{k''',w_{2}}$ are the Hamiltonians of the hot, cold and work baths with mode frequencies $\Omega_{k,h}$, $\Omega_{k',c}$, $\Omega_{k',w_{1}}$ and $\Omega_{k',w_{2}}$ respectively, and $H_{SB_hB_cB_{w_1}B_{w_2}}$ represents the interaction between the qutrit and the baths. The interaction between the system and baths $H_{SB_hB_cB_{w_1}B_{w_2}}$ is given as
\begin{equation}
H_{SB_hB_hB_w}^I= H_{SB_hB_c}+ H_{SB_{w_1}B_{w_2}},
\end{equation}
where 
\[
H_{SB_hB_c}= g_{hc} \sum_{k,k'} (a_{k,h} a_{k',c}^\dag b_{hc}^\dag + a_{k,h}^\dag a_{k',c} b_{hc}), \ \mbox{and} \ 
H_{SB_{w_1}B_{w_2}}= g_{w_1w_2} \sum_{k'',k'''} (a_{k,w_1} a_{k''',w_2}^\dag b_{w_1w_2}^\dag + a_{k'',w_1}^\dag a_{k''',w_2} b_{w_1w_2}).
\]
with $b_{hc}=|0\rangle \langle 1|$ and $b_{w_1w_2}=|0\rangle \langle 1|$, and their conjugates. For very weak system-baths couplings ($g_{hc}$, and $g_{w_1w_2}$), the local dynamics of the four-level system reduces to~\cite{Mohan2024}
\begin{align}\label{QCN}
     \dot{\rho}& = -i[H_{S},\rho] +  \gamma_{1}(b_{hc}{\rho} b_{hc}^{\dagger} 
    - \frac{1}{2}\{b_{hc}^{\dagger}    b_{hc} ,{\rho}\}) +  \gamma_{2}(b_{hc}^{\dagger}{\rho}(t) b_{hc} - \frac{1}{2}\{b_{hc}b_{hc}^{\dagger},{\rho}\}) \nonumber \\
    &+  \gamma'_{3}(b_{w_1w_2}{\rho} b_{w_1w_2}^{\dagger} 
    - \frac{1}{2}\{b_{w_1w_2}^{\dagger}    b_{w_1w_2} ,{\rho}\}) +  \gamma'_{4}(b_{w_1w_2}^{\dagger}{\rho} b_{w_1w_2} - \frac{1}{2}\{b_{w_1w_2}b_{w_1w_2}^{\dagger},{\rho}\}).
\end{align}
with decay rates $\gamma_1=\gamma_{hc}  n_c (n_h+1)$, $\gamma_2=\gamma_{hc}  (n_c+1)n_{h}$, $\gamma_3'=\gamma_{w_1w_2} n_{w_2}(n_{w_1}+1)$, $\gamma_{4}'=\gamma_0n_{w_1}(n_{w_2}+1)$, and $\gamma_{x}$ (where $x\in\{hc, w_{1}w_{2}\}$) is Weiskopf-Wigner decay constant. Note that we have considered $\gamma_{hc}=\gamma_{w_1w_2}=\gamma_0$ in this work. The steady-state solution of the above master equation can be obtained by solving $\dot{\rho}=0$ (we denote the steady state by $\sigma_{SC}$), and it is 
\begin{equation}
\sigma_{SC} = 
\frac{\gamma'_2 + \gamma'_4}{\gamma'_1 + \gamma'_2 + \gamma'_3 + \gamma'_4} \op{1} + \frac{\gamma'_1 + \gamma'_3}{\gamma'_1 + \gamma'_2 + \gamma'_3 + \gamma'_4} \op{0}
\end{equation}

To determine the current statistics in this refrigerator model, we again use the Full Counting Statistics (FCS) technique, which includes counting fields in the master equation. Let $\chi_{h}$, $\chi_{c}$, and $\chi_{w}=\chi_{w_1}=\chi_{w_2}$ be counting fields for the hot, cold, and work baths, respectively. The dressed Lindblad master equation~\eqref{QCN} can be written as 
\begin{align}
     \frac{d{\rho}(\chi,t)}{dt}& = -i[H_{S},\rho(t)] +  \gamma_{1}(e^{i(\omega_h\chi_h-\omega_c\chi_c)}b_{hc}{\rho}(t) b_{hc}^{\dagger} 
    - \frac{1}{2}\{b_{hc}^{\dagger}    b_{hc} ,{\rho}(t)\}) +  \gamma_{2}(e^{-i(\omega_h\chi_h-\omega_c\chi_c)}b_{hc}^{\dagger}{\rho}(t) b_{hc} - \frac{1}{2}\{b_{hc}b_{hc}^{\dagger},{\rho}(t)\}), \nonumber \\
    & +  \gamma_{3}'(e^{i((\omega_{w_1}-\omega_{w_2})\chi_w)}b_{w_1w_2}{\rho}(t) b_{w_1w_2}^{\dagger} 
    - \frac{1}{2}\{b_{w_1w_2}^{\dagger}    b_{w_1w_2} ,{\rho}(t)\}) +  \gamma_{4}'(e^{-i((\omega_{w_1}-\omega_{w_2})\chi_w)}b_{w_{1}w_{2}}^{\dagger}{\rho}(t) b_{w_1w_2} - \frac{1}{2}\{b_{w_1w_2}b_{w_1w_2}^{\dagger},{\rho}(t)\}).
\end{align}

Utilizing FCS like QRIs, the average and the variance of cold bath current become
\begin{align}
\langle J^{SC}_c \rangle &=
\langle \dot{N}^{SC} \rangle \omega_c
, \mbox{and}  \   \Delta J^{SC}_c=\Delta \dot{N}^{SC} \omega_c^2,
\end{align}
where
\begin{equation}
  \langle  \dot{N}^{SC} \rangle =   \frac{  \gamma_1 \gamma'_4-\gamma_2 \gamma'_3}{\gamma_1 + \gamma_2 + \gamma'_3 + \gamma'_4},  
    \quad 
    \Delta \dot{N}^{SC} = {\alpha'}{\langle\dot{N}^{SC}\rangle} \left(1- \frac{2}{p'}\langle\dot{N}^{SC}\rangle^2 \right),
\end{equation}  
with \( \alpha' = p'/m' \), \( {m}' = \gamma_1 \gamma_4'-\gamma_2 \gamma_3' \), and \( p' = \gamma_1 \gamma_4'+\gamma_2 \gamma_3' \).

It is important to note that the currents and variance corresponding to work and hot baths in this model of refrigerator can be obtained similarly. It is worthwhile to mention that the model discussed in the section can be thought of as QRCs with synthetic temperature work baths. This is due to the composition of two work baths, which can be effectively thought of as a single work bath with synthetic temperature attached to level $|0\rangle$ and $|1\rangle$ with energy spacing $\omega_w=\omega_{w_1}-\omega_{w_2}$. We refer to this model as QRCNs (QRCs with synthetic negative-temperature work baths) and compare its performance with QRCs that use conventional work baths in the next section~\ref{compQRCsVsQRCsNT}. QRCs with synthetic positive-temperature work baths are equivalent to QRCs with conventional work baths. Note that the currents corresponding to the cold bath is positive, as they represent energy flowing out of the cold bath, while the current corresponding to the hot bath is negative, as it represents energy flowing into the bath in the refrigeration regime. Moreover, this fact  yields the refrigeration condition given in Eq.~\eqref{SWL} of the main text.

\section{Comparison of cooling powers and noise-to-Signal ratios of QRCs with positive temperature work bath vs QRCs with negative temperature work bath (QRCNs)}\label{compQRCsVsQRCsNT}

Let us assume the following condition
$\gamma'_4\geq\gamma_4$, $\gamma_3\geq\gamma'_3$ and by multiplying these inequalities we can obtain $\gamma'_4\gamma_3\geq\gamma_4\gamma'_3$, that implies  ${\gamma'_4}/{\gamma'_3}\geq{\gamma_4}/{\gamma_3}$. Now, by adding 1 on both the sides of inequality ${\gamma'_4}/{\gamma'_3}\geq{\gamma_4}/{\gamma_3}$, we obtain 

\begin{equation}
    {(\gamma'_4+\gamma'_3)}/{\gamma'_3}\geq{(\gamma_4+\gamma_3)}/{\gamma_3}
\end{equation}

Let us recall the inequality
$\gamma'_3\leq\gamma_3$, which can be rewritten as $ {(\gamma_1+\gamma_2)}/{\gamma'_3}\geq{(\gamma_1+\gamma_2)}/{\gamma_3}$.
Now by utilizing this inequality with the above we obtain 

\begin{equation}
    {(\gamma_1+\gamma_2+\gamma'_3+\gamma'_4)}/{\gamma'_3}\geq{(\gamma_1+\gamma_2+\gamma_3+\gamma_4)}/{\gamma_3} \implies  {\gamma'_3}/{(\gamma_1+\gamma_2+\gamma'_3+\gamma'_4)}\leq{\gamma_3}/{(\gamma_1+\gamma_2+\gamma_3+\gamma_4)}.
\end{equation}

similarly, we can prove 
\begin{equation}
    {\gamma'_4}/{(\gamma_1+\gamma_2+\gamma'_3+\gamma'_4)}\geq{\gamma_4}/{(\gamma_1+\gamma_2+\gamma_3+\gamma_4)}.
\end{equation}
Utilizing these inequalities, we obtain the inequalities between average photon flux in QRCs and QRCNs as
\begin{equation}
    \langle \dot{N}^{C} \rangle =   \frac{  \gamma_1 \gamma_4-\gamma_2 \gamma_3}{\gamma_1 + \gamma_2 + \gamma_3 + \gamma_4} \leq \langle \dot{N}^{SC} \rangle =   \frac{  \gamma_1 \gamma'_4-\gamma_2 \gamma'_3}{\gamma_1 + \gamma_2 + \gamma'_3 + \gamma'_4}.
\end{equation}
which implies that average photon flux is larger for QRCN than the case for QRC. 

The noise-to-signal ratios in cooling powers of QRCs and QRCNs can be written as
\begin{equation}
    \mathcal{N}^{C}_c = \frac{\alpha}{\langle\dot{N}^{C}\rangle} (1- \frac{2}{p}\langle\dot{N}^{C}\rangle^2),\ \mbox{and} \ \mathcal{N}^{SC}_c = \frac{\alpha'}{\langle\dot{N}^{SC}\rangle} (1- \frac{2}{p'}\langle\dot{N}^{SC}\rangle^2).
\end{equation}
 where $\alpha=p/m$, ${m}=\gamma_1 \gamma_4-\gamma_2 \gamma_3$, $p=\gamma_1 \gamma_4+\gamma_2 \gamma_3$, $\alpha'=p'/m'$, ${m}'=\gamma_1 \gamma_4'-\gamma_2 \gamma_3'$, and $p'=\gamma_1 \gamma_4'+\gamma_2 \gamma_3'$.
 Note that, the inequalities $\gamma'_4\geq\gamma_4$, $\gamma_3\geq\gamma'_3$ leads to $m'\geq m$. We can show that the inequality $p'<p$ holds if following condition is satisfied
\begin{equation}
    \frac{\gamma_2}{\gamma_1}\leq\frac{\gamma'_4-\gamma_4}{\gamma_3-\gamma'_3}.
\end{equation}
Moreover, the inequalities  $m'\geq m$ and $p'<p$ leads to $\alpha \geq \alpha'$. Furthermore, implying the inequalities $m'\geq m$, $p'<p$, $\alpha \geq \alpha'$ and $\langle \dot{N}^{C} \rangle  \leq \langle \dot{N}^{SC} \rangle$. we can infer that 
\begin{equation}
    \mathcal{N}^{C}_c = \frac{\alpha}{\langle\dot{N}^{C}\rangle} (1- \frac{2}{p}\langle\dot{N}^{C}\rangle^2) \geq \mathcal{N}^{SC}_c = \frac{\alpha'}{\langle\dot{N}^{SC}\rangle} (1- \frac{2}{p'}\langle\dot{N}^{SC}\rangle^2).
\end{equation}
which implies that the reliability for the QRCN case is more than that of the QRC case.

\section{Thermodynamic uncertainty relation (TUR)}\label{CTUR}
Classical steady-state heat engines always exhibit trade-off relationships between relative fluctuation in current and the thermodynamic cost (quantified by the rate of entropy production $\dot{S}$), which is given as~\cite{Barato2015}
\begin{align}
\mathcal{Q}_{X} &=\dot{S} \frac{ \Delta {J}^{X}_{x}}{\langle {J}^{X}_{x} \rangle ^2} \geq 2\label{appeq:cTUR},
 \end{align}
where $\dot{S}$ the rate of entropy production. Note, Eq.~\eqref{appeq:cTUR} is referred to as the classical thermodynamic uncertainty relation (cTUR)~\cite{Barato2015}. 
The entropy production rate $\dot{S}$ for QRI and QRC can be written as (for $X=C,I$)
\begin{equation}
  \dot{S}_{X}=- \beta_{h} \langle {J}_{h}^{X} \rangle - \beta_{c}\langle {J}^{X}_{c} \rangle- \beta_{w} \langle {J}_{w}^{X} \rangle=( \beta_{s}  - \beta_{c})\omega_c \langle \dot{N}^{X} \rangle > 0,
\end{equation}
where ${\langle\dot{N}^{X}\rangle}={|\langle{J}^{X}\rangle}|/(\omega_x)$ is the average photon number flux and \(
\beta_s = \frac{\beta_h \omega_h - \beta_{w} \omega_w}{\omega_h - \omega_w}
\).
Using above relations, we can show that
\begin{equation}~\label{Unified}
    \mathcal{Q}_{X}= (\beta_{s}  - \beta_{c})\omega_cF_{X}>2.
\end{equation}
Here $F_{X}=\frac{\Delta \dot{N}^X}{\langle\dot{N}^X\rangle}$ is known as the Fano factor of photon number current ($\dot{N}$), where ${\langle\dot{N}^{X}\rangle}={|\langle{J}^{X}_{x}\rangle}|/(\omega_x)$ and ${\Delta \dot{N}^{X}}=\Delta{J^{X}_{x}}/(\omega_{x})^2$ (with $x=h,c,w$) are variance and average of photon number current for the steady state dynamics. 

\section{Experimental feasibility of autonomous refrigerators with correlated  heat transfer}\label{exp}
The key difference between QRCs and QRIs lies in the interaction between three-level systems (working systems) and hot and cold baths. In QRCs, the working system, hot bath, and cold bath interact via three-body interaction (see Eq.~\eqref{RTT}), while in QRIs, interaction with the hot or cold bath is via two-body interaction independently (see Eq.~\eqref{TQC}). Both of these refrigerator models can be realized with three-level atoms ($\Lambda$ type atoms) interacting with three external quantized electromagnetic fields at unequal temperatures. Important to note that these $\Lambda$ type atoms are extensively studied in the standard quantum optics literature for both cases when $\Lambda$ type atoms interact with two different electric fields independently via two-body interaction (one photon transition) or collectively via three-body interaction (for two-photon transition, i.e., referred as Raman transition), for details see Refs.~\citep{Gerry1990, Gerry1992, Wu1996,Cohen1998, *Haroche2006, *Meystre2021, *Larson2021}. Note that to obtain a three-body interaction in a $\Lambda$ system via the Schrieffer--Wolff transformation (or adiabatic elimination).  
The first field couples $\ket{0} \leftrightarrow \ket{2}$ with Rabi frequency $\Omega_{1}$ and detuning $\Delta$, while the second field couples $\ket{1} \leftrightarrow \ket{2}$ with Rabi frequency $\Omega_{2}$ and the same detuning $\Delta$ (satisfying the two-photon resonance condition).  
If $|\Delta| \gg \Omega_{1,2}$, the excited state $\ket{2}$ can be adiabatically eliminated, yielding an effective Raman interaction between levels $\ket{0}$ and $\ket{1}$ and fields, which corresponds to the desired three-body interaction Hamiltonian (see Refs.~\cite{Gerry1990, Gerry1992, Wu1996, Cohen1998}). Moreover, such a setup can also be experimentally realized on various experimental platforms, such as atom-optical setup~\cite{Zanon2005, Chang2023} and superconducting circuits~\cite{Novikov2016, Kumar2016}. Therefore, both refrigerators are experimentally feasible and can be realized on the same experimental platform. Additionally, the QRCs with synthetic temperature work baths (QRCNs) can be experimentally realized with four-level systems (working systems), which interact with four different external quantized electromagnetic fields at unequal temperatures. The working system interacts with a hot bath and cold bath via a three-body interaction, and at the same time, the working system interacts with two distinct work baths via another three-body interaction. Therefore, two independent two-photon channels are utilized.

\twocolumngrid

\bibliography{che}

\begin{thebibliography}{75}%
\makeatletter
\providecommand \@ifxundefined [1]{%
 \@ifx{#1\undefined}
}%
\providecommand \@ifnum [1]{%
 \ifnum #1\expandafter \@firstoftwo
 \else \expandafter \@secondoftwo
 \fi
}%
\providecommand \@ifx [1]{%
 \ifx #1\expandafter \@firstoftwo
 \else \expandafter \@secondoftwo
 \fi
}%
\providecommand \natexlab [1]{#1}%
\providecommand \enquote  [1]{``#1''}%
\providecommand \bibnamefont  [1]{#1}%
\providecommand \bibfnamefont [1]{#1}%
\providecommand \citenamefont [1]{#1}%
\providecommand \href@noop [0]{\@secondoftwo}%
\providecommand \href [0]{\begingroup \@sanitize@url \@href}%
\providecommand \@href[1]{\@@startlink{#1}\@@href}%
\providecommand \@@href[1]{\endgroup#1\@@endlink}%
\providecommand \@sanitize@url [0]{\catcode `\\12\catcode `\$12\catcode `\&12\catcode `\#12\catcode `\^12\catcode `\_12\catcode `\%12\relax}%
\providecommand \@@startlink[1]{}%
\providecommand \@@endlink[0]{}%
\providecommand \url  [0]{\begingroup\@sanitize@url \@url }%
\providecommand \@url [1]{\endgroup\@href {#1}{\urlprefix }}%
\providecommand \urlprefix  [0]{URL }%
\providecommand \Eprint [0]{\href }%
\providecommand \doibase [0]{http://dx.doi.org/}%
\providecommand \selectlanguage [0]{\@gobble}%
\providecommand \bibinfo  [0]{\@secondoftwo}%
\providecommand \bibfield  [0]{\@secondoftwo}%
\providecommand \translation [1]{[#1]}%
\providecommand \BibitemOpen [0]{}%
\providecommand \bibitemStop [0]{}%
\providecommand \bibitemNoStop [0]{.\EOS\space}%
\providecommand \EOS [0]{\spacefactor3000\relax}%
\providecommand \BibitemShut  [1]{\csname bibitem#1\endcsname}%
\let\auto@bib@innerbib\@empty
\bibitem [{\citenamefont {Binder}\ \emph {et~al.}(2018)\citenamefont {Binder}, \citenamefont {Correa}, \citenamefont {Gogolin}, \citenamefont {Anders},\ and\ \citenamefont {Adesso}}]{Binder2018}%
  \BibitemOpen
  \bibfield  {author} {\bibinfo {author} {\bibfnamefont {Felix}\ \bibnamefont {Binder}}, \bibinfo {author} {\bibfnamefont {Luis~A.}\ \bibnamefont {Correa}}, \bibinfo {author} {\bibfnamefont {Christian}\ \bibnamefont {Gogolin}}, \bibinfo {author} {\bibfnamefont {Janet}\ \bibnamefont {Anders}}, \ and\ \bibinfo {author} {\bibfnamefont {Gerardo}\ \bibnamefont {Adesso}},\ }\href {\doibase 10.1007/978-3-319-99046-0} {\emph {\bibinfo {title} {Thermodynamics in the Quantum Regime}}},\ Vol.\ \bibinfo {volume} {195}\ (\bibinfo  {publisher} {Springer International Publishing},\ \bibinfo {year} {2018})\BibitemShut {NoStop}%
\bibitem [{\citenamefont {Deffner}\ and\ \citenamefont {Campbell}(2019)}]{DeffnerBook2019}%
  \BibitemOpen
  \bibfield  {author} {\bibinfo {author} {\bibfnamefont {Sebastian}\ \bibnamefont {Deffner}}\ and\ \bibinfo {author} {\bibfnamefont {Steve}\ \bibnamefont {Campbell}},\ }\href {\doibase 10.1088/2053-2571/ab21c6} {\emph {\bibinfo {title} {Quantum Thermodynamics}}},\ 2053-2571\ (\bibinfo  {publisher} {Morgan \& Claypool Publishers},\ \bibinfo {year} {2019})\BibitemShut {NoStop}%
\bibitem [{\citenamefont {Auff\`eves}(2022)}]{Alexia2022}%
  \BibitemOpen
  \bibfield  {author} {\bibinfo {author} {\bibfnamefont {Alexia}\ \bibnamefont {Auff\`eves}},\ }\bibfield  {title} {\enquote {\bibinfo {title} {Quantum technologies need a quantum energy initiative},}\ }\href {\doibase 10.1103/PRXQuantum.3.020101} {\bibfield  {journal} {\bibinfo  {journal} {PRX Quantum}\ }\textbf {\bibinfo {volume} {3}},\ \bibinfo {pages} {020101} (\bibinfo {year} {2022})}\BibitemShut {NoStop}%
\bibitem [{\citenamefont {Myers}\ \emph {et~al.}(2022)\citenamefont {Myers}, \citenamefont {Abah},\ and\ \citenamefont {Deffner}}]{Myers2022}%
  \BibitemOpen
  \bibfield  {author} {\bibinfo {author} {\bibfnamefont {Nathan~M.}\ \bibnamefont {Myers}}, \bibinfo {author} {\bibfnamefont {Obinna}\ \bibnamefont {Abah}}, \ and\ \bibinfo {author} {\bibfnamefont {Sebastian}\ \bibnamefont {Deffner}},\ }\bibfield  {title} {\enquote {\bibinfo {title} {{Quantum thermodynamic devices: From theoretical proposals to experimental reality}},}\ }\href {\doibase 10.1116/5.0083192} {\bibfield  {journal} {\bibinfo  {journal} {AVS Quantum Science}\ }\textbf {\bibinfo {volume} {4}},\ \bibinfo {pages} {027101} (\bibinfo {year} {2022})}\BibitemShut {NoStop}%
\bibitem [{\citenamefont {Mukherjee}\ and\ \citenamefont {Divakaran}(2024)}]{Mukherjee2024}%
  \BibitemOpen
  \bibfield  {author} {\bibinfo {author} {\bibfnamefont {Victor}\ \bibnamefont {Mukherjee}}\ and\ \bibinfo {author} {\bibfnamefont {Uma}\ \bibnamefont {Divakaran}},\ }\bibfield  {title} {\enquote {\bibinfo {title} {The promises and challenges of many-body quantum technologies: A focus on quantum engines},}\ }\href {\doibase 10.1038/s41467-024-47638-1} {\bibfield  {journal} {\bibinfo  {journal} {Nature Communications}\ }\textbf {\bibinfo {volume} {15}},\ \bibinfo {pages} {3170} (\bibinfo {year} {2024})}\BibitemShut {NoStop}%
\bibitem [{\citenamefont {Vinjanampathy}\ and\ \citenamefont {and}(2016)}]{Vinjanampathy01102016}%
  \BibitemOpen
  \bibfield  {author} {\bibinfo {author} {\bibfnamefont {Sai}\ \bibnamefont {Vinjanampathy}}\ and\ \bibinfo {author} {\bibfnamefont {Janet~Anders}\ \bibnamefont {and}},\ }\bibfield  {title} {\enquote {\bibinfo {title} {Quantum thermodynamics},}\ }\href {\doibase 10.1080/00107514.2016.1201896} {\bibfield  {journal} {\bibinfo  {journal} {Contemporary Physics}\ }\textbf {\bibinfo {volume} {57}},\ \bibinfo {pages} {545} (\bibinfo {year} {2016})}\BibitemShut {NoStop}%
\bibitem [{\citenamefont {Alicki}\ and\ \citenamefont {Kosloff}(2018)}]{Alicki2018}%
  \BibitemOpen
  \bibfield  {author} {\bibinfo {author} {\bibfnamefont {Robert}\ \bibnamefont {Alicki}}\ and\ \bibinfo {author} {\bibfnamefont {Ronnie}\ \bibnamefont {Kosloff}},\ }\enquote {\bibinfo {title} {Introduction to quantum thermodynamics: History and prospects},}\ in\ \href {\doibase 10.1007/978-3-319-99046-0_1} {\emph {\bibinfo {booktitle} {Thermodynamics in the Quantum Regime: Fundamental Aspects and New Directions}}},\ \bibinfo {editor} {edited by\ \bibinfo {editor} {\bibfnamefont {Felix}\ \bibnamefont {Binder}}, \bibinfo {editor} {\bibfnamefont {Luis~A.}\ \bibnamefont {Correa}}, \bibinfo {editor} {\bibfnamefont {Christian}\ \bibnamefont {Gogolin}}, \bibinfo {editor} {\bibfnamefont {Janet}\ \bibnamefont {Anders}}, \ and\ \bibinfo {editor} {\bibfnamefont {Gerardo}\ \bibnamefont {Adesso}}}\ (\bibinfo  {publisher} {Springer International Publishing},\ \bibinfo {address} {Cham},\ \bibinfo {year} {2018})\ pp.\ \bibinfo {pages} {1--33}\BibitemShut {NoStop}%
\bibitem [{\citenamefont {Chattopadhyay}\ \emph {et~al.}(2025)\citenamefont {Chattopadhyay}, \citenamefont {Misra}, \citenamefont {Pandit},\ and\ \citenamefont {Paul}}]{Chattopadhyay2025}%
  \BibitemOpen
  \bibfield  {author} {\bibinfo {author} {\bibfnamefont {Pritam}\ \bibnamefont {Chattopadhyay}}, \bibinfo {author} {\bibfnamefont {Avijit}\ \bibnamefont {Misra}}, \bibinfo {author} {\bibfnamefont {Tanmoy}\ \bibnamefont {Pandit}}, \ and\ \bibinfo {author} {\bibfnamefont {Goutam}\ \bibnamefont {Paul}},\ }\bibfield  {title} {\enquote {\bibinfo {title} {Landauer principle and thermodynamics of computation},}\ }\href {\doibase 10.1088/1361-6633/add6b3} {\bibfield  {journal} {\bibinfo  {journal} {Reports on Progress in Physics}\ }\textbf {\bibinfo {volume} {88}},\ \bibinfo {pages} {086001} (\bibinfo {year} {2025})}\BibitemShut {NoStop}%
\bibitem [{\citenamefont {Scovil}\ and\ \citenamefont {Schulz-DuBois}(1959)}]{Scovil1959}%
  \BibitemOpen
  \bibfield  {author} {\bibinfo {author} {\bibfnamefont {H.~E.~D.}\ \bibnamefont {Scovil}}\ and\ \bibinfo {author} {\bibfnamefont {E.~O.}\ \bibnamefont {Schulz-DuBois}},\ }\bibfield  {title} {\enquote {\bibinfo {title} {Three-level masers as heat engines},}\ }\href {\doibase 10.1103/PhysRevLett.2.262} {\bibfield  {journal} {\bibinfo  {journal} {Physical Review Letters}\ }\textbf {\bibinfo {volume} {2}},\ \bibinfo {pages} {262} (\bibinfo {year} {1959})}\BibitemShut {NoStop}%
\bibitem [{\citenamefont {Kosloff}\ and\ \citenamefont {Levy}(2014)}]{Kosloff2014}%
  \BibitemOpen
  \bibfield  {author} {\bibinfo {author} {\bibfnamefont {Ronnie}\ \bibnamefont {Kosloff}}\ and\ \bibinfo {author} {\bibfnamefont {Amikam}\ \bibnamefont {Levy}},\ }\bibfield  {title} {\enquote {\bibinfo {title} {Quantum heat engines and refrigerators: Continuous devices},}\ }\href {\doibase 10.1146/annurev-physchem-040513-103724} {\bibfield  {journal} {\bibinfo  {journal} {Annual Review of Physical Chemistry}\ }\textbf {\bibinfo {volume} {65}},\ \bibinfo {pages} {365} (\bibinfo {year} {2014})}\BibitemShut {NoStop}%
\bibitem [{\citenamefont {Mohan}\ \emph {et~al.}(2025)\citenamefont {Mohan}, \citenamefont {Gangwar}, \citenamefont {Pandit}, \citenamefont {Bera}, \citenamefont {Lewenstein},\ and\ \citenamefont {Bera}}]{Mohan2024}%
  \BibitemOpen
  \bibfield  {author} {\bibinfo {author} {\bibfnamefont {Brij}\ \bibnamefont {Mohan}}, \bibinfo {author} {\bibfnamefont {Rajeev}\ \bibnamefont {Gangwar}}, \bibinfo {author} {\bibfnamefont {Tanmoy}\ \bibnamefont {Pandit}}, \bibinfo {author} {\bibfnamefont {Mohit~Lal}\ \bibnamefont {Bera}}, \bibinfo {author} {\bibfnamefont {Maciej}\ \bibnamefont {Lewenstein}}, \ and\ \bibinfo {author} {\bibfnamefont {Manabendra~Nath}\ \bibnamefont {Bera}},\ }\bibfield  {title} {\enquote {\bibinfo {title} {Coherent heat transfer leads to genuine quantum enhancement in the performances of continuous engines},}\ }\href {\doibase 10.1103/PhysRevApplied.23.044050} {\bibfield  {journal} {\bibinfo  {journal} {Physical Review Applied}\ }\textbf {\bibinfo {volume} {23}},\ \bibinfo {pages} {044050} (\bibinfo {year} {2025})}\BibitemShut {NoStop}%
\bibitem [{\citenamefont {Palao}\ \emph {et~al.}(2001)\citenamefont {Palao}, \citenamefont {Kosloff},\ and\ \citenamefont {Gordon}}]{Palao2001}%
  \BibitemOpen
  \bibfield  {author} {\bibinfo {author} {\bibfnamefont {Jos\'e~P.}\ \bibnamefont {Palao}}, \bibinfo {author} {\bibfnamefont {Ronnie}\ \bibnamefont {Kosloff}}, \ and\ \bibinfo {author} {\bibfnamefont {Jeffrey~M.}\ \bibnamefont {Gordon}},\ }\bibfield  {title} {\enquote {\bibinfo {title} {Quantum thermodynamic cooling cycle},}\ }\href {\doibase 10.1103/PhysRevE.64.056130} {\bibfield  {journal} {\bibinfo  {journal} {Physical Review E}\ }\textbf {\bibinfo {volume} {64}},\ \bibinfo {pages} {056130} (\bibinfo {year} {2001})}\BibitemShut {NoStop}%
\bibitem [{\citenamefont {Linden}\ \emph {et~al.}(2010)\citenamefont {Linden}, \citenamefont {Popescu},\ and\ \citenamefont {Skrzypczyk}}]{Linden2010}%
  \BibitemOpen
  \bibfield  {author} {\bibinfo {author} {\bibfnamefont {Noah}\ \bibnamefont {Linden}}, \bibinfo {author} {\bibfnamefont {Sandu}\ \bibnamefont {Popescu}}, \ and\ \bibinfo {author} {\bibfnamefont {Paul}\ \bibnamefont {Skrzypczyk}},\ }\bibfield  {title} {\enquote {\bibinfo {title} {How small can thermal machines be? the smallest possible refrigerator},}\ }\href {\doibase 10.1103/PhysRevLett.105.130401} {\bibfield  {journal} {\bibinfo  {journal} {Physical Review Letters}\ }\textbf {\bibinfo {volume} {105}},\ \bibinfo {pages} {130401} (\bibinfo {year} {2010})}\BibitemShut {NoStop}%
\bibitem [{\citenamefont {Levy}\ \emph {et~al.}(2012)\citenamefont {Levy}, \citenamefont {Alicki},\ and\ \citenamefont {Kosloff}}]{Levy2012}%
  \BibitemOpen
  \bibfield  {author} {\bibinfo {author} {\bibfnamefont {Amikam}\ \bibnamefont {Levy}}, \bibinfo {author} {\bibfnamefont {Robert}\ \bibnamefont {Alicki}}, \ and\ \bibinfo {author} {\bibfnamefont {Ronnie}\ \bibnamefont {Kosloff}},\ }\bibfield  {title} {\enquote {\bibinfo {title} {Quantum refrigerators and the third law of thermodynamics},}\ }\href {\doibase 10.1103/PhysRevE.85.061126} {\bibfield  {journal} {\bibinfo  {journal} {Physical Review E}\ }\textbf {\bibinfo {volume} {85}},\ \bibinfo {pages} {061126} (\bibinfo {year} {2012})}\BibitemShut {NoStop}%
\bibitem [{\citenamefont {Correa}\ \emph {et~al.}(2014)\citenamefont {Correa}, \citenamefont {Palao}, \citenamefont {Alonso},\ and\ \citenamefont {Adesso}}]{Correa2014}%
  \BibitemOpen
  \bibfield  {author} {\bibinfo {author} {\bibfnamefont {Luis~A.}\ \bibnamefont {Correa}}, \bibinfo {author} {\bibfnamefont {Jos{\'e}~P.}\ \bibnamefont {Palao}}, \bibinfo {author} {\bibfnamefont {Daniel}\ \bibnamefont {Alonso}}, \ and\ \bibinfo {author} {\bibfnamefont {Gerardo}\ \bibnamefont {Adesso}},\ }\bibfield  {title} {\enquote {\bibinfo {title} {Quantum-enhanced absorption refrigerators},}\ }\href {\doibase 10.1038/srep03949} {\bibfield  {journal} {\bibinfo  {journal} {Scientific Reports}\ }\textbf {\bibinfo {volume} {4}},\ \bibinfo {pages} {3949} (\bibinfo {year} {2014})}\BibitemShut {NoStop}%
\bibitem [{\citenamefont {Singh}\ \emph {et~al.}(2020)\citenamefont {Singh}, \citenamefont {Pandit},\ and\ \citenamefont {Johal}}]{Singh2020T}%
  \BibitemOpen
  \bibfield  {author} {\bibinfo {author} {\bibfnamefont {Varinder}\ \bibnamefont {Singh}}, \bibinfo {author} {\bibfnamefont {Tanmoy}\ \bibnamefont {Pandit}}, \ and\ \bibinfo {author} {\bibfnamefont {Ramandeep~S.}\ \bibnamefont {Johal}},\ }\bibfield  {title} {\enquote {\bibinfo {title} {Optimal performance of a three-level quantum refrigerator},}\ }\href {\doibase 10.1103/PhysRevE.101.062121} {\bibfield  {journal} {\bibinfo  {journal} {Physical Review E}\ }\textbf {\bibinfo {volume} {101}},\ \bibinfo {pages} {062121} (\bibinfo {year} {2020})}\BibitemShut {NoStop}%
\bibitem [{\citenamefont {Ghoshal}\ \emph {et~al.}(2021)\citenamefont {Ghoshal}, \citenamefont {Das}, \citenamefont {Pal}, \citenamefont {Sen(De)},\ and\ \citenamefont {Sen}}]{Ghoshal2021}%
  \BibitemOpen
  \bibfield  {author} {\bibinfo {author} {\bibfnamefont {Ahana}\ \bibnamefont {Ghoshal}}, \bibinfo {author} {\bibfnamefont {Sreetama}\ \bibnamefont {Das}}, \bibinfo {author} {\bibfnamefont {Amit~Kumar}\ \bibnamefont {Pal}}, \bibinfo {author} {\bibfnamefont {Aditi}\ \bibnamefont {Sen(De)}}, \ and\ \bibinfo {author} {\bibfnamefont {Ujjwal}\ \bibnamefont {Sen}},\ }\bibfield  {title} {\enquote {\bibinfo {title} {Three qubits in less than three baths: Beyond two-body system-bath interactions in quantum refrigerators},}\ }\href {\doibase 10.1103/PhysRevA.104.042208} {\bibfield  {journal} {\bibinfo  {journal} {Physical Review A}\ }\textbf {\bibinfo {volume} {104}},\ \bibinfo {pages} {042208} (\bibinfo {year} {2021})}\BibitemShut {NoStop}%
\bibitem [{\citenamefont {Mohanta}\ \emph {et~al.}(2022)\citenamefont {Mohanta}, \citenamefont {Saryal},\ and\ \citenamefont {Agarwalla}}]{Mohanta2022}%
  \BibitemOpen
  \bibfield  {author} {\bibinfo {author} {\bibfnamefont {Sandipan}\ \bibnamefont {Mohanta}}, \bibinfo {author} {\bibfnamefont {Sushant}\ \bibnamefont {Saryal}}, \ and\ \bibinfo {author} {\bibfnamefont {Bijay~Kumar}\ \bibnamefont {Agarwalla}},\ }\bibfield  {title} {\enquote {\bibinfo {title} {Universal bounds on cooling power and cooling efficiency for autonomous absorption refrigerators},}\ }\href {\doibase 10.1103/PhysRevE.105.034127} {\bibfield  {journal} {\bibinfo  {journal} {Physical Review E}\ }\textbf {\bibinfo {volume} {105}},\ \bibinfo {pages} {034127} (\bibinfo {year} {2022})}\BibitemShut {NoStop}%
\bibitem [{\citenamefont {Yang}\ \emph {et~al.}(2023)\citenamefont {Yang}, \citenamefont {Liu},\ and\ \citenamefont {Yu}}]{Yang2023}%
  \BibitemOpen
  \bibfield  {author} {\bibinfo {author} {\bibfnamefont {Yi-jia}\ \bibnamefont {Yang}}, \bibinfo {author} {\bibfnamefont {Yu-qiang}\ \bibnamefont {Liu}}, \ and\ \bibinfo {author} {\bibfnamefont {Chang-shui}\ \bibnamefont {Yu}},\ }\bibfield  {title} {\enquote {\bibinfo {title} {Quantum thermal diode dominated by pure classical correlation via three triangular-coupled qubits},}\ }\href {\doibase 10.1103/PhysRevE.107.064125} {\bibfield  {journal} {\bibinfo  {journal} {Physical Review E}\ }\textbf {\bibinfo {volume} {107}},\ \bibinfo {pages} {064125} (\bibinfo {year} {2023})}\BibitemShut {NoStop}%
\bibitem [{\citenamefont {Joulain}\ \emph {et~al.}(2016)\citenamefont {Joulain}, \citenamefont {Drevillon}, \citenamefont {Ezzahri},\ and\ \citenamefont {Ordonez-Miranda}}]{Joulain2016}%
  \BibitemOpen
  \bibfield  {author} {\bibinfo {author} {\bibfnamefont {Karl}\ \bibnamefont {Joulain}}, \bibinfo {author} {\bibfnamefont {J\'er\'emie}\ \bibnamefont {Drevillon}}, \bibinfo {author} {\bibfnamefont {Youn\`es}\ \bibnamefont {Ezzahri}}, \ and\ \bibinfo {author} {\bibfnamefont {Jose}\ \bibnamefont {Ordonez-Miranda}},\ }\bibfield  {title} {\enquote {\bibinfo {title} {Quantum thermal transistor},}\ }\href {\doibase 10.1103/PhysRevLett.116.200601} {\bibfield  {journal} {\bibinfo  {journal} {Physical Review Letters}\ }\textbf {\bibinfo {volume} {116}},\ \bibinfo {pages} {200601} (\bibinfo {year} {2016})}\BibitemShut {NoStop}%
\bibitem [{\citenamefont {Ghosh}\ \emph {et~al.}(2021)\citenamefont {Ghosh}, \citenamefont {Ghoshal},\ and\ \citenamefont {Sen}}]{Ghosh2021}%
  \BibitemOpen
  \bibfield  {author} {\bibinfo {author} {\bibfnamefont {Riddhi}\ \bibnamefont {Ghosh}}, \bibinfo {author} {\bibfnamefont {Ahana}\ \bibnamefont {Ghoshal}}, \ and\ \bibinfo {author} {\bibfnamefont {Ujjwal}\ \bibnamefont {Sen}},\ }\bibfield  {title} {\enquote {\bibinfo {title} {Quantum thermal transistors: Operation characteristics in steady state versus transient regimes},}\ }\href {\doibase 10.1103/PhysRevA.103.052613} {\bibfield  {journal} {\bibinfo  {journal} {Physical Review A}\ }\textbf {\bibinfo {volume} {103}},\ \bibinfo {pages} {052613} (\bibinfo {year} {2021})}\BibitemShut {NoStop}%
\bibitem [{\citenamefont {Gupt}\ \emph {et~al.}(2022)\citenamefont {Gupt}, \citenamefont {Bhattacharyya}, \citenamefont {Das}, \citenamefont {Datta}, \citenamefont {Mukherjee},\ and\ \citenamefont {Ghosh}}]{Gupt2022}%
  \BibitemOpen
  \bibfield  {author} {\bibinfo {author} {\bibfnamefont {Nikhil}\ \bibnamefont {Gupt}}, \bibinfo {author} {\bibfnamefont {Srijan}\ \bibnamefont {Bhattacharyya}}, \bibinfo {author} {\bibfnamefont {Bikash}\ \bibnamefont {Das}}, \bibinfo {author} {\bibfnamefont {Subhadeep}\ \bibnamefont {Datta}}, \bibinfo {author} {\bibfnamefont {Victor}\ \bibnamefont {Mukherjee}}, \ and\ \bibinfo {author} {\bibfnamefont {Arnab}\ \bibnamefont {Ghosh}},\ }\bibfield  {title} {\enquote {\bibinfo {title} {Floquet quantum thermal transistor},}\ }\href {\doibase 10.1103/PhysRevE.106.024110} {\bibfield  {journal} {\bibinfo  {journal} {Physical Review E}\ }\textbf {\bibinfo {volume} {106}},\ \bibinfo {pages} {024110} (\bibinfo {year} {2022})}\BibitemShut {NoStop}%
\bibitem [{\citenamefont {Erker}\ \emph {et~al.}(2017)\citenamefont {Erker}, \citenamefont {Mitchison}, \citenamefont {Silva}, \citenamefont {Woods}, \citenamefont {Brunner},\ and\ \citenamefont {Huber}}]{Erker2017}%
  \BibitemOpen
  \bibfield  {author} {\bibinfo {author} {\bibfnamefont {Paul}\ \bibnamefont {Erker}}, \bibinfo {author} {\bibfnamefont {Mark~T.}\ \bibnamefont {Mitchison}}, \bibinfo {author} {\bibfnamefont {Ralph}\ \bibnamefont {Silva}}, \bibinfo {author} {\bibfnamefont {Mischa~P.}\ \bibnamefont {Woods}}, \bibinfo {author} {\bibfnamefont {Nicolas}\ \bibnamefont {Brunner}}, \ and\ \bibinfo {author} {\bibfnamefont {Marcus}\ \bibnamefont {Huber}},\ }\bibfield  {title} {\enquote {\bibinfo {title} {Autonomous quantum clocks: Does thermodynamics limit our ability to measure time?}}\ }\href {\doibase 10.1103/PhysRevX.7.031022} {\bibfield  {journal} {\bibinfo  {journal} {Physical Review X}\ }\textbf {\bibinfo {volume} {7}},\ \bibinfo {pages} {031022} (\bibinfo {year} {2017})}\BibitemShut {NoStop}%
\bibitem [{\citenamefont {Santiago-García}\ \emph {et~al.}(2025)\citenamefont {Santiago-García}, \citenamefont {Pusuluk}, \citenamefont {Müstecaplıoğlu}, \citenamefont {Çakmak},\ and\ \citenamefont {Román-Ancheyta}}]{Santiago-García2025}%
  \BibitemOpen
  \bibfield  {author} {\bibinfo {author} {\bibfnamefont {M}~\bibnamefont {Santiago-García}}, \bibinfo {author} {\bibfnamefont {O}~\bibnamefont {Pusuluk}}, \bibinfo {author} {\bibfnamefont {Ö~E}\ \bibnamefont {Müstecaplıoğlu}}, \bibinfo {author} {\bibfnamefont {B}~\bibnamefont {Çakmak}}, \ and\ \bibinfo {author} {\bibfnamefont {R}~\bibnamefont {Román-Ancheyta}},\ }\bibfield  {title} {\enquote {\bibinfo {title} {Quantum thermal machine as a rectifier},}\ }\href {\doibase 10.1088/2058-9565/adaee0} {\bibfield  {journal} {\bibinfo  {journal} {Quantum Science and Technology}\ }\textbf {\bibinfo {volume} {10}},\ \bibinfo {pages} {025018} (\bibinfo {year} {2025})}\BibitemShut {NoStop}%
\bibitem [{\citenamefont {Juli\`a-Farr\'e}\ \emph {et~al.}(2020)\citenamefont {Juli\`a-Farr\'e}, \citenamefont {Salamon}, \citenamefont {Riera}, \citenamefont {Bera},\ and\ \citenamefont {Lewenstein}}]{Sergi2020}%
  \BibitemOpen
  \bibfield  {author} {\bibinfo {author} {\bibfnamefont {Sergi}\ \bibnamefont {Juli\`a-Farr\'e}}, \bibinfo {author} {\bibfnamefont {Tymoteusz}\ \bibnamefont {Salamon}}, \bibinfo {author} {\bibfnamefont {Arnau}\ \bibnamefont {Riera}}, \bibinfo {author} {\bibfnamefont {Manabendra~N.}\ \bibnamefont {Bera}}, \ and\ \bibinfo {author} {\bibfnamefont {Maciej}\ \bibnamefont {Lewenstein}},\ }\bibfield  {title} {\enquote {\bibinfo {title} {Bounds on the capacity and power of quantum batteries},}\ }\href {\doibase 10.1103/PhysRevResearch.2.023113} {\bibfield  {journal} {\bibinfo  {journal} {Physical Review Research}\ }\textbf {\bibinfo {volume} {2}},\ \bibinfo {pages} {023113} (\bibinfo {year} {2020})}\BibitemShut {NoStop}%
\bibitem [{\citenamefont {Mohan}\ and\ \citenamefont {Pati}(2021)}]{Mohan2021}%
  \BibitemOpen
  \bibfield  {author} {\bibinfo {author} {\bibfnamefont {Brij}\ \bibnamefont {Mohan}}\ and\ \bibinfo {author} {\bibfnamefont {Arun~K.}\ \bibnamefont {Pati}},\ }\bibfield  {title} {\enquote {\bibinfo {title} {Reverse quantum speed limit: How slowly a quantum battery can discharge},}\ }\href {\doibase 10.1103/PhysRevA.104.042209} {\bibfield  {journal} {\bibinfo  {journal} {Physical Review A}\ }\textbf {\bibinfo {volume} {104}},\ \bibinfo {pages} {042209} (\bibinfo {year} {2021})}\BibitemShut {NoStop}%
\bibitem [{\citenamefont {Mohan}\ and\ \citenamefont {Pati}(2022)}]{Mohan2022}%
  \BibitemOpen
  \bibfield  {author} {\bibinfo {author} {\bibfnamefont {Brij}\ \bibnamefont {Mohan}}\ and\ \bibinfo {author} {\bibfnamefont {Arun~Kumar}\ \bibnamefont {Pati}},\ }\bibfield  {title} {\enquote {\bibinfo {title} {Quantum speed limits for observables},}\ }\href {\doibase 10.1103/PhysRevA.106.042436} {\bibfield  {journal} {\bibinfo  {journal} {Physical Review A}\ }\textbf {\bibinfo {volume} {106}},\ \bibinfo {pages} {042436} (\bibinfo {year} {2022})}\BibitemShut {NoStop}%
\bibitem [{\citenamefont {Mitchison}(2019)}]{Mitchison2019}%
  \BibitemOpen
  \bibfield  {author} {\bibinfo {author} {\bibfnamefont {Mark~T.}\ \bibnamefont {Mitchison}},\ }\bibfield  {title} {\enquote {\bibinfo {title} {Quantum thermal absorption machines: refrigerators, engines and clocks},}\ }\href {\doibase 10.1080/00107514.2019.1631555} {\bibfield  {journal} {\bibinfo  {journal} {Contemporary Physics}\ }\textbf {\bibinfo {volume} {60}},\ \bibinfo {pages} {164} (\bibinfo {year} {2019})}\BibitemShut {NoStop}%
\bibitem [{\citenamefont {Cangemi}\ \emph {et~al.}(2024)\citenamefont {Cangemi}, \citenamefont {Bhadra},\ and\ \citenamefont {Levy}}]{Cangemi2024}%
  \BibitemOpen
  \bibfield  {author} {\bibinfo {author} {\bibfnamefont {Loris~Maria}\ \bibnamefont {Cangemi}}, \bibinfo {author} {\bibfnamefont {Chitrak}\ \bibnamefont {Bhadra}}, \ and\ \bibinfo {author} {\bibfnamefont {Amikam}\ \bibnamefont {Levy}},\ }\bibfield  {title} {\enquote {\bibinfo {title} {Quantum engines and refrigerators},}\ }\href {\doibase https://doi.org/10.1016/j.physrep.2024.07.001} {\bibfield  {journal} {\bibinfo  {journal} {Physics Reports}\ }\textbf {\bibinfo {volume} {1087}},\ \bibinfo {pages} {1} (\bibinfo {year} {2024})}\BibitemShut {NoStop}%
\bibitem [{\citenamefont {Antonio Marín~Guzmán}\ \emph {et~al.}(2024)\citenamefont {Antonio Marín~Guzmán}, \citenamefont {Erker}, \citenamefont {Gasparinetti}, \citenamefont {Huber},\ and\ \citenamefont {Yunger~Halpern}}]{Guzmán2024}%
  \BibitemOpen
  \bibfield  {author} {\bibinfo {author} {\bibfnamefont {José}\ \bibnamefont {Antonio Marín~Guzmán}}, \bibinfo {author} {\bibfnamefont {Paul}\ \bibnamefont {Erker}}, \bibinfo {author} {\bibfnamefont {Simone}\ \bibnamefont {Gasparinetti}}, \bibinfo {author} {\bibfnamefont {Marcus}\ \bibnamefont {Huber}}, \ and\ \bibinfo {author} {\bibfnamefont {Nicole}\ \bibnamefont {Yunger~Halpern}},\ }\bibfield  {title} {\enquote {\bibinfo {title} {Key issues review: useful autonomous quantum machines},}\ }\href {\doibase 10.1088/1361-6633/ad8803} {\bibfield  {journal} {\bibinfo  {journal} {Reports on Progress in Physics}\ }\textbf {\bibinfo {volume} {87}},\ \bibinfo {pages} {122001} (\bibinfo {year} {2024})}\BibitemShut {NoStop}%
\bibitem [{\citenamefont {Levy}\ and\ \citenamefont {Kosloff}(2012)}]{AR_PRL_2012}%
  \BibitemOpen
  \bibfield  {author} {\bibinfo {author} {\bibfnamefont {Amikam}\ \bibnamefont {Levy}}\ and\ \bibinfo {author} {\bibfnamefont {Ronnie}\ \bibnamefont {Kosloff}},\ }\bibfield  {title} {\enquote {\bibinfo {title} {Quantum absorption refrigerator},}\ }\href {\doibase 10.1103/PhysRevLett.108.070604} {\bibfield  {journal} {\bibinfo  {journal} {Physical Review Letters}\ }\textbf {\bibinfo {volume} {108}},\ \bibinfo {pages} {070604} (\bibinfo {year} {2012})}\BibitemShut {NoStop}%
\bibitem [{\citenamefont {Mu}\ \emph {et~al.}(2017)\citenamefont {Mu}, \citenamefont {Agarwalla}, \citenamefont {Schaller},\ and\ \citenamefont {Segal}}]{Mu_2017}%
  \BibitemOpen
  \bibfield  {author} {\bibinfo {author} {\bibfnamefont {Anqi}\ \bibnamefont {Mu}}, \bibinfo {author} {\bibfnamefont {Bijay~K.}\ \bibnamefont {Agarwalla}}, \bibinfo {author} {\bibfnamefont {Gernot}\ \bibnamefont {Schaller}}, \ and\ \bibinfo {author} {\bibfnamefont {Dvira}\ \bibnamefont {Segal}},\ }\bibfield  {title} {\enquote {\bibinfo {title} {Qubit absorption refrigerator at strong coupling},}\ }\href {\doibase 10.1088/1367-2630/aa9b75} {\bibfield  {journal} {\bibinfo  {journal} {New Journal of Physics}\ }\textbf {\bibinfo {volume} {19}},\ \bibinfo {pages} {123034} (\bibinfo {year} {2017})}\BibitemShut {NoStop}%
\bibitem [{\citenamefont {Ivander}\ \emph {et~al.}(2022)\citenamefont {Ivander}, \citenamefont {Anto-Sztrikacs},\ and\ \citenamefont {Segal}}]{ISS_PRE_2022}%
  \BibitemOpen
  \bibfield  {author} {\bibinfo {author} {\bibfnamefont {Felix}\ \bibnamefont {Ivander}}, \bibinfo {author} {\bibfnamefont {Nicholas}\ \bibnamefont {Anto-Sztrikacs}}, \ and\ \bibinfo {author} {\bibfnamefont {Dvira}\ \bibnamefont {Segal}},\ }\bibfield  {title} {\enquote {\bibinfo {title} {Strong system-bath coupling effects in quantum absorption refrigerators},}\ }\href {\doibase 10.1103/PhysRevE.105.034112} {\bibfield  {journal} {\bibinfo  {journal} {Physical Review E}\ }\textbf {\bibinfo {volume} {105}},\ \bibinfo {pages} {034112} (\bibinfo {year} {2022})}\BibitemShut {NoStop}%
\bibitem [{\citenamefont {Kilgour}\ and\ \citenamefont {Segal}(2018)}]{KS_PRE_2018}%
  \BibitemOpen
  \bibfield  {author} {\bibinfo {author} {\bibfnamefont {Michael}\ \bibnamefont {Kilgour}}\ and\ \bibinfo {author} {\bibfnamefont {Dvira}\ \bibnamefont {Segal}},\ }\bibfield  {title} {\enquote {\bibinfo {title} {Coherence and decoherence in quantum absorption refrigerators},}\ }\href {\doibase 10.1103/PhysRevE.98.012117} {\bibfield  {journal} {\bibinfo  {journal} {Physical Review E}\ }\textbf {\bibinfo {volume} {98}},\ \bibinfo {pages} {012117} (\bibinfo {year} {2018})}\BibitemShut {NoStop}%
\bibitem [{\citenamefont {González}\ \emph {et~al.}(2017)\citenamefont {González}, \citenamefont {Palao},\ and\ \citenamefont {Alonso}}]{Gonzalez_2017}%
  \BibitemOpen
  \bibfield  {author} {\bibinfo {author} {\bibfnamefont {J~Onam}\ \bibnamefont {González}}, \bibinfo {author} {\bibfnamefont {José~P}\ \bibnamefont {Palao}}, \ and\ \bibinfo {author} {\bibfnamefont {Daniel}\ \bibnamefont {Alonso}},\ }\bibfield  {title} {\enquote {\bibinfo {title} {Relation between topology and heat currents in multilevel absorption machines},}\ }\href {\doibase 10.1088/1367-2630/aa8647} {\bibfield  {journal} {\bibinfo  {journal} {New Journal of Physics}\ }\textbf {\bibinfo {volume} {19}},\ \bibinfo {pages} {113037} (\bibinfo {year} {2017})}\BibitemShut {NoStop}%
\bibitem [{\citenamefont {Nimmrichter}\ \emph {et~al.}(2017)\citenamefont {Nimmrichter}, \citenamefont {Dai}, \citenamefont {Roulet},\ and\ \citenamefont {Scarani}}]{Nimmrichter2017quantumclassical}%
  \BibitemOpen
  \bibfield  {author} {\bibinfo {author} {\bibfnamefont {Stefan}\ \bibnamefont {Nimmrichter}}, \bibinfo {author} {\bibfnamefont {Jibo}\ \bibnamefont {Dai}}, \bibinfo {author} {\bibfnamefont {Alexandre}\ \bibnamefont {Roulet}}, \ and\ \bibinfo {author} {\bibfnamefont {Valerio}\ \bibnamefont {Scarani}},\ }\bibfield  {title} {\enquote {\bibinfo {title} {Quantum and classical dynamics of a three-mode absorption refrigerator},}\ }\href {\doibase 10.22331/q-2017-12-11-37} {\bibfield  {journal} {\bibinfo  {journal} {{Quantum}}\ }\textbf {\bibinfo {volume} {1}},\ \bibinfo {pages} {37} (\bibinfo {year} {2017})}\BibitemShut {NoStop}%
\bibitem [{\citenamefont {Manikandan}\ \emph {et~al.}(2020)\citenamefont {Manikandan}, \citenamefont {Jussiau},\ and\ \citenamefont {Jordan}}]{SEJ_PRB_2020}%
  \BibitemOpen
  \bibfield  {author} {\bibinfo {author} {\bibfnamefont {Sreenath~K.}\ \bibnamefont {Manikandan}}, \bibinfo {author} {\bibfnamefont {\'Etienne}\ \bibnamefont {Jussiau}}, \ and\ \bibinfo {author} {\bibfnamefont {Andrew~N.}\ \bibnamefont {Jordan}},\ }\bibfield  {title} {\enquote {\bibinfo {title} {Autonomous quantum absorption refrigerators},}\ }\href {\doibase 10.1103/PhysRevB.102.235427} {\bibfield  {journal} {\bibinfo  {journal} {Physical Review B}\ }\textbf {\bibinfo {volume} {102}},\ \bibinfo {pages} {235427} (\bibinfo {year} {2020})}\BibitemShut {NoStop}%
\bibitem [{\citenamefont {Brask}\ and\ \citenamefont {Brunner}(2015)}]{Brask2015}%
  \BibitemOpen
  \bibfield  {author} {\bibinfo {author} {\bibfnamefont {Jonatan~Bohr}\ \bibnamefont {Brask}}\ and\ \bibinfo {author} {\bibfnamefont {Nicolas}\ \bibnamefont {Brunner}},\ }\bibfield  {title} {\enquote {\bibinfo {title} {Small quantum absorption refrigerator in the transient regime: Time scales, enhanced cooling, and entanglement},}\ }\href {\doibase 10.1103/PhysRevE.92.062101} {\bibfield  {journal} {\bibinfo  {journal} {Physical Review E}\ }\textbf {\bibinfo {volume} {92}},\ \bibinfo {pages} {062101} (\bibinfo {year} {2015})}\BibitemShut {NoStop}%
\bibitem [{\citenamefont {Correa}\ \emph {et~al.}(2013)\citenamefont {Correa}, \citenamefont {Palao}, \citenamefont {Adesso},\ and\ \citenamefont {Alonso}}]{Correa2013}%
  \BibitemOpen
  \bibfield  {author} {\bibinfo {author} {\bibfnamefont {Luis~A.}\ \bibnamefont {Correa}}, \bibinfo {author} {\bibfnamefont {José~P.}\ \bibnamefont {Palao}}, \bibinfo {author} {\bibfnamefont {Gerardo}\ \bibnamefont {Adesso}}, \ and\ \bibinfo {author} {\bibfnamefont {Daniel}\ \bibnamefont {Alonso}},\ }\bibfield  {title} {\enquote {\bibinfo {title} {Performance bound for quantum absorption refrigerators},}\ }\href {\doibase 10.1103/PhysRevE.87.042131} {\bibfield  {journal} {\bibinfo  {journal} {Physical Review E}\ }\textbf {\bibinfo {volume} {87}},\ \bibinfo {pages} {042131} (\bibinfo {year} {2013})}\BibitemShut {NoStop}%
\bibitem [{\citenamefont {Maslennikov}\ \emph {et~al.}(2019)\citenamefont {Maslennikov}, \citenamefont {Ding}, \citenamefont {Habl{\"u}tzel}, \citenamefont {Gan}, \citenamefont {Roulet}, \citenamefont {Nimmrichter}, \citenamefont {Dai}, \citenamefont {Scarani},\ and\ \citenamefont {Matsukevich}}]{Maslennikov2019}%
  \BibitemOpen
  \bibfield  {author} {\bibinfo {author} {\bibfnamefont {Gleb}\ \bibnamefont {Maslennikov}}, \bibinfo {author} {\bibfnamefont {Shiqian}\ \bibnamefont {Ding}}, \bibinfo {author} {\bibfnamefont {Roland}\ \bibnamefont {Habl{\"u}tzel}}, \bibinfo {author} {\bibfnamefont {Jaren}\ \bibnamefont {Gan}}, \bibinfo {author} {\bibfnamefont {Alexandre}\ \bibnamefont {Roulet}}, \bibinfo {author} {\bibfnamefont {Stefan}\ \bibnamefont {Nimmrichter}}, \bibinfo {author} {\bibfnamefont {Jibo}\ \bibnamefont {Dai}}, \bibinfo {author} {\bibfnamefont {Valerio}\ \bibnamefont {Scarani}}, \ and\ \bibinfo {author} {\bibfnamefont {Dzmitry}\ \bibnamefont {Matsukevich}},\ }\bibfield  {title} {\enquote {\bibinfo {title} {Quantum absorption refrigerator with trapped ions},}\ }\href {\doibase 10.1038/s41467-018-08090-0} {\bibfield  {journal} {\bibinfo  {journal} {Nature Communications}\ }\textbf {\bibinfo {volume} {10}},\ \bibinfo {pages} {202} (\bibinfo {year} {2019})}\BibitemShut {NoStop}%
\bibitem [{\citenamefont {Aamir}\ \emph {et~al.}(2025)\citenamefont {Aamir}, \citenamefont {Jamet~Suria}, \citenamefont {Mar{\'i}n~Guzm{\'a}n}, \citenamefont {Castillo-Moreno}, \citenamefont {Epstein}, \citenamefont {Yunger~Halpern},\ and\ \citenamefont {Gasparinetti}}]{Aamir2025}%
  \BibitemOpen
  \bibfield  {author} {\bibinfo {author} {\bibfnamefont {Mohammed~Ali}\ \bibnamefont {Aamir}}, \bibinfo {author} {\bibfnamefont {Paul}\ \bibnamefont {Jamet~Suria}}, \bibinfo {author} {\bibfnamefont {Jos{\'e}~Antonio}\ \bibnamefont {Mar{\'i}n~Guzm{\'a}n}}, \bibinfo {author} {\bibfnamefont {Claudia}\ \bibnamefont {Castillo-Moreno}}, \bibinfo {author} {\bibfnamefont {Jeffrey~M.}\ \bibnamefont {Epstein}}, \bibinfo {author} {\bibfnamefont {Nicole}\ \bibnamefont {Yunger~Halpern}}, \ and\ \bibinfo {author} {\bibfnamefont {Simone}\ \bibnamefont {Gasparinetti}},\ }\bibfield  {title} {\enquote {\bibinfo {title} {Thermally driven quantum refrigerator autonomously resets a superconducting qubit},}\ }\href {https://doi.org/10.1038/s41567-024-02708-5} {\bibfield  {journal} {\bibinfo  {journal} {Nature Physics}\ }\textbf {\bibinfo {volume} {21}},\ \bibinfo {pages} {318} (\bibinfo {year} {2025})}\BibitemShut {NoStop}%
\bibitem [{\citenamefont {Felce}\ and\ \citenamefont {Vedral}(2020)}]{Felce2020}%
  \BibitemOpen
  \bibfield  {author} {\bibinfo {author} {\bibfnamefont {David}\ \bibnamefont {Felce}}\ and\ \bibinfo {author} {\bibfnamefont {Vlatko}\ \bibnamefont {Vedral}},\ }\bibfield  {title} {\enquote {\bibinfo {title} {Quantum refrigeration with indefinite causal order},}\ }\href {\doibase 10.1103/PhysRevLett.125.070603} {\bibfield  {journal} {\bibinfo  {journal} {Phys. Rev. Lett.}\ }\textbf {\bibinfo {volume} {125}},\ \bibinfo {pages} {070603} (\bibinfo {year} {2020})}\BibitemShut {NoStop}%
\bibitem [{\citenamefont {Gerry}\ and\ \citenamefont {Eberly}(1990)}]{Gerry1990}%
  \BibitemOpen
  \bibfield  {author} {\bibinfo {author} {\bibfnamefont {Christopher~C.}\ \bibnamefont {Gerry}}\ and\ \bibinfo {author} {\bibfnamefont {J.~H.}\ \bibnamefont {Eberly}},\ }\bibfield  {title} {\enquote {\bibinfo {title} {Dynamics of a raman coupled model interacting with two quantized cavity fields},}\ }\href {\doibase 10.1103/PhysRevA.42.6805} {\bibfield  {journal} {\bibinfo  {journal} {Physical Review A}\ }\textbf {\bibinfo {volume} {42}},\ \bibinfo {pages} {6805} (\bibinfo {year} {1990})}\BibitemShut {NoStop}%
\bibitem [{\citenamefont {Gerry}\ and\ \citenamefont {Huang}(1992)}]{Gerry1992}%
  \BibitemOpen
  \bibfield  {author} {\bibinfo {author} {\bibfnamefont {Christopher~C.}\ \bibnamefont {Gerry}}\ and\ \bibinfo {author} {\bibfnamefont {H.}~\bibnamefont {Huang}},\ }\bibfield  {title} {\enquote {\bibinfo {title} {Dynamics of a two-atom raman coupled model interacting with two quantized cavity fields},}\ }\href {\doibase 10.1103/PhysRevA.45.8037} {\bibfield  {journal} {\bibinfo  {journal} {Physical Review A}\ }\textbf {\bibinfo {volume} {45}},\ \bibinfo {pages} {8037} (\bibinfo {year} {1992})}\BibitemShut {NoStop}%
\bibitem [{\citenamefont {Wu}(1996)}]{Wu1996}%
  \BibitemOpen
  \bibfield  {author} {\bibinfo {author} {\bibfnamefont {Ying}\ \bibnamefont {Wu}},\ }\bibfield  {title} {\enquote {\bibinfo {title} {Effective raman theory for a three-level atom in the \ensuremath{\Lambda} configuration},}\ }\href {\doibase 10.1103/PhysRevA.54.1586} {\bibfield  {journal} {\bibinfo  {journal} {Physical Review A}\ }\textbf {\bibinfo {volume} {54}},\ \bibinfo {pages} {1586} (\bibinfo {year} {1996})}\BibitemShut {NoStop}%
\bibitem [{\citenamefont {Barato}\ and\ \citenamefont {Seifert}(2015)}]{Barato2015}%
  \BibitemOpen
  \bibfield  {author} {\bibinfo {author} {\bibfnamefont {Andre~C.}\ \bibnamefont {Barato}}\ and\ \bibinfo {author} {\bibfnamefont {Udo}\ \bibnamefont {Seifert}},\ }\bibfield  {title} {\enquote {\bibinfo {title} {Thermodynamic uncertainty relation for biomolecular processes},}\ }\href {\doibase 10.1103/PhysRevLett.114.158101} {\bibfield  {journal} {\bibinfo  {journal} {Physical Review Letters}\ }\textbf {\bibinfo {volume} {114}},\ \bibinfo {pages} {158101} (\bibinfo {year} {2015})}\BibitemShut {NoStop}%
\bibitem [{\citenamefont {Gingrich}\ \emph {et~al.}(2016)\citenamefont {Gingrich}, \citenamefont {Horowitz}, \citenamefont {Perunov},\ and\ \citenamefont {England}}]{Gingrich2016}%
  \BibitemOpen
  \bibfield  {author} {\bibinfo {author} {\bibfnamefont {Todd~R.}\ \bibnamefont {Gingrich}}, \bibinfo {author} {\bibfnamefont {Jordan~M.}\ \bibnamefont {Horowitz}}, \bibinfo {author} {\bibfnamefont {Nikolay}\ \bibnamefont {Perunov}}, \ and\ \bibinfo {author} {\bibfnamefont {Jeremy~L.}\ \bibnamefont {England}},\ }\bibfield  {title} {\enquote {\bibinfo {title} {Dissipation bounds all steady-state current fluctuations},}\ }\href {\doibase 10.1103/PhysRevLett.116.120601} {\bibfield  {journal} {\bibinfo  {journal} {Physical Review Letters}\ }\textbf {\bibinfo {volume} {116}},\ \bibinfo {pages} {120601} (\bibinfo {year} {2016})}\BibitemShut {NoStop}%
\bibitem [{\citenamefont {Seifert}(2019)}]{Seifert2019}%
  \BibitemOpen
  \bibfield  {author} {\bibinfo {author} {\bibfnamefont {Udo}\ \bibnamefont {Seifert}},\ }\bibfield  {title} {\enquote {\bibinfo {title} {From stochastic thermodynamics to thermodynamic inference},}\ }\href {\doibase 10.1146/annurev-conmatphys-031218-013554} {\bibfield  {journal} {\bibinfo  {journal} {Annual Review of Condensed Matter Physics}\ }\textbf {\bibinfo {volume} {10}},\ \bibinfo {pages} {171} (\bibinfo {year} {2019})}\BibitemShut {NoStop}%
\bibitem [{\citenamefont {Agarwalla}\ and\ \citenamefont {Segal}(2018)}]{AS_2018}%
  \BibitemOpen
  \bibfield  {author} {\bibinfo {author} {\bibfnamefont {Bijay~Kumar}\ \bibnamefont {Agarwalla}}\ and\ \bibinfo {author} {\bibfnamefont {Dvira}\ \bibnamefont {Segal}},\ }\bibfield  {title} {\enquote {\bibinfo {title} {Assessing the validity of the thermodynamic uncertainty relation in quantum systems},}\ }\href {\doibase 10.1103/PhysRevB.98.155438} {\bibfield  {journal} {\bibinfo  {journal} {Physical Review B}\ }\textbf {\bibinfo {volume} {98}},\ \bibinfo {pages} {155438} (\bibinfo {year} {2018})}\BibitemShut {NoStop}%
\bibitem [{\citenamefont {Kalaee}\ \emph {et~al.}(2021)\citenamefont {Kalaee}, \citenamefont {Wacker},\ and\ \citenamefont {Potts}}]{Kalaee2021}%
  \BibitemOpen
  \bibfield  {author} {\bibinfo {author} {\bibfnamefont {Alex Arash~Sand}\ \bibnamefont {Kalaee}}, \bibinfo {author} {\bibfnamefont {Andreas}\ \bibnamefont {Wacker}}, \ and\ \bibinfo {author} {\bibfnamefont {Patrick~P.}\ \bibnamefont {Potts}},\ }\bibfield  {title} {\enquote {\bibinfo {title} {Violating the thermodynamic uncertainty relation in the three-level maser},}\ }\href {\doibase 10.1103/PhysRevE.104.L012103} {\bibfield  {journal} {\bibinfo  {journal} {Physical Review E}\ }\textbf {\bibinfo {volume} {104}},\ \bibinfo {pages} {L012103} (\bibinfo {year} {2021})}\BibitemShut {NoStop}%
\bibitem [{\citenamefont {Guarnieri}\ \emph {et~al.}(2019)\citenamefont {Guarnieri}, \citenamefont {Landi}, \citenamefont {Clark},\ and\ \citenamefont {Goold}}]{Guarnieri2019}%
  \BibitemOpen
  \bibfield  {author} {\bibinfo {author} {\bibfnamefont {Giacomo}\ \bibnamefont {Guarnieri}}, \bibinfo {author} {\bibfnamefont {Gabriel~T.}\ \bibnamefont {Landi}}, \bibinfo {author} {\bibfnamefont {Stephen~R.}\ \bibnamefont {Clark}}, \ and\ \bibinfo {author} {\bibfnamefont {John}\ \bibnamefont {Goold}},\ }\bibfield  {title} {\enquote {\bibinfo {title} {Thermodynamics of precision in quantum nonequilibrium steady states},}\ }\href {\doibase 10.1103/PhysRevResearch.1.033021} {\bibfield  {journal} {\bibinfo  {journal} {Physical Review Research}\ }\textbf {\bibinfo {volume} {1}},\ \bibinfo {pages} {033021} (\bibinfo {year} {2019})}\BibitemShut {NoStop}%
\bibitem [{\citenamefont {Falasco}\ \emph {et~al.}(2020)\citenamefont {Falasco}, \citenamefont {Esposito},\ and\ \citenamefont {Delvenne}}]{Falasco2020}%
  \BibitemOpen
  \bibfield  {author} {\bibinfo {author} {\bibfnamefont {Gianmaria}\ \bibnamefont {Falasco}}, \bibinfo {author} {\bibfnamefont {Massimiliano}\ \bibnamefont {Esposito}}, \ and\ \bibinfo {author} {\bibfnamefont {Jean-Charles}\ \bibnamefont {Delvenne}},\ }\bibfield  {title} {\enquote {\bibinfo {title} {Unifying thermodynamic uncertainty relations},}\ }\href {\doibase 10.1088/1367-2630/ab8679} {\bibfield  {journal} {\bibinfo  {journal} {New Journal of Physics}\ }\textbf {\bibinfo {volume} {22}},\ \bibinfo {pages} {053046} (\bibinfo {year} {2020})}\BibitemShut {NoStop}%
\bibitem [{\citenamefont {Hasegawa}(2021)}]{Hasegawa2021}%
  \BibitemOpen
  \bibfield  {author} {\bibinfo {author} {\bibfnamefont {Yoshihiko}\ \bibnamefont {Hasegawa}},\ }\bibfield  {title} {\enquote {\bibinfo {title} {Thermodynamic uncertainty relation for general open quantum systems},}\ }\href {\doibase 10.1103/PhysRevLett.126.010602} {\bibfield  {journal} {\bibinfo  {journal} {Physical Review Letters}\ }\textbf {\bibinfo {volume} {126}},\ \bibinfo {pages} {010602} (\bibinfo {year} {2021})}\BibitemShut {NoStop}%
\bibitem [{\citenamefont {Saryal}\ \emph {et~al.}(2019)\citenamefont {Saryal}, \citenamefont {Friedman}, \citenamefont {Segal},\ and\ \citenamefont {Agarwalla}}]{Sushant_TUR_thermal}%
  \BibitemOpen
  \bibfield  {author} {\bibinfo {author} {\bibfnamefont {Sushant}\ \bibnamefont {Saryal}}, \bibinfo {author} {\bibfnamefont {Hava~Meira}\ \bibnamefont {Friedman}}, \bibinfo {author} {\bibfnamefont {Dvira}\ \bibnamefont {Segal}}, \ and\ \bibinfo {author} {\bibfnamefont {Bijay~Kumar}\ \bibnamefont {Agarwalla}},\ }\bibfield  {title} {\enquote {\bibinfo {title} {Thermodynamic uncertainty relation in thermal transport},}\ }\href {\doibase 10.1103/PhysRevE.100.042101} {\bibfield  {journal} {\bibinfo  {journal} {Physical Review E}\ }\textbf {\bibinfo {volume} {100}},\ \bibinfo {pages} {042101} (\bibinfo {year} {2019})}\BibitemShut {NoStop}%
\bibitem [{\citenamefont {Pal}\ \emph {et~al.}(2020)\citenamefont {Pal}, \citenamefont {Saryal}, \citenamefont {Segal}, \citenamefont {Mahesh},\ and\ \citenamefont {Agarwalla}}]{expt_TUR}%
  \BibitemOpen
  \bibfield  {author} {\bibinfo {author} {\bibfnamefont {Soham}\ \bibnamefont {Pal}}, \bibinfo {author} {\bibfnamefont {Sushant}\ \bibnamefont {Saryal}}, \bibinfo {author} {\bibfnamefont {Dvira}\ \bibnamefont {Segal}}, \bibinfo {author} {\bibfnamefont {T.~S.}\ \bibnamefont {Mahesh}}, \ and\ \bibinfo {author} {\bibfnamefont {Bijay~Kumar}\ \bibnamefont {Agarwalla}},\ }\bibfield  {title} {\enquote {\bibinfo {title} {Experimental study of the thermodynamic uncertainty relation},}\ }\href {\doibase 10.1103/PhysRevResearch.2.022044} {\bibfield  {journal} {\bibinfo  {journal} {Physical Review Research}\ }\textbf {\bibinfo {volume} {2}},\ \bibinfo {pages} {022044} (\bibinfo {year} {2020})}\BibitemShut {NoStop}%
\bibitem [{\citenamefont {Liu}\ and\ \citenamefont {Segal}(2021)}]{LS_PRE_2021}%
  \BibitemOpen
  \bibfield  {author} {\bibinfo {author} {\bibfnamefont {Junjie}\ \bibnamefont {Liu}}\ and\ \bibinfo {author} {\bibfnamefont {Dvira}\ \bibnamefont {Segal}},\ }\bibfield  {title} {\enquote {\bibinfo {title} {Coherences and the thermodynamic uncertainty relation: Insights from quantum absorption refrigerators},}\ }\href {\doibase 10.1103/PhysRevE.103.032138} {\bibfield  {journal} {\bibinfo  {journal} {Physical Review E}\ }\textbf {\bibinfo {volume} {103}},\ \bibinfo {pages} {032138} (\bibinfo {year} {2021})}\BibitemShut {NoStop}%
\bibitem [{\citenamefont {Friedman}\ and\ \citenamefont {Segal}(2019)}]{FS_PRE_2019}%
  \BibitemOpen
  \bibfield  {author} {\bibinfo {author} {\bibfnamefont {Hava~Meira}\ \bibnamefont {Friedman}}\ and\ \bibinfo {author} {\bibfnamefont {Dvira}\ \bibnamefont {Segal}},\ }\bibfield  {title} {\enquote {\bibinfo {title} {Cooling condition for multilevel quantum absorption refrigerators},}\ }\href {\doibase 10.1103/PhysRevE.100.062112} {\bibfield  {journal} {\bibinfo  {journal} {Physical Review E}\ }\textbf {\bibinfo {volume} {100}},\ \bibinfo {pages} {062112} (\bibinfo {year} {2019})}\BibitemShut {NoStop}%
\bibitem [{\citenamefont {Segal}(2018)}]{S_PRE_2018}%
  \BibitemOpen
  \bibfield  {author} {\bibinfo {author} {\bibfnamefont {Dvira}\ \bibnamefont {Segal}},\ }\bibfield  {title} {\enquote {\bibinfo {title} {Current fluctuations in quantum absorption refrigerators},}\ }\href {\doibase 10.1103/PhysRevE.97.052145} {\bibfield  {journal} {\bibinfo  {journal} {Physical Review E}\ }\textbf {\bibinfo {volume} {97}},\ \bibinfo {pages} {052145} (\bibinfo {year} {2018})}\BibitemShut {NoStop}%
\bibitem [{\citenamefont {Friedman}\ \emph {et~al.}(2018)\citenamefont {Friedman}, \citenamefont {Agarwalla},\ and\ \citenamefont {Segal}}]{Friedman_2018}%
  \BibitemOpen
  \bibfield  {author} {\bibinfo {author} {\bibfnamefont {Hava~Meira}\ \bibnamefont {Friedman}}, \bibinfo {author} {\bibfnamefont {Bijay~Kumar}\ \bibnamefont {Agarwalla}}, \ and\ \bibinfo {author} {\bibfnamefont {Dvira}\ \bibnamefont {Segal}},\ }\bibfield  {title} {\enquote {\bibinfo {title} {Quantum energy exchange and refrigeration: a full-counting statistics approach},}\ }\href {\doibase 10.1088/1367-2630/aad5fc} {\bibfield  {journal} {\bibinfo  {journal} {New Journal of Physics}\ }\textbf {\bibinfo {volume} {20}},\ \bibinfo {pages} {083026} (\bibinfo {year} {2018})}\BibitemShut {NoStop}%
\bibitem [{\citenamefont {Bera}\ \emph {et~al.}(2024)\citenamefont {Bera}, \citenamefont {Pandit}, \citenamefont {Chatterjee}, \citenamefont {Singh}, \citenamefont {Lewenstein}, \citenamefont {Bhattacharya},\ and\ \citenamefont {Bera}}]{Bera2024}%
  \BibitemOpen
  \bibfield  {author} {\bibinfo {author} {\bibfnamefont {Mohit~Lal}\ \bibnamefont {Bera}}, \bibinfo {author} {\bibfnamefont {Tanmoy}\ \bibnamefont {Pandit}}, \bibinfo {author} {\bibfnamefont {Kaustav}\ \bibnamefont {Chatterjee}}, \bibinfo {author} {\bibfnamefont {Varinder}\ \bibnamefont {Singh}}, \bibinfo {author} {\bibfnamefont {Maciej}\ \bibnamefont {Lewenstein}}, \bibinfo {author} {\bibfnamefont {Utso}\ \bibnamefont {Bhattacharya}}, \ and\ \bibinfo {author} {\bibfnamefont {Manabendra~Nath}\ \bibnamefont {Bera}},\ }\bibfield  {title} {\enquote {\bibinfo {title} {Steady-state quantum thermodynamics with synthetic negative temperatures},}\ }\href {\doibase 10.1103/PhysRevResearch.6.013318} {\bibfield  {journal} {\bibinfo  {journal} {Physical Review Research}\ }\textbf {\bibinfo {volume} {6}},\ \bibinfo {pages} {013318} (\bibinfo {year} {2024})}\BibitemShut {NoStop}%
\bibitem [{\citenamefont {Ramsey}(1956)}]{Ramsey1956}%
  \BibitemOpen
  \bibfield  {author} {\bibinfo {author} {\bibfnamefont {Norman~F.}\ \bibnamefont {Ramsey}},\ }\bibfield  {title} {\enquote {\bibinfo {title} {Thermodynamics and statistical mechanics at negative absolute temperatures},}\ }\href {\doibase 10.1103/PhysRev.103.20} {\bibfield  {journal} {\bibinfo  {journal} {Physical Review}\ }\textbf {\bibinfo {volume} {103}},\ \bibinfo {pages} {20} (\bibinfo {year} {1956})}\BibitemShut {NoStop}%
\bibitem [{\citenamefont {Aamir}\ \emph {et~al.}(2022)\citenamefont {Aamir}, \citenamefont {Moreno}, \citenamefont {Sundelin}, \citenamefont {Bizn\'arov\'a}, \citenamefont {Scigliuzzo}, \citenamefont {Patel}, \citenamefont {Osman}, \citenamefont {Lozano}, \citenamefont {Strandberg},\ and\ \citenamefont {Gasparinetti}}]{Aamir2022}%
  \BibitemOpen
  \bibfield  {author} {\bibinfo {author} {\bibfnamefont {Mohammed~Ali}\ \bibnamefont {Aamir}}, \bibinfo {author} {\bibfnamefont {Claudia~Castillo}\ \bibnamefont {Moreno}}, \bibinfo {author} {\bibfnamefont {Simon}\ \bibnamefont {Sundelin}}, \bibinfo {author} {\bibfnamefont {Janka}\ \bibnamefont {Bizn\'arov\'a}}, \bibinfo {author} {\bibfnamefont {Marco}\ \bibnamefont {Scigliuzzo}}, \bibinfo {author} {\bibfnamefont {Kowshik~Erappaji}\ \bibnamefont {Patel}}, \bibinfo {author} {\bibfnamefont {Amr}\ \bibnamefont {Osman}}, \bibinfo {author} {\bibfnamefont {D.~P.}\ \bibnamefont {Lozano}}, \bibinfo {author} {\bibfnamefont {Ingrid}\ \bibnamefont {Strandberg}}, \ and\ \bibinfo {author} {\bibfnamefont {Simone}\ \bibnamefont {Gasparinetti}},\ }\bibfield  {title} {\enquote {\bibinfo {title} {Engineering symmetry-selective couplings of a superconducting artificial molecule to microwave waveguides},}\ }\href {\doibase 10.1103/PhysRevLett.129.123604} {\bibfield  {journal} {\bibinfo  {journal} {Physical Review Letters}\
  }\textbf {\bibinfo {volume} {129}},\ \bibinfo {pages} {123604} (\bibinfo {year} {2022})}\BibitemShut {NoStop}%
\bibitem [{\citenamefont {Esposito}\ \emph {et~al.}(2009)\citenamefont {Esposito}, \citenamefont {Harbola},\ and\ \citenamefont {Mukamel}}]{Esposito2009}%
  \BibitemOpen
  \bibfield  {author} {\bibinfo {author} {\bibfnamefont {Massimiliano}\ \bibnamefont {Esposito}}, \bibinfo {author} {\bibfnamefont {Upendra}\ \bibnamefont {Harbola}}, \ and\ \bibinfo {author} {\bibfnamefont {Shaul}\ \bibnamefont {Mukamel}},\ }\bibfield  {title} {\enquote {\bibinfo {title} {Nonequilibrium fluctuations, fluctuation theorems, and counting statistics in quantum systems},}\ }\href {\doibase 10.1103/RevModPhys.81.1665} {\bibfield  {journal} {\bibinfo  {journal} {Reviews of Modern Physics}\ }\textbf {\bibinfo {volume} {81}},\ \bibinfo {pages} {1665} (\bibinfo {year} {2009})}\BibitemShut {NoStop}%
\bibitem [{\citenamefont {Bruderer}\ \emph {et~al.}(2014)\citenamefont {Bruderer}, \citenamefont {Contreras-Pulido}, \citenamefont {Thaller}, \citenamefont {Sironi}, \citenamefont {Obreschkow},\ and\ \citenamefont {Plenio}}]{Bruderer2014}%
  \BibitemOpen
  \bibfield  {author} {\bibinfo {author} {\bibfnamefont {M}~\bibnamefont {Bruderer}}, \bibinfo {author} {\bibfnamefont {L~D}\ \bibnamefont {Contreras-Pulido}}, \bibinfo {author} {\bibfnamefont {M}~\bibnamefont {Thaller}}, \bibinfo {author} {\bibfnamefont {L}~\bibnamefont {Sironi}}, \bibinfo {author} {\bibfnamefont {D}~\bibnamefont {Obreschkow}}, \ and\ \bibinfo {author} {\bibfnamefont {M~B}\ \bibnamefont {Plenio}},\ }\bibfield  {title} {\enquote {\bibinfo {title} {Inverse counting statistics for stochastic and open quantum systems: the characteristic polynomial approach},}\ }\href {\doibase 10.1088/1367-2630/16/3/033030} {\bibfield  {journal} {\bibinfo  {journal} {New Journal of Physics}\ }\textbf {\bibinfo {volume} {16}},\ \bibinfo {pages} {033030} (\bibinfo {year} {2014})}\BibitemShut {NoStop}%
\bibitem [{\citenamefont {Prech}\ \emph {et~al.}(2023)\citenamefont {Prech}, \citenamefont {Johansson}, \citenamefont {Nyholm}, \citenamefont {Landi}, \citenamefont {Verdozzi}, \citenamefont {Samuelsson},\ and\ \citenamefont {Potts}}]{Prech2023}%
  \BibitemOpen
  \bibfield  {author} {\bibinfo {author} {\bibfnamefont {Kacper}\ \bibnamefont {Prech}}, \bibinfo {author} {\bibfnamefont {Philip}\ \bibnamefont {Johansson}}, \bibinfo {author} {\bibfnamefont {Elias}\ \bibnamefont {Nyholm}}, \bibinfo {author} {\bibfnamefont {Gabriel~T.}\ \bibnamefont {Landi}}, \bibinfo {author} {\bibfnamefont {Claudio}\ \bibnamefont {Verdozzi}}, \bibinfo {author} {\bibfnamefont {Peter}\ \bibnamefont {Samuelsson}}, \ and\ \bibinfo {author} {\bibfnamefont {Patrick~P.}\ \bibnamefont {Potts}},\ }\bibfield  {title} {\enquote {\bibinfo {title} {Entanglement and thermokinetic uncertainty relations in coherent mesoscopic transport},}\ }\href {\doibase 10.1103/PhysRevResearch.5.023155} {\bibfield  {journal} {\bibinfo  {journal} {Physical Review Research}\ }\textbf {\bibinfo {volume} {5}},\ \bibinfo {pages} {023155} (\bibinfo {year} {2023})}\BibitemShut {NoStop}%
\bibitem [{\citenamefont {Landi}\ \emph {et~al.}(2024)\citenamefont {Landi}, \citenamefont {Kewming}, \citenamefont {Mitchison},\ and\ \citenamefont {Potts}}]{Landi2024}%
  \BibitemOpen
  \bibfield  {author} {\bibinfo {author} {\bibfnamefont {Gabriel~T.}\ \bibnamefont {Landi}}, \bibinfo {author} {\bibfnamefont {Michael~J.}\ \bibnamefont {Kewming}}, \bibinfo {author} {\bibfnamefont {Mark~T.}\ \bibnamefont {Mitchison}}, \ and\ \bibinfo {author} {\bibfnamefont {Patrick~P.}\ \bibnamefont {Potts}},\ }\bibfield  {title} {\enquote {\bibinfo {title} {Current fluctuations in open quantum systems: Bridging the gap between quantum continuous measurements and full counting statistics},}\ }\href {\doibase 10.1103/PRXQuantum.5.020201} {\bibfield  {journal} {\bibinfo  {journal} {PRX Quantum}\ }\textbf {\bibinfo {volume} {5}},\ \bibinfo {pages} {020201} (\bibinfo {year} {2024})}\BibitemShut {NoStop}%
\bibitem [{\citenamefont {Agarwalla}\ \emph {et~al.}(2015)\citenamefont {Agarwalla}, \citenamefont {Jiang},\ and\ \citenamefont {Segal}}]{Agarwalla2015}%
  \BibitemOpen
  \bibfield  {author} {\bibinfo {author} {\bibfnamefont {Bijay~Kumar}\ \bibnamefont {Agarwalla}}, \bibinfo {author} {\bibfnamefont {Jian-Hua}\ \bibnamefont {Jiang}}, \ and\ \bibinfo {author} {\bibfnamefont {Dvira}\ \bibnamefont {Segal}},\ }\bibfield  {title} {\enquote {\bibinfo {title} {Full counting statistics of vibrationally-assisted electronic conduction: Transport and fluctuations of the thermoelectric efficiency},}\ }\href {\doibase 10.1103/PhysRevB.92.245418} {\bibfield  {journal} {\bibinfo  {journal} {Physical Review B}\ }\textbf {\bibinfo {volume} {92}},\ \bibinfo {pages} {245418} (\bibinfo {year} {2015})}\BibitemShut {NoStop}%
\bibitem [{\citenamefont {Cohen-Tannoudji}\ \emph {et~al.}(1998)\citenamefont {Cohen-Tannoudji}, \citenamefont {Dupont-Roc},\ and\ \citenamefont {Grynberg}}]{Cohen1998}%
  \BibitemOpen
  \bibfield  {author} {\bibinfo {author} {\bibfnamefont {Claude}\ \bibnamefont {Cohen-Tannoudji}}, \bibinfo {author} {\bibfnamefont {Jacques}\ \bibnamefont {Dupont-Roc}}, \ and\ \bibinfo {author} {\bibfnamefont {Gilbert}\ \bibnamefont {Grynberg}},\ }\href@noop {} {\emph {\bibinfo {title} {Atom-photon interactions: basic processes and applications}}}\ (\bibinfo  {publisher} {John Wiley \& Sons},\ \bibinfo {year} {1998})\BibitemShut {NoStop}%
\bibitem [{\citenamefont {Haroche}\ and\ \citenamefont {Raimond}(2006)}]{Haroche2006}%
  \BibitemOpen
  \bibfield  {author} {\bibinfo {author} {\bibfnamefont {Serge}\ \bibnamefont {Haroche}}\ and\ \bibinfo {author} {\bibfnamefont {J-M}\ \bibnamefont {Raimond}},\ }\href@noop {} {\emph {\bibinfo {title} {Exploring the quantum: atoms, cavities, and photons}}}\ (\bibinfo  {publisher} {Oxford University Press},\ \bibinfo {year} {2006})\BibitemShut {NoStop}%
\bibitem [{\citenamefont {Meystre}\ and\ \citenamefont {Scully}(2021)}]{Meystre2021}%
  \BibitemOpen
  \bibfield  {author} {\bibinfo {author} {\bibfnamefont {Pierre}\ \bibnamefont {Meystre}}\ and\ \bibinfo {author} {\bibfnamefont {Marlan~O}\ \bibnamefont {Scully}},\ }\href@noop {} {\emph {\bibinfo {title} {Quantum optics}}}\ (\bibinfo  {publisher} {Springer},\ \bibinfo {year} {2021})\BibitemShut {NoStop}%
\bibitem [{\citenamefont {Larson}\ and\ \citenamefont {Mavrogordatos}(2021)}]{Larson2021}%
  \BibitemOpen
  \bibfield  {author} {\bibinfo {author} {\bibfnamefont {Jonas}\ \bibnamefont {Larson}}\ and\ \bibinfo {author} {\bibfnamefont {Themistoklis}\ \bibnamefont {Mavrogordatos}},\ }\href@noop {} {\emph {\bibinfo {title} {The Jaynes--Cummings model and its descendants: modern research directions}}}\ (\bibinfo  {publisher} {IoP Publishing},\ \bibinfo {year} {2021})\BibitemShut {NoStop}%
\bibitem [{\citenamefont {Zanon}\ \emph {et~al.}(2005)\citenamefont {Zanon}, \citenamefont {Guerandel}, \citenamefont {de~Clercq}, \citenamefont {Holleville}, \citenamefont {Dimarcq},\ and\ \citenamefont {Clairon}}]{Zanon2005}%
  \BibitemOpen
  \bibfield  {author} {\bibinfo {author} {\bibfnamefont {T.}~\bibnamefont {Zanon}}, \bibinfo {author} {\bibfnamefont {S.}~\bibnamefont {Guerandel}}, \bibinfo {author} {\bibfnamefont {E.}~\bibnamefont {de~Clercq}}, \bibinfo {author} {\bibfnamefont {D.}~\bibnamefont {Holleville}}, \bibinfo {author} {\bibfnamefont {N.}~\bibnamefont {Dimarcq}}, \ and\ \bibinfo {author} {\bibfnamefont {A.}~\bibnamefont {Clairon}},\ }\bibfield  {title} {\enquote {\bibinfo {title} {High contrast ramsey fringes with coherent-population-trapping pulses in a double lambda atomic system},}\ }\href {\doibase 10.1103/PhysRevLett.94.193002} {\bibfield  {journal} {\bibinfo  {journal} {Physical Review Letters}\ }\textbf {\bibinfo {volume} {94}},\ \bibinfo {pages} {193002} (\bibinfo {year} {2005})}\BibitemShut {NoStop}%
\bibitem [{\citenamefont {Chang}\ \emph {et~al.}(2023)\citenamefont {Chang}, \citenamefont {Huang}, \citenamefont {Xu}, \citenamefont {Kan}, \citenamefont {Huang}, \citenamefont {Huang}, \citenamefont {Fu}, \citenamefont {Liu},\ and\ \citenamefont {Kuan}}]{Chang2023}%
  \BibitemOpen
  \bibfield  {author} {\bibinfo {author} {\bibfnamefont {Jia-Wei}\ \bibnamefont {Chang}}, \bibinfo {author} {\bibfnamefont {Yu-Chieh}\ \bibnamefont {Huang}}, \bibinfo {author} {\bibfnamefont {Zhen-Wei}\ \bibnamefont {Xu}}, \bibinfo {author} {\bibfnamefont {Yuan-Hung}\ \bibnamefont {Kan}}, \bibinfo {author} {\bibfnamefont {Hsing-Han}\ \bibnamefont {Huang}}, \bibinfo {author} {\bibfnamefont {Jia-Zhun}\ \bibnamefont {Huang}}, \bibinfo {author} {\bibfnamefont {Kai-Jun}\ \bibnamefont {Fu}}, \bibinfo {author} {\bibfnamefont {Xu-Dong}\ \bibnamefont {Liu}}, \ and\ \bibinfo {author} {\bibfnamefont {Pei-Chen}\ \bibnamefont {Kuan}},\ }\bibfield  {title} {\enquote {\bibinfo {title} {Multiphoton hyperfine raman transitions with different quantization axes},}\ }\href {\doibase 10.1103/PhysRevA.108.013308} {\bibfield  {journal} {\bibinfo  {journal} {Physical Review A}\ }\textbf {\bibinfo {volume} {108}},\ \bibinfo {pages} {013308} (\bibinfo {year} {2023})}\BibitemShut {NoStop}%
\bibitem [{\citenamefont {Novikov}\ \emph {et~al.}(2016)\citenamefont {Novikov}, \citenamefont {Sweeney}, \citenamefont {Robinson}, \citenamefont {Premaratne}, \citenamefont {Suri}, \citenamefont {Wellstood},\ and\ \citenamefont {Palmer}}]{Novikov2016}%
  \BibitemOpen
  \bibfield  {author} {\bibinfo {author} {\bibfnamefont {S.}~\bibnamefont {Novikov}}, \bibinfo {author} {\bibfnamefont {T.}~\bibnamefont {Sweeney}}, \bibinfo {author} {\bibfnamefont {J.~E.}\ \bibnamefont {Robinson}}, \bibinfo {author} {\bibfnamefont {S.~P.}\ \bibnamefont {Premaratne}}, \bibinfo {author} {\bibfnamefont {B.}~\bibnamefont {Suri}}, \bibinfo {author} {\bibfnamefont {F.~C.}\ \bibnamefont {Wellstood}}, \ and\ \bibinfo {author} {\bibfnamefont {B.~S.}\ \bibnamefont {Palmer}},\ }\bibfield  {title} {\enquote {\bibinfo {title} {Raman coherence in a circuit quantum electrodynamics lambda system},}\ }\href {\doibase 10.1038/nphys3537} {\bibfield  {journal} {\bibinfo  {journal} {Nature Physics}\ }\textbf {\bibinfo {volume} {12}},\ \bibinfo {pages} {75} (\bibinfo {year} {2016})}\BibitemShut {NoStop}%
\bibitem [{\citenamefont {Kumar}\ \emph {et~al.}(2016)\citenamefont {Kumar}, \citenamefont {Veps{\"a}l{\"a}inen}, \citenamefont {Danilin},\ and\ \citenamefont {Paraoanu}}]{Kumar2016}%
  \BibitemOpen
  \bibfield  {author} {\bibinfo {author} {\bibfnamefont {K.~S.}\ \bibnamefont {Kumar}}, \bibinfo {author} {\bibfnamefont {A.}~\bibnamefont {Veps{\"a}l{\"a}inen}}, \bibinfo {author} {\bibfnamefont {S.}~\bibnamefont {Danilin}}, \ and\ \bibinfo {author} {\bibfnamefont {G.~S.}\ \bibnamefont {Paraoanu}},\ }\bibfield  {title} {\enquote {\bibinfo {title} {Stimulated raman adiabatic passage in a three-level superconducting circuit},}\ }\href {\doibase 10.1038/ncomms10628} {\bibfield  {journal} {\bibinfo  {journal} {Nature Communications}\ }\textbf {\bibinfo {volume} {7}},\ \bibinfo {pages} {10628} (\bibinfo {year} {2016})}\BibitemShut {NoStop}%
\end{thebibliography}%

\end{document}